\documentclass[preprint,11pt,3p,times]{elsarticle}
	
\usepackage[normalem]{ulem}

\usepackage{amssymb}
\usepackage{amsmath}

\usepackage[allcolors=blue,colorlinks=true,linkcolor=blue,menucolor=blue,urlcolor=blue,bookmarksnumbered]{hyperref}

\usepackage{subfig}
\usepackage{booktabs}
\usepackage{float}
\usepackage{nicefrac}
\usepackage{siunitx}
\DeclareSIUnit\cpm{cpm}
\DeclareSIUnit\mmHg{mmHg}

\usepackage{tikz}
\usetikzlibrary{shapes.geometric, positioning}
\usetikzlibrary{calc}
\definecolor{myred}{RGB}{220,120,110}

\usepackage[noabbrev,capitalise]{cleveref}
\crefname{equation}{Eq.}{Eqs.} 
\Crefname{equation}{Eq.}{Eqs.}  

\renewcommand{\vec}[1]{\boldsymbol{#1}}
\newcommand{\mat}[1]{\boldsymbol{#1}}
\newcommand{\dd}{\mathrm{d}}
\newcommand{\pd}{\partial}
\newcommand{\Dprod}{:}
\newcommand{\pfrac}[2]{\frac{\pd #1}{\pd #2}}
\renewcommand{\dfrac}[2]{\frac{\dd #1}{\dd #2}}
\newcommand{\norm}[1]{\| #1 \|}
\newcommand{\abs}[1]{| #1 |}

\newcommand{\defgrad}{\mat{F}}
\newcommand{\rightCG}{\mat{C}}
\newcommand{\fPK}{\mat{P}}

\newcommand{\ca}{\ensuremath{Ca^{2+}}}

\usepackage[acronyms,nopostdot,section=section,shortcuts]{glossaries}
\glsdisablehyper

\newacronym[longplural={interstitial cells of Cajal}]{ac:icc}{ICC}{interstitial cell of Cajal}
\newcommand{\icc}{\gls{ac:icc}}   
\newcommand{\iccs}{\glspl{ac:icc}}    
\newacronym[longplural={smooth muscle cells}]{ac:smc}{SMC}{smooth muscle cell}
\newcommand{\smc}{\gls{ac:smc}}
\newcommand{\smcs}{\glspl{ac:smc}}
\newacronym{ac:vdcc}{VDCC}{voltage-dependent calcium channel}
\newcommand{\vdcc}{\gls{ac:vdcc}}
\newacronym{ac:mri}{MRI}{magnetic resonance imaging}
\newcommand{\mri}{\gls{ac:mri}}
\newacronym{ac:eas}{EAS}{enhanced assumed strain}
\newcommand{\eas}{\gls{ac:eas}}
\newacronym{ac:ans}{ANS}{assumed natural strain}
\newcommand{\ans}{\gls{ac:ans}}
\newcommand{\ICC}{\textup{icc}}
\newcommand{\SMC}{\textup{smc}}
\newcommand{\fp}{\textup{fp}}
\newcommand{\ep}{\textup{ep}}
\newcommand{\gl}{\textup{gl}}
\newcommand{\cf}{\textup{c}}
\newcommand{\lf}{\textup{l}}
\newcommand{\gm}{\textup{gm}}

\makeatletter
\def\ps@pprintTitle{%
 \let\@oddhead\@empty
 \let\@evenhead\@empty
 \def\@oddfoot{}%
 \let\@evenfoot\@oddfoot}
\makeatother

\begin{document}

\begin{frontmatter}

\title{Electromechanical computational model of the human stomach}

\author[1]{Maire S. Henke\fnref{label1}}
\author[2]{Sebastian Brandstaeter\corref{cor1}\fnref{label1}}\ead{sebastian.brandstaeter@unibw.de}
\author[1,3]{Sebastian L. Fuchs}
\author[1,4]{Roland C. Aydin}
\author[5]{Alessio Gizzi}
\author[1,4]{Christian J. Cyron\corref{cor1}}\ead{christian.cyron@tuhh.de}

\fntext[label1]{These authors contributed equally to this work.}
\cortext[cor1]{Corresponding authors}

%% Author affiliation
\affiliation[1]{organization={Institute for Continuum and Material Mechanics, Hamburg University of Technology},
            addressline={Eißendorfer Straße 42},
            city={Hamburg},
            postcode={21073},
            %state={},
            country={Germany}}
            
\affiliation[2]{organization={Institute for Mathematics and Computer-Based Simulation, University of the Bundeswehr Munich},
            addressline={Werner-Heisenberg-Weg 39},
            city={Neubiberg},
            postcode={85577},
            %state={},
            country={Germany}}         
            
\affiliation[3]{organization={Institute for Computational Mechanics, Technical University of Munich},
            addressline={Boltzmannstraße 15},
            city={Garching},
            postcode={85748},
            %state={},
            country={Germany}}         
            
\affiliation[4]{organization={Institute of Material Systems Modeling, Helmholtz-Zentrum Hereon},
            addressline={Max–Planck-Straße 1},
            city={Geesthacht},
            postcode={21502},
            %state={x},
            country={Germany}}   
            
\affiliation[5]{organization={
Theoretical and Computational Biomechanics Lab,
Department of Engineering, Università Campus Bio-Medico di Roma},
            addressline={Via A. del Portillo 21},
            city={Rome},
            postcode={00128},
            %state={},
            country={Italy}}

\begin{abstract}
The stomach plays a central role in digestion through coordinated muscle contractions, known as gastric peristalsis, driven by slow-wave electrophysiology.
Understanding this process is critical for treating motility disorders such as gastroparesis, dyspepsia, and gastroesophageal reflux disease.
Computer simulations can be a valuable tool to deepen our understanding of these disorders and help to develop new therapies. However, existing approaches often neglect spatial heterogeneity, fail to capture large anisotropic deformations, or rely on computationally expensive three-dimensional formulations. We present here a computational framework of human gastric electromechanics, that combines a nonlinear, rotation-free shell formulation with a constrained mixture material model. The formulation incorporates active-strain, constituent-specific prestress, and spatially non-uniform parameter fields. Numerical examples demonstrate that the framework can reproduce characteristic features of gastric motility, including slow-wave entrainment, conduction velocity gradients, and large peristaltic contractions with physiologically realistic amplitudes. 
The proposed framework enables robust electromechanical simulations of the whole stomach at the organ scale. It thus provides a promising basis for future in silico studies of both physiological function and pathological motility disorders.
\end{abstract}


\end{frontmatter}

\section{Introduction}
\label{sec:Introduction}
The stomach plays a central role in the gastrointestinal tract. It is concerned with storing, mixing, and grinding food. Fundamental to this functionality are rhythmic gastric contractions, which generate peristaltic waves that mix and propel food through the digestive tract. Perturbations of this system are associated with a broad spectrum of diseases and disorders, including obesity, which itself is a major risk factor for conditions such as cardiovascular disease. Other gastric motility disorders include gastroparesis, gastroesophageal reflux disease (GERD), and functional dyspepsia, all of which are linked to abnormal gastric motility~\cite{Streutker-2007-ICCHealthDisease,Grady-2021-GastricConductionReview}.
A detailed understanding of gastric motility is therefore essential for the effective diagnosis and treatment.

\begin{figure}[t!]
    \centering
    \includegraphics[width=0.9\linewidth]{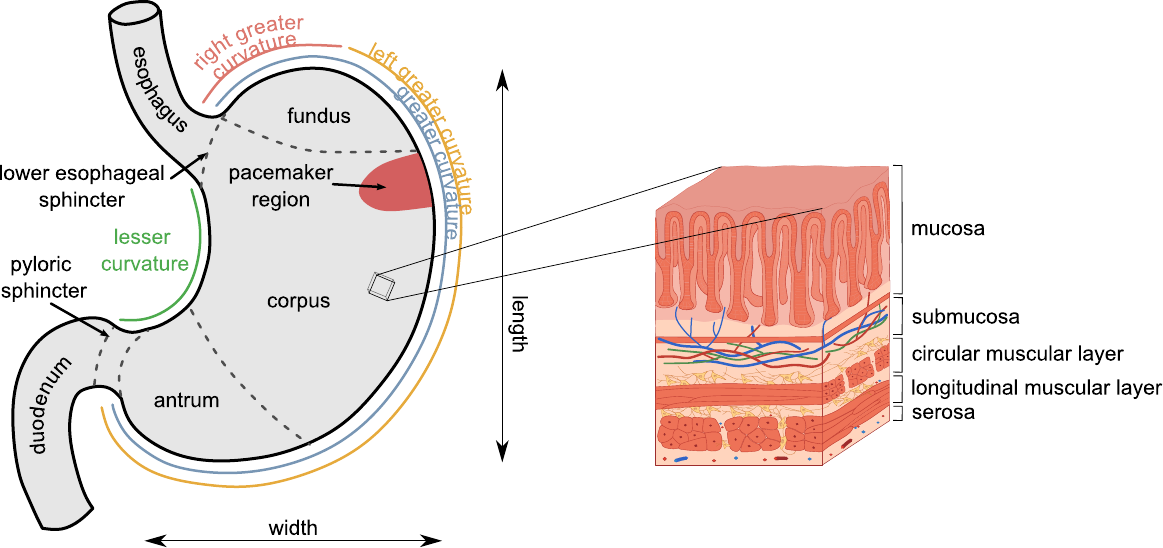}
        \caption{Anatomy and microstructure of the stomach. Left: Schematic  of the stomach showing key anatomical regions. Right: Cross-sectional view of the gastric wall highlighting its layered structure and the orientation of the circular and longitudinal smooth muscle fibers. This organization informs the fiber architecture in the electromechanical model.}     \label{fig:stomach_anatomics}
\end{figure}

Anatomically, the stomach is divided into three main regions (fundus, corpus, antrum) and exhibits both a so-called greater and lesser curvature as illustrated in~\cref{fig:stomach_anatomics}. Despite its relatively thin wall (on average \SIrange[]{2.8}{4.5}{\milli\meter} in humans~\cite{Friis-2023-GastricTissue,Holzer-Stock2025}) compared to its overall size (approximately \SIrange[]{15}{25}{\centi\meter} in length~\cite{Friis-2023-GastricTissue,Holzer-Stock2025}), the gastric wall consists of distinct layers. The muscular layer, primarily responsible for gastric contractions, comprises two smooth muscle sheets: one oriented circumferentially and the other longitudinally. An additional oblique smooth muscle layer is restricted to the upper corpus near the esophagus. It is neglected in this study, as it plays a limited role in gastric peristalsis, which occurs mainly in the lower body and antrum~\cite{DiNatale-2023-FunctionalAndAnatomicalRegions}. 

The stomach exhibits intricate patterns of mechanical contraction~\cite{DiNatale-2023-FunctionalAndAnatomicalRegions}. Gastric motility is governed by complex electromechanical couplings comprising at least two cell types, \iccs{} and \smcs{}~\cite{huizinga2009a}. \iccs{}, arranged in a distributed network, act as pacemaker cells generating electrical activity in the form of slow waves~\cite{Sanders-2016-RegulationSMCFunction}. Being self-excitatory, \iccs{} show peculiar features of nonlinear coupled oscillators~\cite{Winfree1987,Glass2001}; i.e., they generate electrical activity at specific intrinsic frequencies while the integrated network synchronizes at a unified emerging frequency. This so-called entrainment generates ordered waves at the highest intrinsic frequency of approximately \SI{3}{\cpm} (cycles per minute).
The highest intrinsic frequency is observed near the pacemaker region at the mid-corpus along the greater curvature. The intrinsic frequency decreases longitudinally and circumferentially and assumes its minimum near the pylorus at the lesser curvature~\cite{Grady-2021-GastricConductionReview}. 
The \smcs{} are electrically coupled to \iccs{}, such that the electrical slow waves coordinate the contractions of the \smcs{} via an elaborate excitation--contraction mechanism~\cite{CorriasBuist-2007-SMCActivation,sanders2012b}.
These contractions can produce substantial gastric wall deformations, with amplitudes measured in the range of \SIrange[]{7}{20}{\mm} \cite{Schulze-2006-ImagingStomach,Hosseini2023,Wang2024}. 

Despite its physiological importance, the biomechanics of the gastric motility remains poorly understood, both in terms of the underlying biophysical mechanisms and their theoretical and computational modeling. The latter, integrated with medical imaging, has emerged as a powerful, non-invasive tool for investigating biomechanical systems. Robust, physiologically accurate whole-organ electromechanical models hold great promise for advancing our understanding of gastric (dys-)function; however, such models remain scarce in the literature, and numerous critical computational challenges persist.

Early whole-organ modeling approaches of the stomach focused primarily on its electrophysiological system~\cite{du2013c}.
More recently, models incorporating mechanical phenomena have also been proposed.
Our previous work~\cite{brandstaeter2018a} was the first to simulate gastric electromechanics on an idealized whole-organ geometry, reproducing key phenomena. However, it relied on a highly simplified geometry and material model of the tissue and could deal only with very limited contraction amplitudes due to reduced numerical robustness that resulted from the use of membrane finite elements to discretize the thin wall of the stomach.
Subsequent developments have advanced mechanical modeling toward anatomically realistic stomach geometries. \cite{Fontanella2019} examined gastric deformation in animal-specific models but were restricted to passive tissue mechanics. Similarly,~\cite{Toniolo2022a,Toniolo2024} applied a passive mechanical framework to analyze patient-specific bariatric surgery, precluding the simulation of active motility.
In contrast,~\cite{Klemm2020a,Klemm2023} proposed an electro-chemomechanical model for the porcine stomach, using an active-stress formulation and incorporating mechano-electric feedback to capture propagating contraction waves. While their approach was particularly advanced in terms of the electro-chemical components, several limitations remained such as an unrealistic antrum activation and mild contraction intensity.

The thin-walled structure of the stomach lends itself naturally to reduced-dimensional formulations, such as membrane or shell models. While membrane models are computationally efficient, they neglect bending stiffness, limiting their application in regions with high curvature and large contractions. Nonlinear shell formulations capture both membrane and bending effects under large deformations and are thus preferable for accurate gastric biomechanics. Shell models have proven effective for studying the biomechanics of thin-walled structures like arteries. For example, \cite{Laubrie2020} introduced a layer-specific axisymmetric shell formulation within a constrained mixture model to study arterial growth and remodeling. \cite{Nitti2021} used a Kirchhoff–Love shell in an isogeometric framework for cardiac atrial electromechanics, leveraging assumptions of inextensible directors and negligible transverse shear. \cite{Nama2020,Nama2023} developed a nonlinear, rotation-free shell formulation for vascular biomechanics based on the Kirchhoff–Love theory, utilizing midsurface displacement and a director vector to capture thickness stretch, which is statically condensed to enforce incompressibility via a plane stress condition. \cite{Nama2023} also compared their shell formulation with membrane and solid elements in vascular bifurcations, demonstrating that the shell approach offers an optimal balance between computational cost and accuracy. 

However, these cardiovascular findings cannot be directly translated to gastric tissue. 
Although the stomach wall is thicker than vascular tissues---where shell models have been successfully applied---it remains thinner than the ventricular wall of the heart, which typically requires solid finite element formulations. Cardiac ventricular models must capture thickness- and sheet-dependent features such as fiber rotational anisotropy, necessitating advanced, structure-based modeling frameworks~\cite{Holzapfel-2009-ConstitutiveMyocard,Eriksson2013,Sommer2015,Gltekin2016}. The substantial wall thickness of the cardiac chambers, particularly the left ventricle, demands solid finite element to achieve physiologically accurate and mechanically reliable simulations~\cite{Land2015,Viola2020,Brown2025,Aróstica-2025-CardiacBenchmark}.

In summary, the persisting challenges in whole-organ coupled electromechanical modeling of the human stomach include the treatment of large nonlinear deformations, the treatment of the multiple spatially heterogeneous tissue properties, and the accommodation of substantial inter-individual variability.
These variabilities encompass patient-specific geometries and complex, heterogeneous spatial propagation and motility patterns~\cite{EgbujiGrady-2010-OriginPropagationSW}. 

Therefore, addressing these challenges requires a robust yet flexible description of: 
(1) the anatomical location and functional distribution of pacemaking regions, 
(2) the amplitude and velocity of slow waves across different subregions, and 
(3) the coupling of a spatiotemporal reaction-diffusion process with a fully nonlinear finite elasticity framework, capable for both geometric and material nonlinearities to capture very large deformations within complex computational domains.

As in most biomechanical applications, the intrinsic multiscale nature of the system requires ad hoc simplifying assumptions, which allow the construction of a reliable computational framework to advance understanding of a specific subproblem.
Accordingly, in this paper, we focus on the electromechanical modeling and computational implementation of gastric motility, while disregarding fluid-structure interaction.
Building on our previous work~\cite{brandstaeter2018a}, which introduced a preliminary electromechanical framework limited to basic material models, simplified geometries, and reduced contractions, we substantially extend the approach in the present study. Here, we generalize the formulation within the framework of constrained mixture theory~\cite{Humphrey-2021-CMMOverview}, incorporate an anisotropic active strain model~\cite{Brandstaeter2019,Patel-Gizzi-2022-GastrointestinalTissueReview}, and integrate spatially varying electrophysiological and mechanical fields derived automatically from the organ geometry.  
To address the challenges posed by large deformations and nonlinear material behavior in the gastric wall, we transition from the previously employed membrane-based formulation to a $7$-parameter shell model~\cite{Braun-1994-Shell, Büchter-1994-3DExtensionEAS}, which captures thickness stretch and transverse shear. Additionally, we introduce a whole-organ gastric motility model that accounts for heterogeneous parameter fields. These fields encode regional variations in key physiological features, including intrinsic slow wave frequency, electrical conductivity, tissue stiffness, and active contractility, across the full gastric domain. Unlike previous approaches that rely on discrete anatomical subregions or piecewise-constant parameters, our framework defines smooth parameter gradients informed by anatomy and physiology. This enables realistic simulation of entrainment, wave propagation, and peristaltic contractions, while preserving continuity and anatomical fidelity throughout the domain.

We demonstrate the capabilities of our model by simulating a realistic human stomach geometry, incorporating prestress, regionally varying material properties, and non-uniform boundary conditions. The study highlights the improved performance of the $7$-parameter shell compared to membrane and solid element formulations through a rigorous computational comparison of electromechanical motility and demonstrates how regionally varying heterogeneous fields enable spatially accurate simulation of contraction patterns.
We provide an open-source implementation of our formulation within the multiphysics simulation framework 4C~\cite{4C}.

The remainder of this paper is organized as follows. 
\cref{sec:electromechanical_modeling_of_gastric_peristalsis} introduces the electromechanical model, comprising the $7$-parameter shell formulation, the constrained mixture material model with prestress and active strain, the electrophysiology model, the non-uniform parameter distributions, and the applied boundary and initial conditions. 
\cref{sec:numerical_implementation} discusses numerical aspects, including convergence studies and a comparison of  membrane, shell, and solid element formulations.
\cref{sec:examples} presents numerical examples, examining the physiological coordination and activation intensities of longitudinal and circular muscle layers and demonstrating whole-organ simulations of gastric motility. 
Finally,~\cref{sec:conclusions} summarizes conclusions, discusses limitations, and outlines directions for future work.

%%%%%%%%%%%%%%%%%%%%%%%%%%%%%%%%%%%%%%%%%%%%%%%%%%%%%%%%%%%
\section{Electromechanical model}
\label{sec:electromechanical_modeling_of_gastric_peristalsis}
Herein, we indicate the scalar product with $(\cdot)$, the cross product with $(\times)$, the double contraction with $(\Dprod)$, and the dyadic product with $(\otimes)$. Additionally, $(\nabla)$ and $(\Delta=\nabla^2)$ represent the Nabla and Laplace operators, respectively. The determinant is denoted as $(\det(\cdot))$. Einstein index notation is adopted along with covariant and contravariant basis functions.

We start by setting the mathematical and mechanical framework which is based on the general theory of nonlinear continuum mechanics (e.g.~\cite{Holzapfel-2001-NonlinearSolidMechanics}).
A material point $\vec{X}$ in the reference configuration $\Omega_\textup{R}$ is mapped to a spatial position $\mat{x}$ in the current configuration $\Omega_\textup{t}$ at time $t$ via
\begin{align}
    \vec{x}: \Omega_\textup{R} \times [0,\infty) \rightarrow \Omega_\textup{t} ; \,\, (\vec{X}, t) \mapsto \vec{x}(\vec{X},t)\, .
\end{align}
This results for each point in a displacement
\begin{align}
   \vec{u}=\vec{x}- \vec{X}\, .
\end{align}
The deformation gradient
\begin{align}
   \defgrad =\pfrac{\vec{x}}{\vec{X}}
\end{align}
characterizes the local deformation of the material. If diagonalized, its diagonal elements represent the principal material stretches in the three principal stretch direction.

\subsection{Structural model: nonlinear 7-parameter shell}
\label{sec:shell_formulation}
\newcommand{\ida}{k}
\newcommand{\idb}{l}
\newcommand{\idc}{m}
\newcommand{\idd}{n}
\newcommand{\Xmid}{\overline{\vec{X}}}
\newcommand{\xmid}{\overline{\vec{x}}}
The stomach is a thin-walled, curved structure (see~\cref{fig:stomach_anatomics}), which aligns with the core assumption of shell formulations: one dimension (the thickness) is much smaller than the other two (in-plane) dimensions.
Based on this assumption, we briefly summarize the mathematical formulation of the $7$-parameter shell originally proposed in~\cite{Braun-1994-Shell, Büchter-1994-3DExtensionEAS}.

\subsubsection{Geometry}
A shell can effectively be described by a two-dimensional ($2$D) manifold---the so-called midsurface---embedded in the three-dimensional ($3$D) space.
The shell midsurface is parameterized by a set of curvilinear coordinates $\{\theta^1,\theta^2\} \in [-1,1]$, where $\theta^1$ and $\theta^2$ represent the in-plane directions of the midsurface.
The out-of-plane (thickness) direction is denoted by $\theta^3 \in [-1,1]$. 
The vectors $\Xmid$ and $\xmid$ denote the position vectors of the midsurface projection (i.e., at $\theta^3=0$) of a material point in the reference and current configurations, respectively.
The covariant basis vectors on the shell midsurface ($\theta^3=0$) are defined as:
\begin{align}
    \vec{A}_\ida = \pfrac{\Xmid(\theta^1,\theta^2)}{\theta^\ida}, \qquad \vec{a}_\ida = \pfrac{\xmid(\theta^1,\theta^2)}{\theta^\ida}, \qquad \ida=1,2 \, .
\end{align}
The shell director vector in the reference configuration is perpendicular to $\mat{A}_1$ and $\mat{A}_2$, and is normalized with the initial half-thickness of the shell
\begin{align}
    \vec{A}_3 = \frac{H}{2}\frac{\vec{A}_1\times\vec{A}_2}{\abs{\vec{A}_1\times\vec{A}_2}}\, .
\end{align}
Assuming a linear variation of the displacements across the thickness, the position vectors of an arbitrary point within the shell body in the reference and current configurations reads
\begin{align}
    \vec{X}=\Xmid(\theta^1,\theta^2)+\theta^3\vec{A}_3 \quad \text{and} \quad
    \vec{x}=\xmid(\theta^1,\theta^2)+\theta^3\vec{a}_3\, ,
\end{align}
where $\vec{a}_3$ is the shell director vector in the current configuration. With the previous definitions, the basis vectors of a point within the shell body are defined as
\begin{align}
    \vec{G}_\ida = \pfrac{\vec{X}}{\theta^\ida}, \quad \ida=1,2 \quad \text{and} \quad \vec{G}_3=\vec{A}_3 \, ,
    \\
    \vec{g}_\ida = \pfrac{\vec{x}}{\theta^\ida}, \quad \ida=1,2 \quad \text{and} \quad \vec{g}_3=\vec{a}_3 \, .
\end{align}
The covariant and contravariant basis vectors in the reference ($\vec{G}^\ida$) and in the current ($\vec{g}^\ida$) configurations satisfy the orthonormality relations $\vec{G}_\ida \cdot \vec{G}^\idb=\delta_\ida^\idb$ and $\vec{g}_\ida \cdot \vec{g}^\idb=\delta_\ida^\idb$, where $\delta_\ida^\idb$ denotes the Kronecker delta.

\subsubsection{Kinematics}
The displacement field $\vec{u}$ of a material point is defined in terms of quantities referred to the midsurface as follows:
\begin{align}
   \vec{u}=\vec{x}- \vec{X}=\xmid- \Xmid + \theta^3(\vec{a}_3- \vec{A}_3) =\vec{v} + \theta^3 \vec{w}\, ,
\end{align}
with $\vec{v}$ representing the displacement of the shell midsurface and $\vec{w}$ the difference vector of the shell directors between the reference and deformed configurations.

The deformation gradient associated with the motion is defined as $\defgrad =\partial \vec{x}/ \partial \vec{X}=\mat{g}_\ida \otimes \mat{G}^\ida$, and its Jacobian is $J=\textup{det}\defgrad>0$.
Based on the deformation gradient, one can define the so-called right Cauchy-Green deformation tensor $\rightCG=\defgrad^T\defgrad$, and the material Green-Lagrange strain tensor as $\vec{E}^u=\nicefrac{1}{2}(\rightCG-\vec{I})=E_{\ida\idb}^u\vec{G}^\ida \otimes\vec{G}^\idb$,
where $\vec{I}$ is the second order identity tensor, and $G_{\ida\idb}=\vec{G}_\ida\cdot\vec{G}_\idb$ and $g_{\ida\idb}=\vec{g}_\ida\cdot\vec{g}_\idb$ identify the metric coefficients of the reference and current configurations, respectively. 

According to the \eas{} method~\cite{SimoRifai-1990-EAS,Büchter-1994-3DExtensionEAS}, an independent strain field $\vec{\tilde{E}}$ is introduced to avoid artificial stresses in the thickness direction, resulting in the total Green-Lagrange strain tensor $\vec{E}=\vec{E}^u+\vec{\tilde{E}}$. The additional strains affect only the transversal normal strains, i.e.,
\begin{align}
    E_{\ida\idb}&=E_{\ida\idb}^u\, , \qquad (\ida, \idb) \neq (3,3), \\
    E_{33}&=E_{33}^u+\tilde{E}_{33}=\vec{A}_3 \cdot \vec{w}+ \frac{1}{2} \vec{w}\cdot \vec{w}+\theta^3\tilde{\beta} \, ,
\end{align}
where $\tilde{\beta}$ represents the seventh degree of freedom, which leads to the 7-parameter shell model.

\subsubsection{Constitutive model}
In line with the previous procedure, the constitutive operator of the shell is referred to the shell midsurface. This requires the numerical pre-integration of the constitutive tensor $\mathbb{C}$ over the thickness direction. 
The resulting midsurface-based (pre-integrated) constitutive tensors $\mathbb{D}^{(M)}$, corresponding to the $M$-th thickness moment, are defined as
\begin{align}
\mathbb{D}^{(M)}=\int_{-1}^{+1} (\theta^3)^M \bigg(\frac{H}{2}\bigg)^{M}\, \hat{\mu} \, \mathbb{C}\; \dd \theta^3\, , \quad M \in \{0,1,2\} \quad \text{with }     \hat{\mu} = \frac{(\vec{G}_1 \times \vec{G}_2) \cdot \vec{G}_3}{\norm{\vec{A}_1 \times \vec{A}_2}}\, ,
\end{align}
where $\hat{\mu}$ represents the determinant of the shell shifter tensor~\cite{Braun-1994-Shell, Büchter-1994-3DExtensionEAS}, 
which transforms a volume integral into an integration over the shell midsurface. 

Leveraging these definitions and employing the principle of internal virtual work~\cite{Holzapfel-2001-NonlinearSolidMechanics}, the static mechanical equilibrium of the shell, considering the differential shell volume $\dd V=\hat{\mu}\ \dd \theta^3\dd A$, is expressed as
\begin{align}
\label{eq:StaticMechanicalEquilibrium}
     \delta W = \int_{\Omega_\textup{R}}\fPK \Dprod \delta \defgrad \dd V = \int_{\overline{\Omega}_\textup{R}}\int_{-1}^{+1}\fPK \Dprod \delta \defgrad \ \hat{\mu}\ \dd \theta^3\dd A = 0,
\end{align}
where $\fPK=\partial \Psi/ \partial \defgrad$ represents the first Piola-Kirchhoff stress tensor, $\overline{\Omega}_\textup{R}$ is the shell midsurface in reference configuration, and $\dd A$ is its differential area element.

\subsection{Material model: constrained mixture theory}
\label{sec:constrained_mixture}

\subsubsection{General}
The gastric microstructure is represented as a mixture of three load-bearing constituents: an isotropic ground matrix, mainly elastin, and two fiber families composed of collagen and \smcs{}.
The fiber families are oriented in longitudinal and circumferential direction respectively, cf.~\cref{fig:fiber_long,fig:fiber_cir}.
For simplicity, we do not distinguish between collagen and smooth muscle fibers.
Instead, we assume that the fiber families in our model represent both.
Each constituent is modeled individually and characterized by its mass fraction, material properties, and the prestress it exhibits in the in vivo reference configuration~\cite{Humphrey-2021-CMMOverview}. 
We denote quantities associated with a specific constituent by a superscript $i \in I$, where $I$ is an index set of cardinality $N$, corresponding to the number of constituents.

Following the constrained mixture model, we assume that all constituents are locally entangled and deform together without relative motion between them, such that $\defgrad = \defgrad^i$, where $\defgrad^i$ is the deformation gradient of the $i$-th constituent. Intuitively, $\defgrad$
captures the overall visible deformation of the tissue, for example, how it expands or contracts during digestion.

The deformation of biological tissues, including that of the stomach, can often be decomposed into three major contributions. They can be accounted for by a multiplicative decomposition of the deformation gradient of each constituent as
\begin{align}
\label{eq:Multiplicative_split}
    \defgrad = \defgrad_\textup{e}^i \defgrad_\textup{a}^i \defgrad_\textup{gr}^i\, .
\end{align}
Here, $\defgrad_\textup{e}^i$ represents the elastic part of the deformation. $\vec{F}_\textup{a}^i$ models that gastric tissue exhibits active muscular contractions. The resulting deformation is not an elastic one but an active (inelastic) change of the natural configuration of a tissue volume element. It can be accounted for by a so-called active strain approach~\cite{Brandstaeter2019,Patel-Gizzi-2022-GastrointestinalTissueReview} through an active part $\defgrad_\textup{a}^i$ of the deformation gradient. Ultimately, one has to account for the fact that biological tissues are in general able to grow and remodel. Growth and remodeling induces permanent, inelastic changes of the stress-free natural configuration of a tissue and can be modeled by a related inelastic part of the deformation gradient $\defgrad_\textup{gr}^i$~\cite{Braeu-2017-HCMM}. In the following, we describe in detail how the individual components of the deformation gradient are defined and computed.

%%%%%%%%%%%%%%%%%%%
\subsubsection{Reference and initial configuration}
\label{sec:reference_initial_configuration}
\begin{figure}[!t]
    \centering
    \includegraphics[width=0.8\linewidth]{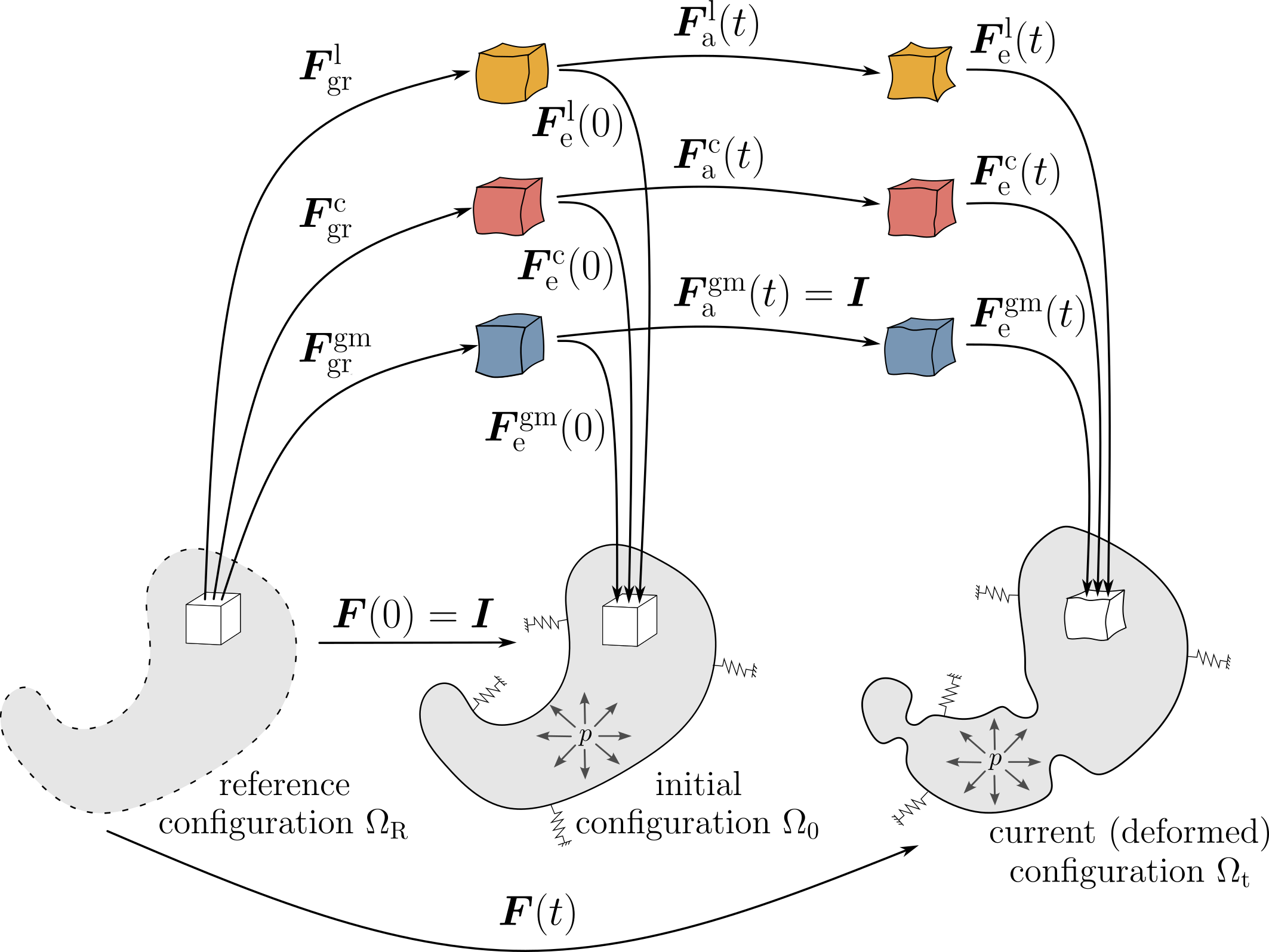}
        \caption{We define a fictitious reference configuration $\Omega_\textup{R}$ whose geometry is identical to the one acquired through medical imaging, but where we assume that the whole material is free of stress. The shape of the initial configuration $\Omega_0$ is identical to the one of $\Omega_\textup{R}$, so that $\defgrad(0) = \vec{I}$. However, in $\Omega_0$, the body is subjected to physiological loading, which introduces some prestress and thus elastic deformation through $\defgrad_\textup{e}^i(0)$. The constant $\defgrad_\textup{gr}^i$ generally represent some inelastic deformation from growth and remodeling before time $t=0$. In subsequent, deformed configurations, the deformation gradient additionally comprises an active part $\defgrad_\textup{a}^i(0)$, representing muscle contraction. At each point, the material is a mixture of collagen fibers and \smcs{} in longitudinal ($i=\lf$) and circumferential  ($i=\cf$) directions, as well as the ground matrix ($i=\gm$).} \label{fig:HCM_configurations}
\end{figure}

First of all, we define an initial configuration $\Omega_0$ at time $t=0$. This initial configuration is acquired from medical imaging. For simplicity, we assume herein that the image was taken in a state where active muscle tension was absent or at least negligibly small, that is, $\defgrad_\textup{a}^i = \vec{0}$. This assumes that the image was taken when the stomach was in a state of rest where peristaltic contractions have not yet started. In fact, such contractions typically start only with some delay after ingestion of food or liquids. In future work also other assumptions about active stress in the initial configuration can be applied without changing the overall framework proposed here. For example, a baseline muscle tone depending on certain control mechanisms in the nervous system could be assumed. Here we skip such advanced hypotheses because at the moment we lack data to translate them into viable model parameters. While the initial configuration $\Omega_0$ may be---at least nearly---free from active stress it will yet exhibit passive stress from the tissue elasticity due to the physiological loading that results from the internal gastric pressure $p$. As imaging of an empty stomach is much harder than of a filled stomach, medical images of the stomach will typically capture a situation with a non-negligible internal pressure $p$. Therefore, both $\defgrad_\textup{e}^i$ and $\defgrad_\textup{gr}^i$ will in general be non-zero in the initial configuration $\Omega_0$. To determine both, we first have to agree on a reference configuration. In principle, in continuum mechanics the reference configuraton can be chosen nearly arbitrarily. However, there are, of course, specific choices that make computations particularly easy. A beneficial choice in that sense is the configuration $\Omega_R$ illustrated on the lower left in~\cref{fig:HCM_configurations}. In this configuration, the stomach has exactly the same shape as in the initial configuration $\Omega_0$ acquired directly from medical imaging. However, the material is assumed to be completely free of any stress which includes also an absence of any external loading. Importantly, this reference configuration $\Omega_\textup{R}$ is a purely fictitious one. That is, it is a mathematical construct that represents a situation that has been present in reality at no point in time. However, it is helpful as a basis for computations. Given that we assume---as discussed above---that $\defgrad_\textup{a}^i$ is zero at time $t=0$, the deformation gradient mapping the fictitious reference configuration $\Omega_\textup{R}$ to the initial configuration $\Omega_0$ is simply 
\begin{align}
\label{eq:prestressing}
    \defgrad(0)=     \defgrad_\textup{e}^i(0)  \defgrad_\textup{gr}^i = \vec{I}.
\end{align}
Here, $\defgrad(0)$ and $\defgrad_\textup{e}^i(0)$ denote the deformation gradient at $t = 0$ and its elastic part, respectively. The requirement $\defgrad(0)=\vec{I}$ results from our definition that in $\Omega_\textup{R}$ and $\Omega_0$ our body exhibits the same geometry. That is, in the fictitious configuration $\Omega_\textup{R}$, each material point lies at the same position as in the initial configuration $\Omega_0$ acquired from medical imaging. Herein, we do not study growth and remodeling of gastric tissue. That is, we assume that such processes do not happen on the time scale of peristaltic contractions that is relevant for this article. This is motivated by the fact that growth and remodeling typically happens on the time scale of weeks to months, whereas peristaltic contractions happen on the time scale of minutes to hours. 
Therefore, the inelastic part of the deformation gradient due to growth and remodeling, $\defgrad_\textup{gr}^i$, captures only effects of previous growth and remodeling in the past. It can thus be assumed in the above~\cref{eq:prestressing} to be a constant in time. To determine this constant for each constituent, we essentially rely on concepts for prestress computation proposed by~\cite{Weisbecker-2014-Prestress} and~\cite{MousaviAvril-2017} and recently used also by~\cite{Gebauer-2023-CMMCardiac}. Details are discussed below. 

\subsubsection{Material elasticity}
\label{sec:passive-material}
The elastic part of the deformation gradient, $\defgrad_\textup{e}^i$, ultimately results from the passive elasticity of the tissue such that it ensures everywhere geometric compatibility and balance of linear momentum. Therefore, to compute it, one needs a model that relates $\defgrad_\textup{e}^i$ to a specific strain energy from which then elastic stresses can be computed.
To model the passive elastic response of the stomach wall in a constrained mixture framework, we define the total strain energy $\Psi$ of the mixture as a sum of the strain energies of the different constituents. For each constituent the strain energy $\Psi^i$ per unit volume can be expressed as a product of the strain energy per unit mass $W^i$ and the (referential) mass density $\rho_0^i$. This results in 
\begin{align}
    \Psi(\defgrad)=\sum_{i=1}^N \Psi^i(\defgrad_\textup{e}^i) &= \sum_{i=1}^N \rho_0^i W^i(\defgrad_\textup{e}^i) \, .
\end{align}
Notably, the strain energy of the $i$-th constituent depends only on the elastic part of deformation gradient $\defgrad_\textup{e}^i$. 

In our stomach model, anisotropic stiffness contributions arise from smooth muscle and collagen fiber families arranged in preferential directions (longitudinal and circumferential), whereas the isotropic ground matrix is attributed to the elastin network. 
For sake of conciseness, the specific form of the strain energy functions $W^i$ used to characterize each constituent are provided in~\ref{ap:strain_energy_functions}. These expressions follow established formulations (e.g.~\cite{HolzapfelGasserOgden-2012-ConstArterial,Holzapfel-2001-NonlinearSolidMechanics}) and are not repeated here to focus on model integration and physiological specificity.

\subsubsection{Prestress}
\label{sec:prestress}
From~\cref{eq:prestressing}, one can see immediately that the effects of growth and remodeling prior to time $t=0$ that are captured by $\defgrad_\textup{gr}^i$ are exactly the inverse of the elastic prestretch $\defgrad_\textup{e}^i(0)$ in the initial configuration. So, determining $\defgrad_\textup{gr}^i$ is basically identical to determining the elastic prestretch or prestress in the initial configuration. To this end, one has to identify $\defgrad_\textup{e}^i$ such it ensures mechanical equilibrium for the given loading in the initial configuration. Such inverse problems are in general ill-posed, especially if the tissue is modeled as a mixture of several constituents whose elastic deformation in the initial configuration can vary independently (as is the case here because in principle also the $\defgrad_\textup{gr}^i$ can vary independently). One typically needs some physiologically reasonable assumptions to overcome the ill-posedness of such problems. To this end, we rely on the concept of tensional homeostasis. It is well-known that a continuous turnover of mass in collagenous tissues ensures that collagen fibers in healthy tissues typically reside in some mechanical target state, the homeostatic state. Although the precise definition of this state remains under debate, many models assume that its hallmark is a specific homeostatic fiber stress~\cite{Eichinger2021a}. Herein, we adopt this assumption. That is, we assume that for the fibers in our material, the inelastic deformation gradient $\defgrad_\textup{gr}^i$ is such that the fibers exhibit a specific homeostatic fiber stress. As this fiber stress is easy to translate in a fiber stretch, which again is easy to translate into the respective $\defgrad_\textup{gr}^i$, both $\defgrad_\textup{e}^i$ and $\defgrad_\textup{gr}^i$ are directly determined by the hypothesis that the collagen and muscle fibers in the stomach are in the initial configuration in their homeostatic state. 
This reduces the inverse problem to determining the $\defgrad_\textup{e}^{\gm}$ and $\defgrad_\textup{gr}^{\gm}$ of the ground matrix. The ground matrix consists to a large extent of elastin. Elastin forms a highly elastic largely isotropic protein network that enables reversible shape changes. In particular the high water content makes the ground matrix nearly incompressible under the conditions relevant herein. Hence, we model the ground matrix as an isotropic, elastic, incompressible material (cf.~\ref{ap:strain_energy_functions}). This directly leads to the assumption that $\defgrad_\textup{gr}^{\gm}$ is isochoric and rotation-free, i.e., symmetric. 

\begin{figure}[!h]
\centering
\tikzstyle{startstop} = [rectangle, rounded corners=15pt, minimum width=2cm, minimum height=1cm,text centered, draw=black, fill=myred!50]
\tikzstyle{process} = [rectangle, minimum width=3.5cm, minimum height=1cm, text centered, draw=black, fill=gray!10]
\tikzstyle{decision} = [diamond, aspect=2, text centered, draw=black, fill=gray!10, inner sep=0pt, minimum size=2cm]
\tikzstyle{arrow} = [thick,->,>=stealth]
\resizebox{0.8\textwidth}{!}{
\begin{tikzpicture}[node distance=1.7cm and 3.3cm]
\node (start) [startstop] {Start};

\node (init) [process, right=of start] {\shortstack{Initialize: $k \gets 0$ \\ ${\defgrad_{\textup{gr},0}^\gm} \gets \mat{I}$}};

\node (checkN) [decision, below of=init, yshift=-0.3cm] {Is $k<k_p$ ?};
\node (ramp) [process, right=3.5cm of checkN] {
  \shortstack{
    Gradually increase\\
    $p$ and $\lambda_\textup{e}^i,\, i \in {\cf,\lf}$
  }
};
\node (fixed) [process, below=1.cm of checkN] {
  \shortstack{
    Keep $p$ and $\lambda_\textup{e}^i,\, i \in {\cf,\lf}$ fixed
  }
};
\node (solve) [process, below of =fixed] {\shortstack{Solve equilibrium: get $\defgrad_k$, $\vec{u}_k$}};
\node (postprocess) [process, below of=solve] {\shortstack{Isochoric deformation
$\overline{\defgrad}_k = \det(\defgrad_k)^{-\nicefrac{1}{3}} \defgrad_k$}};
\node (ftrial) [process, below of=postprocess] {\shortstack{Adopt to reduce deformation in next step
${\overline{\defgrad}_{\textup{gr},k}^\textup{trial}} = \overline{\defgrad}_k^{-1} \, {\defgrad_{\textup{gr},k}^\gm}$}};
\node (polar) [process, below of=ftrial] {\shortstack{Polar decomposition: ${\overline{\defgrad}_{\textup{gr},k}^\textup{trial}} =
{\overline{\mat{R}}_{\textup{gr},k}^\textup{trial}}\,
{\overline{\mat{U}}_{\textup{gr},k}^\textup{trial}}$}};
\node (update) [process, below of=polar] {Update prestretch: ${\defgrad_{\textup{gr},k+1}^\gm} = {\overline{\mat{U}}_{\textup{gr},k}^\textup{trial}}$};
\node (checkconv) [decision, below of=update, yshift=-0.3cm] {Is $\|\vec{u}_k\| < \epsilon$ ?};
\coordinate (posBelowStart) at ($(start.south) $);
\coordinate (posLeftSolve) at ($(solve.west)$);
\node (updatek) [process] at ($(posBelowStart |- posLeftSolve)$) {$k \gets k+1$};
\node (end) [startstop, right=3.5cm of checkconv] {\shortstack{Stop }};
% Arrows
\draw [arrow] (start) -- (init);
\draw [arrow] (init) -- (checkN);
\draw [arrow] (checkN) -- node[above] {Yes} (ramp);
\draw [arrow] (checkN) -- node[right] {No} (fixed);
\draw [arrow] (ramp) |- (solve);
\draw [arrow] (fixed) -- (solve);
\draw [arrow] (solve) -- (postprocess);
\draw [arrow] (postprocess) -- (ftrial);
\draw [arrow] (ftrial) -- (polar);
\draw [arrow] (polar) -- (update);
\draw [arrow] (update) -- (checkconv);
\draw [arrow] (checkconv) -- node[midway, above] {Yes} (end);
\draw [arrow] (checkconv.west) -- ($(start.south |- checkconv.west)$) node[midway, above] {No} -- ($(updatek.south)$);
\draw [arrow] (updatek) |- (checkN);
\end{tikzpicture}}
\caption{Flowchart of the iterative prestress algorithm for computing for the ground matrix the tensor $\defgrad_\textup{gr}^\gm$ that characterizes the inelastic deformation through tissue formation prior to time $t = 0$ and thus determines the elastic prestress in the loaded configuration at $t = 0$. This tensor is adopted iteratively such that the deformation between reference and initial configuration becomes smaller and smaller until it vanishes within a certain tolerance $\epsilon$.}
\label{fig:prestress_algorithm}
\end{figure}
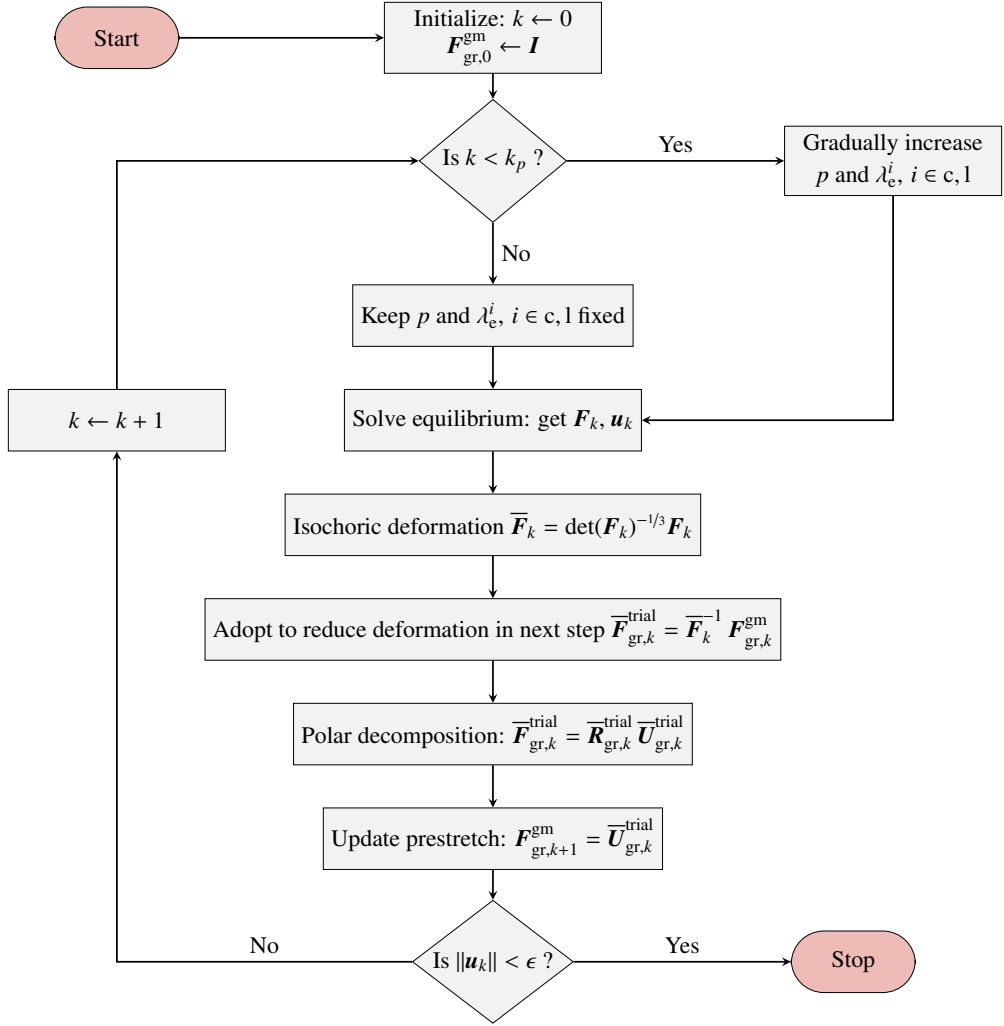
To solve then the inverse problem, we use the iterative prestress algorithm that is illustrated in \cref{fig:prestress_algorithm} and based on \cite{Weisbecker-2014-Prestress,MousaviAvril-2017,Gebauer-2023-CMMCardiac}. We initialize the algorithm with $\defgrad_\textup{gr}^{\gm} = \vec{I}$. During the first $k_p$ iterations, the intraluminal pressure $p$ and fiber prestretches $\lambda_\textup{e}^i,\, i \in {\cf,\lf}$ are gradually increased to the known intragastric pressure in vivo and the homeostatic stretch of collagenous fibers. For all subsequent iterations ($k \geq k_p$), both are held constant. In each iteration, we use the current guess ${\defgrad_{\textup{gr},k}^{\gm}}$ to solve the nonlinear equilibrium equations through a Newton-Raphson method. Subsequently, we evaluate the deformation gradient $\defgrad_k$ and its isochoric part $\overline{\defgrad}_k = \det(\defgrad_k)^{-\nicefrac{1}{3}} \defgrad_k$. Ultimately, we seek to determine ${\defgrad_{\textup{gr},k}^{\gm}}$ such that the deformation gradient is equal to the identity tensor so that the shape of the reference and initial configuration are identical as required by definition. If $\overline{\defgrad}_k$ is not yet equal to the identity tensor within a sufficiently small tolerance, we have to modify $\defgrad_\textup{gr}^{\gm}$ such that we move closer towards this state in the next iteration step. To achieve this, we have to make sure that the stretches implied by ${\defgrad_{\textup{gr},k+1}^{\gm}}$ become smaller than the ones implied by ${\defgrad_{\textup{gr},k}^{\gm}}$ in exactly those principal directions where $\defgrad_k$ has eigenvalues larger than one and vice versa. This can be achieved by multiplying $\defgrad_\textup{gr}^{\gm}$ with the inverse of $\overline{\defgrad}_k$. Therefore, we compute
\begin{align}
    {\overline{\defgrad}_{\textup{gr},k}^{\textup{trial}}} = \overline{\defgrad}_k^{-1} \, {\vec{F}_{\textup{gr},k}^\gm} \, .
\end{align}
To eliminate the rotational part, we perform a polar decomposition
\begin{align}
    {\overline{\defgrad}_{\textup{gr},k}^{\textup{trial}}}=
    {\overline{\vec{R}}_{\textup{gr},k}^{\textup{trial}}}\,
    {\overline{\vec{U}}_{\textup{gr},k}^{\textup{trial}}}\, ,
\end{align}
where ${\overline{\vec{R}}_{\textup{gr},k}^{\textup{trial}}}$ is a proper orthogonal rotation tensor and ${\overline{\vec{U}}_{\textup{gr},k}^{\textup{trial}}}$ a symmetric positive-definite (right) stretch tensor. Finally, we update
\begin{align}
    {\vec{F}_{\textup{gr},k+1}^\gm}={\overline{\vec{U}}_{\textup{gr},k}^{\textup{trial}}}. 
\end{align}
This update ensures that ${\vec{F}_\textup{gr}^\gm}$ remains symmetric and isochoric. The iterations continue until the maximum Euclidean norm of the displacement $\norm{\vec{u}_k}_\infty$ drops below some fixed tolerance $\epsilon$. Once this has been achieved, the algorithm has identified a value of ${\vec{F}_\textup{gr}^\gm}$ that implies an elastic prestress ensuring the balance of linear momentum is satisfied in the initial configuration.

\subsubsection{Active strain}
\label{sec:active-strain_appraoch}
Active deformation of the gastric wall arises from the contraction of \smcs{}. These are directional contractile units that contract only along their fiber direction. In the stomach wall, two primary muscle fiber directions are present, the circumferential and longitudinal direction~\cite{DiNatale-2023-FunctionalAndAnatomicalRegions}, represented by the unit vectors $\vec{f}_\textup{R}^{\cf}$ and $\vec{f}_\textup{R}^{\lf}$ in the reference configuration. For these two fiber families, we adopt a transversely isotropic, isochoric active deformation gradient
\begin{align}
    \defgrad_\textup{a}^i = (1-\gamma \alpha_i) \, \vec{f}_\textup{a}^i \otimes \vec{f}_\textup{gr}^i + \frac{1}{\sqrt{1-\gamma\alpha_i}}\, (\mat{I}- \vec{f}_\textup{a}^i \otimes \vec{f}_\textup{gr}^i)\, ,  \qquad i \in \{\cf,\lf\}.
\end{align}
Here, $\gamma$ denotes the muscle activation, and $\alpha_i$ is a material parameter controlling the magnitude of active contractility along the corresponding fiber directions. The active contraction through $\defgrad_\textup{a}^i$ is assumed to be volume-preserving, that is, the contraction is coupled with a transverse expansion such that $\det(\defgrad_\textup{a}^i) = 1$. The active deformation gradient $\defgrad_\textup{a}^i$ is a two-point tensor relying on the unit basis vectors
\begin{align}
    \vec{f}_\textup{gr}^i= \frac{\defgrad_\textup{gr}^i \vec{f}_\textup{R}^i}{\norm{\defgrad_\textup{gr}^i \vec{f}_\textup{R}^i}}\quad \textup{and}\quad      \vec{f}_\textup{a}^i= \frac{\defgrad_\textup{a}^i \vec{f}_\textup{gr}^i}{\norm{\defgrad_\textup{a}^i \vec{f}_\textup{gr}^i}}\, .
\end{align}
Geometrically, $\vec{f}_\textup{gr}^i$ is the fiber direction in the intermediate configuration that results from applying $\defgrad_\textup{gr}^i$ only, and $\vec{f}_\textup{a}^i$ is the fiber direction in the intermediate configuration resulting from applying first $\defgrad_\textup{gr}^i$ and then additionally $\defgrad_\textup{a}^i$. Here, we focus on physiological time scales where long-term growth and remodeling are negligible compared to the fast electromechanical dynamics. Any fiber rotation from growth and remodeling before $t=0$ can be accounted for by the choice of the reference fiber direction $\vec{f}_\textup{R}^i$ without loss of generality. Hence, we can assume $\defgrad_\textup{gr}^i$ to be a rotation-free prestretch tensor. Moreover, muscular contraction is assumed to shorten the fiber but not to change its direction in space. Therefore, neither $\defgrad_\textup{gr}^i$ nor $\defgrad_\textup{a}^i$ rotate the fiber direction in space. Hence, $\vec{f}_\textup{a}^i = \vec{f}_\textup{gr}^i= \vec{f}_\textup{R}^i$ so that 
\begin{align}
\label{eq:active_deformation_gradient}
    \defgrad_\textup{a}^i = (1-\gamma \alpha_i) \, \vec{f}_\textup{R}^i \otimes \vec{f}_\textup{R}^i + \frac{1}{\sqrt{1-\gamma\alpha_i}}\, (\mat{I}- \vec{f}_\textup{R}^i \otimes \vec{f}_\textup{R}^i)\, ,  \qquad i \in \{\cf,\lf\}\, .
\end{align}
In accordance with \cite{brandstaeter2018a}, we assume that the muscle activation $\gamma$ is a direct function of the \smcs{} transmembrane voltage $v^\SMC$, i.e., $\gamma = \gamma(v^\SMC)$. 
While this represents a simplification of the complex excitation–contraction coupling mechanism in gastric \smcs{}, \cite{brandstaeter2018a} demonstrated that it yields a sufficiently accurate model of electromechanical coupling for organ-scale simulations.
The activation function is given by
\begin{align}
    \gamma(v^\SMC) = \big(1 - e^{-\beta_1(v^\SMC - v^{\textup{thr}})}\big) \big(1 - e^{-\beta_2(v^\SMC - v^{\textup{thr}})}\big) H(v^\SMC - v^{\textup{thr}}) \, ,
    \label{eq:Threshold}
\end{align}
where $\beta_1$ and $\beta_2$ govern the \ca{} dynamics and the opening of \vdcc{}, respectively.
The Heaviside function ${H}(v^\SMC-v^{\textup{thr}})$ ensures contraction is triggered only when the $v^\SMC$ exceeds a threshold value $v^{\textup{thr}}$, representing the opening voltage of \vdcc{}
\begin{align}
   {H}(v^\SMC-v^{\textup{thr}}) =
   \begin{cases}
        0, &  v^\SMC\leq v^{\textup{thr}} \, ,\\
        1, &  v^\SMC>v^{\textup{thr}} \, .
    \end{cases}
\end{align}

The isotropic ground matrix, primarily composed of elastin, cannot contract actively, i.e., $\defgrad_\textup{a}^\gm= \mat{I}$. Due to kinematic coupling in the constrained mixture, elastin is impelled to undergo the same deformation as the surrounding smooth muscle fibers. Since it lacks an active mechanism, its response is purely elastic through $\defgrad_\textup{e}^\gm$, ensuring that the total deformation $\defgrad^i$ remains identical across all constituents.

%%%%%%%%%%%%%%%%%%%%%%%%%%%%%%%%
\subsection{Electrophysiological model}
\label{sec:gastric_electrophysiological_moedl}
We employ a phenomenological approach to model gastric electrophysiology~\cite{Keener-2009-MathematicalPhysiology,Pullan-2006-ModellingElectricalActivityHeart}.
The normalized transmembrane potentials of \iccs{} and \smcs{} are represented by the spatio-temporal scalar functions $v^\ICC: \Omega_0 \times [0,\infty) \rightarrow [0, 1]$ and $ v^\SMC: \Omega_0 \times [0,\infty) \rightarrow [0, 1]$, respectively.
The propagation of these potentials across the gastric tissue is governed by two coupled monodomain formulations
\begin{align}\label{eq:RDsys}
    \left\{ \begin{aligned} 
    \chi^\ICC \pfrac{v^\ICC}{t}  &= \nabla \cdot (\sigma^\ICC \nabla v^\ICC) - \chi^\textup{gap} D_\textup{gap}(v^\ICC - v^\SMC) + \chi^\ICC I_{\textup{ion}}^\ICC
    \\
   \chi^\SMC  \pfrac{v^\SMC}{t} &=  \nabla\cdot (\sigma^\SMC \nabla v^\SMC) + \chi^\textup{gap} D_\textup{gap}(v^\ICC - v^\SMC) +\chi^\SMC  I_{\textup{ion}}^\SMC
    \end{aligned} \right. \, ,
\end{align}
where $\sigma^\ICC$ and $\sigma^\SMC$ denote the isotropic diffusion coefficients associated with \iccs{} and \smcs{}, respectively, with units in \si{\mm^2\per\s}. These coefficients are modeled as spatially heterogeneous, isotropic scalar fields, reflecting that electrical conduction is dominated by \icc{}-mediated propagation in the myenteric plexus, which lacks a preferred fiber orientation~\cite{huizinga2009a,Sanders-2016-RegulationSMCFunction}, while \smc{} activation is mainly driven by coupling from \iccs{} rather than direct fiber-aligned conduction~\cite{OGrady-2012-AbnormalInitiationSW,du2013c}. The term $D_\textup{gap}$ models the homogenized coupling resistance between the two cell types via gap junctions, and $I_{\textup{ion}}^\ICC$ and $I_{\textup{ion}}^\SMC$ denote the total transmembrane ionic currents in \iccs{} and \smcs{}, respectively.
The coefficients $\chi^\ICC$, $\chi^\SMC$, and $\chi^\textup{gap}$ act as dimensionless scaling factors that modulate the relative contributions of transmembrane ($\chi^\ICC$, $\chi^\SMC$) and intercellular coupling terms ($\chi^\textup{gap}$) in the tissue-scale model.
In biophysically motivated monodomain formulations, these coefficients represent the surface-to-volume ratio, which naturally emerges from analyzing how membrane surface area contributes to transmembrane ionic and capacitive currents per unit tissue volume, bridging microscopic cellular physiology and macroscopic electrical wave propagation.
In contrast, the present phenomenological formulation employs normalized transmembrane potentials and effective ionic current terms.
Thus, the coefficients $\chi^\ICC$ and $\chi^\SMC$ are interpreted as spatially varying weights that reflect regional heterogeneity in cell density.
This flexible representation preserves the fundamental ionic kinetics as well-known from cardiac electrophysiology~\cite{Keener-2009-MathematicalPhysiology}.
For the definition of $\chi^\ICC$, $\chi^\SMC$, and $\chi^\textup{gap}$, refer to \cref{sec:regional-contributions-iccs-smcs}.
The formulations of the ionic currents ($I_{\textup{ion}}^\ICC$ and $I_{\textup{ion}}^\SMC$) are taken from~\cite{brandstaeter2018a} and are recalled in~\ref{ap:electrophysiology_cell_model}. In particular, each cell type (\icc{} and \smc{}) is modeled using a modified two-variable Mitchell–Schaeffer cell model~\cite{MitchellSchaeffer-2003-TwoCurrentModel}, which provides a sufficient description of cellular dynamics, while allowing stable and robust long-term simulations of slow-wave propagation, as demonstrated in our previous work~\cite{brandstaeter2018a}.

%%%%%%%%%%%%%%%%%%%%%%%%%%%%%%%%%%%%%%%%%%%%%%%%%%%%%%%%%%%
\subsection{Spatially non-uniform parametrization}
\label{sec:heterogenous-parameter-modeling}
The stomach exhibits highly spatially varying features such as fiber orientations~\cite{Patel-Gizzi-2022-GastrointestinalTissueReview,DiNatale-2023-FunctionalAndAnatomicalRegions,Friis-2023-GastricTissue}, intrinsic frequencies~\cite{vanHelden-2009_GenerationSW,OGrady-2012-AbnormalInitiationSW,EgbujiGrady-2010-OriginPropagationSW}, and passive material properties~\cite{Friis-2023-GastricTissue,Patel-Gizzi-2022-GastrointestinalTissueReview,Holzer-Stock2025}. 
To accommodate this heterogeneity in computational models, particularly in patient-specific geometries, we move from a purely anatomical description based on the fundus, corpus, and antrum to a functionally driven representation of the stomach~\cite{DiNatale-2023-FunctionalAndAnatomicalRegions}. 
Functional regions, such as the proximal and distal stomach, are strongly associated with physiological roles in motility and electrical activity, though they do not map directly onto anatomical landmarks (cf.~\cref{fig:stomach_anatomics,fig:stomach_boundary_conditions}). In practice, patient-specific geometries often lack detailed microstructural data~\cite{Patel-Gizzi-2022-GastrointestinalTissueReview}, which limits direct spatial assignment of physiological parameters. To address this, we propose a framework that builds functionally meaningful coordinate systems derived from harmonic scalar fields. These fields are generated by solving Laplace-Dirichlet problems~\cite{Bayer-2005-LDFiber} over the reference configuration $\Omega_\textup{R}$ and clearly defined anatomical landmarks. They serve as normalized control coordinates for assigning spatially varying model parameters in a way that reflects the underlying functional architecture. This approach enables consistent parameterization across realistic stomach geometries, a critical feature given the substantial inter-subject variability in anatomical shape. Similar approaches have been successfully applied in cardiac modeling to define fiber orientations~\cite{Piersanti-2021-CardiacFibers,Baier-2020-CardiacFiber,Bayer-2012-RuleBasedFiberHEart} and principle conduction axes~\cite{sathar2015a}.
Here, we extend this method to address a broader range of heterogeneities in the stomach, including, but not limited to, fiber directions in gastric tissue. To this end, we define three harmonic scalar fields as smooth functions of the material point $\vec{X} \in \Omega_\textup{R}$, each aligned with a distinct physiological axis: (i) $\phi_\ep(\vec{X})$, from the esophagus to the pyloric sphincter, representing the longitudinal direction; (ii) $\phi_\fp(\vec{X})$, spanning the fundus apex to the pyloric sphincter; and (iii) $\phi_\gl(\vec{X})$, from the greater to the lesser curvature, aligned with the circumferential direction.  

\subsubsection{Principal fiber directions}
\label{sec:fiber_distribution}
Peristaltic contractions are chiefly driven by the longitudinal and circumferential muscle fibers. The longitudinal layer extends from the esophagus to the pyloric sphincter, aligning with esophageal muscles and crossing the fundus apex. In contrast, the circumferential fibers run perpendicular to the axis from the fundus to the duodenum~\cite{DiNatale-2023-FunctionalAndAnatomicalRegions}.
\begin{figure}[tp]
    \centering
    \subfloat[]{%
        \includegraphics[trim={0cm 0cm 0cm 0cm},width=0.32\textwidth]{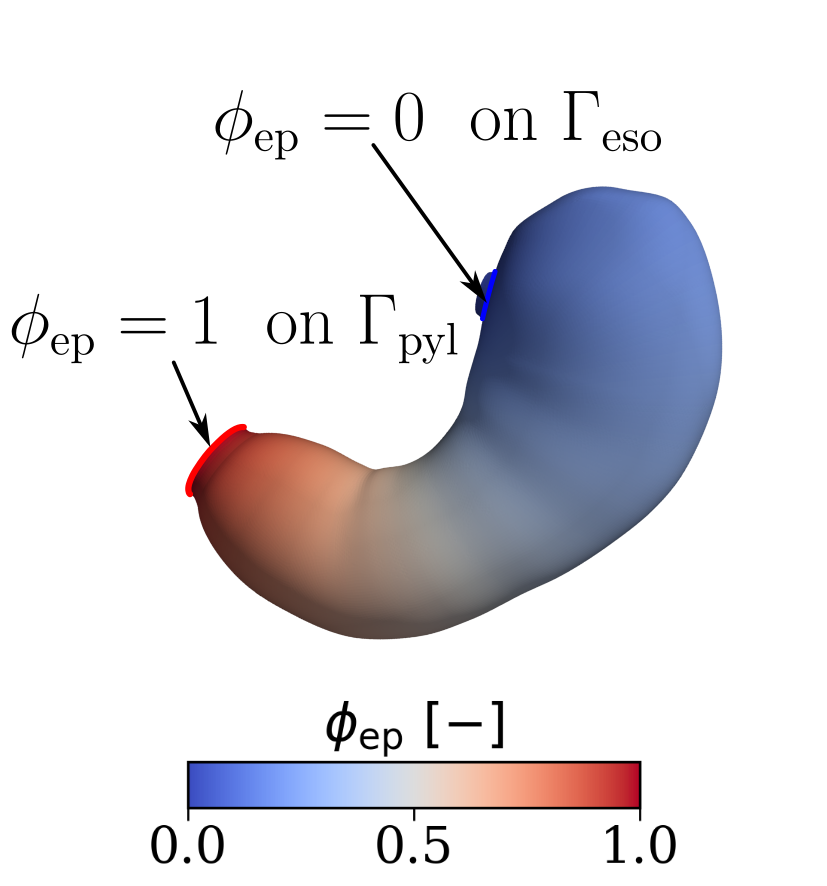}\label{fig:stomach_field_1}}
    \subfloat[]{%
        \includegraphics[trim={0cm 0cm 0cm 0cm},width=0.32\linewidth]{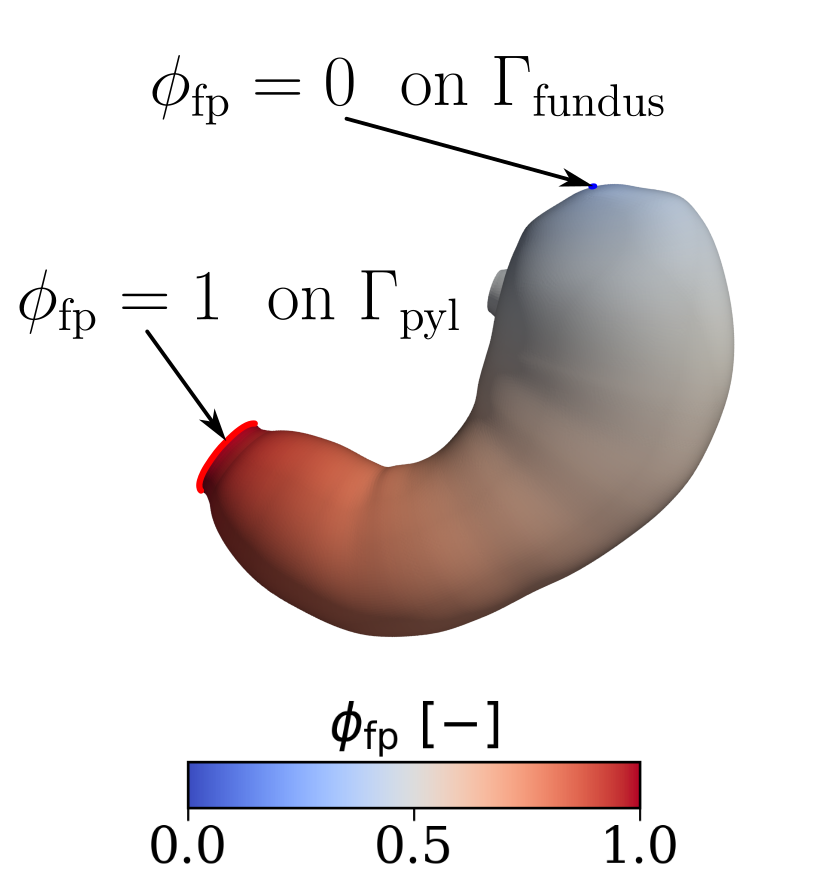}\label{fig:stomach_field_2}}
    \subfloat[]{%
        \includegraphics[trim={0cm 0cm 0cm 0cm},width=0.32\linewidth]{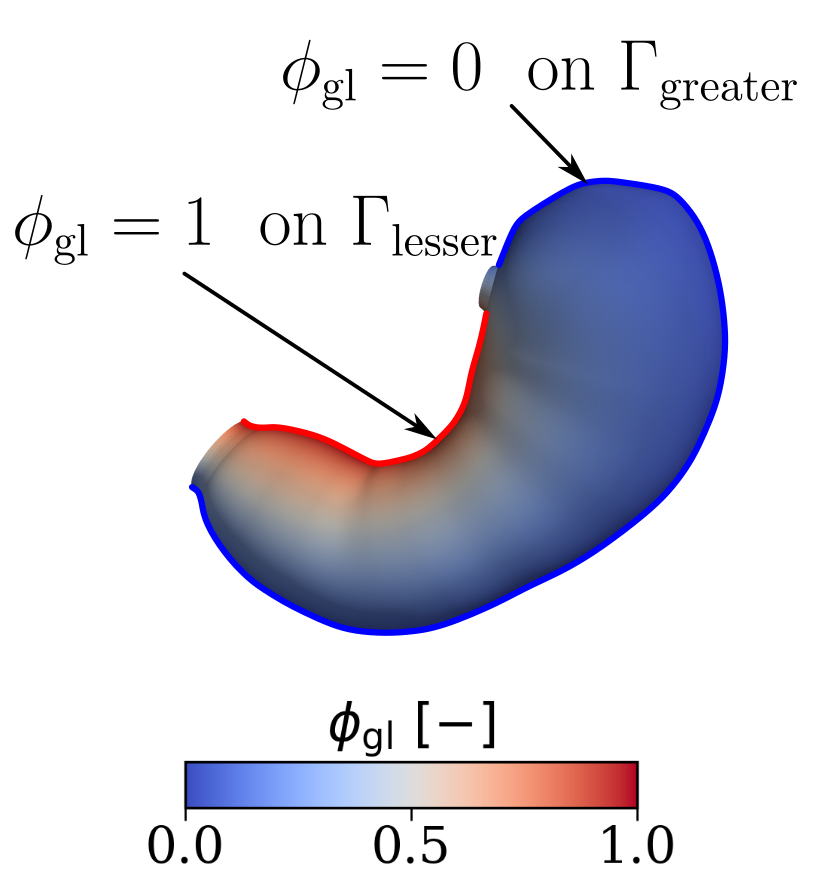}\label{fig:stomach_field_3}}\\
     \subfloat[]{%
        \includegraphics[trim={0cm 1cm 0cm 1cm},width=0.32\linewidth]{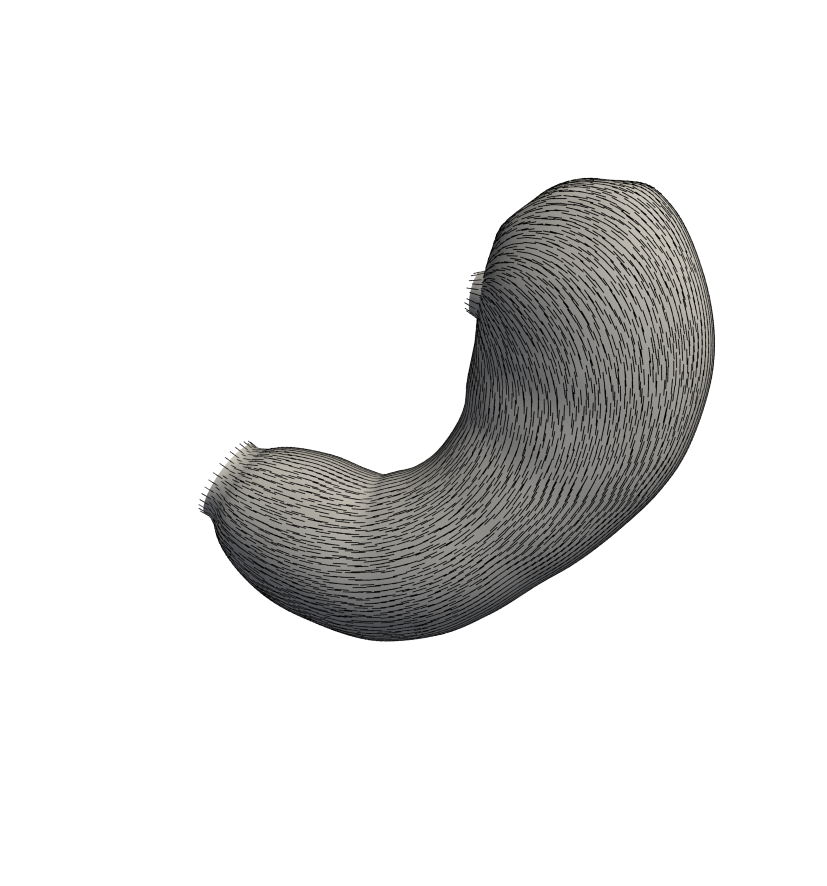}\label{fig:fiber_long}}
    \subfloat[]{%
        \includegraphics[trim={0cm 1cm 0cm 1cm},width=0.32\linewidth]{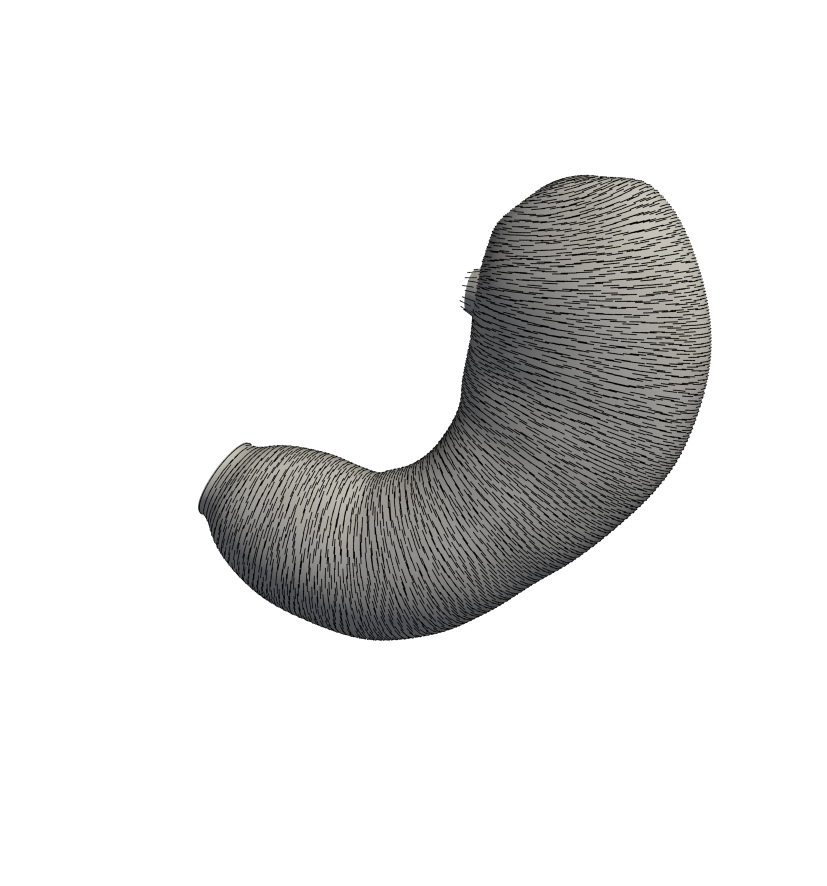}\label{fig:fiber_cir}}   
    \subfloat[]{%
        \includegraphics[trim={0cm 1cm 0cm 1cm},width=0.32\linewidth]{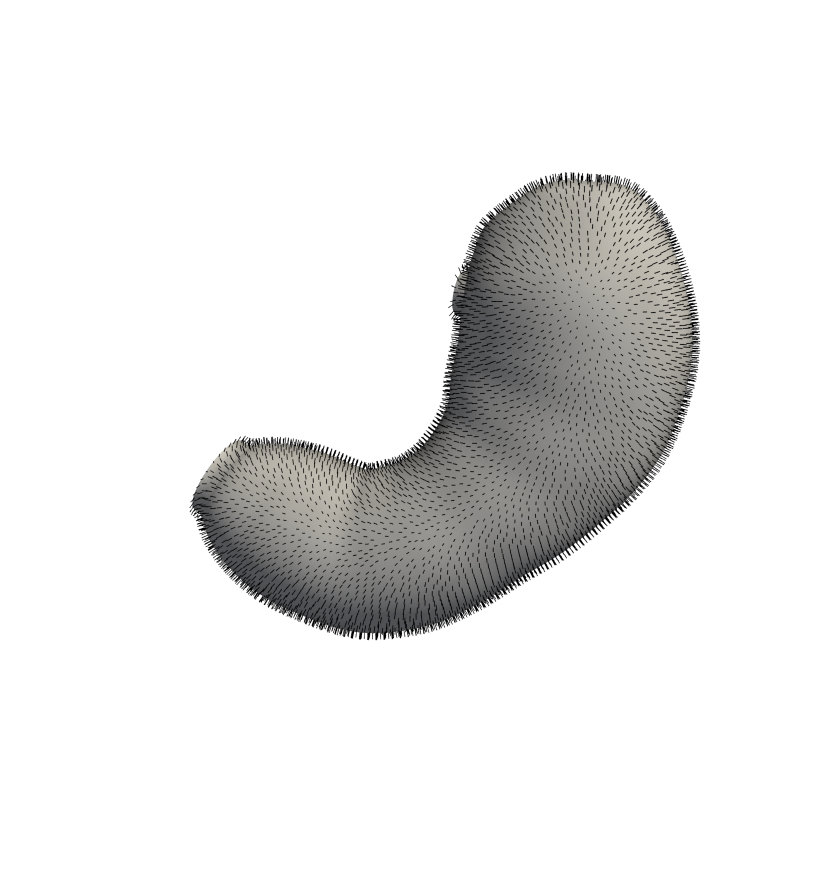}\label{fig:fiber_perp}} \\    
    \subfloat[]{%
        \includegraphics[trim={0cm 0cm 0cm 1cm},width=0.32\linewidth]{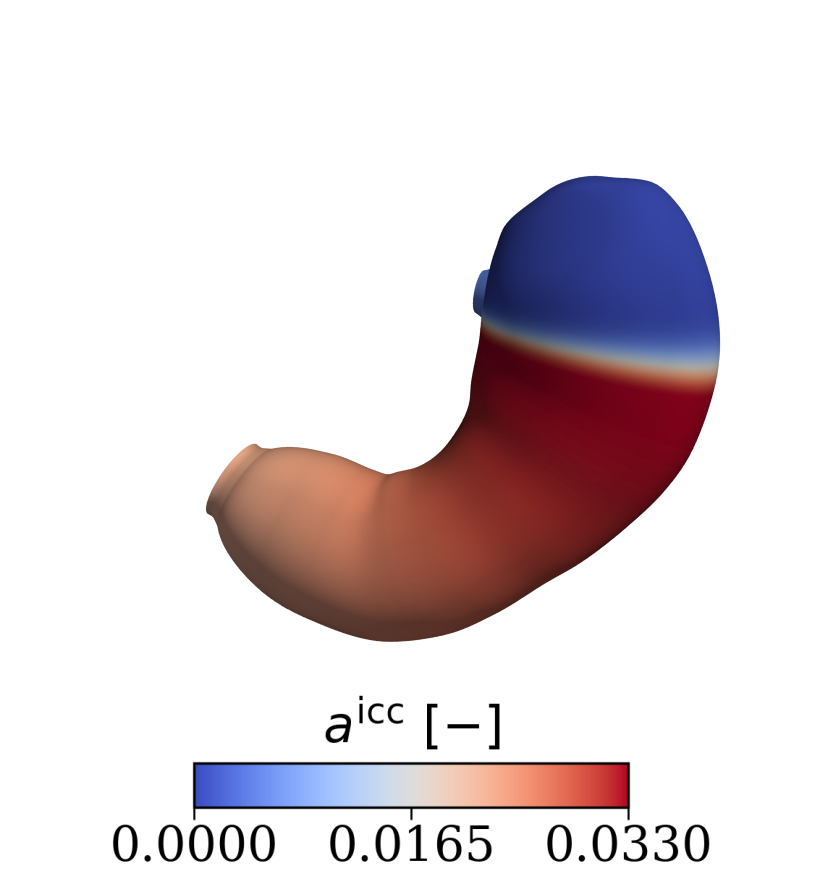}\label{fig:a_i}}
    \subfloat[]{%
        \includegraphics[trim={0cm 0cm 0cm 1cm},width=0.32\linewidth]{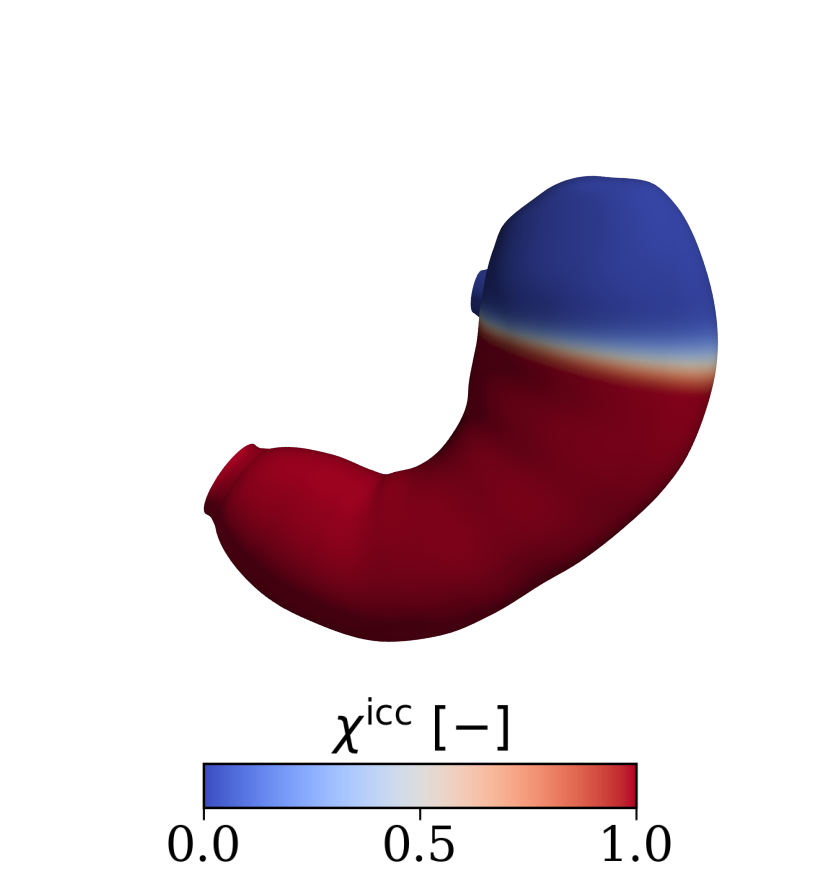}\label{fig:chi_i}}
    \subfloat[]{%
        \includegraphics[trim={0cm 0cm 0cm 1cm},width=0.32\linewidth]{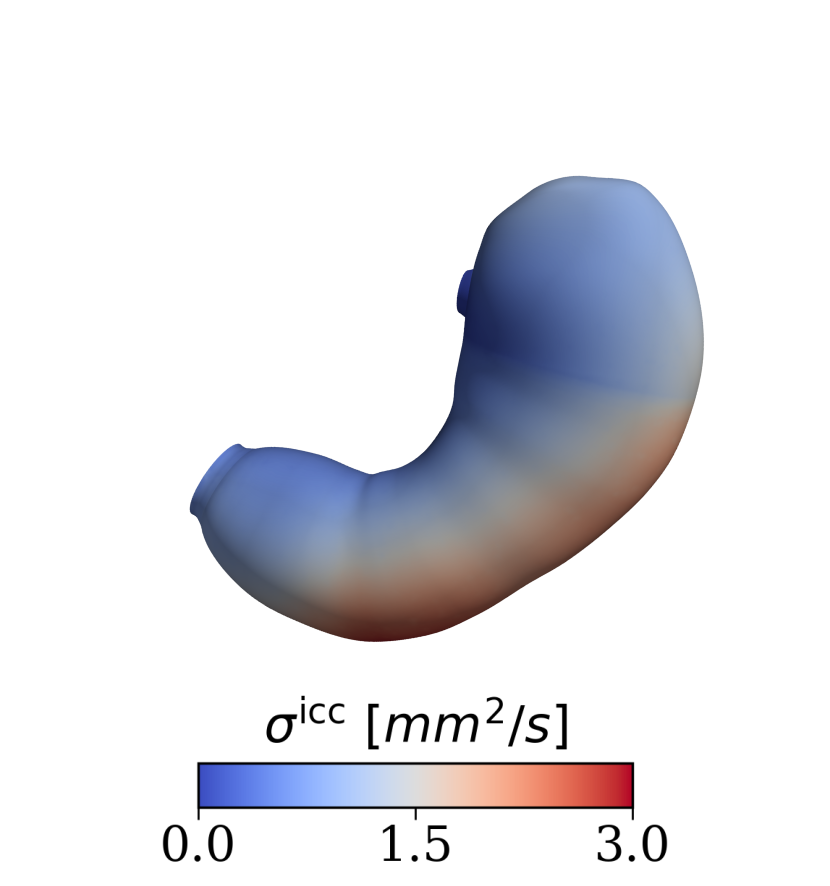}\label{fig:sigma_i}}    
    \caption{Definition of stomach-specific harmonic fields, computed fiber directions, and spatially heterogeneous \icc{} parameters.
     (a–c)~Harmonic scalar fields computed via Laplace-Dirichlet problems, used to construct stomach-specific coordinate axes: (a)~$\phi_\ep$ from the lower esophageal sphincter ($\Gamma_\textup{eso}$) to the pyloric sphincter ($\Gamma_\textup{pyl}$), (b)~$\phi_\fp$ from the fundus apex ($\Gamma_\textup{fundus}$) to $\Gamma_\textup{pyl}$, and (c)~$\phi_\gl$ from the greater ($\Gamma_\textup{greater}$) to the lesser curvature ($\Gamma_\textup{lesser}$).
    (d–f)~Surface vector fields representing fiber directions and the surface normal direction: (d)~longitudinal $\vec{f}_\textup{R}^\lf$, (e)~circumferential $\vec{f}_\textup{R}^\cf$, and (f)~outward normal direction $\vec{n}$.
    (g-i)~Spatial distributions of key \icc{} parameters: (g)~the excitability parameter $a^\ICC$, (f)~the weighting factor $\chi^\ICC$, and (i)~the diffusion coefficient $\sigma^\ICC$. These parameter fields are mapped based on the anatomical coordinate system defined by the harmonic fields.
    }  \label{fig:hetereogenous_field}
\end{figure}
These fiber directions are defined by solving Laplace-Dirichlet problems on the reference domain $\Omega_\textup{R}$:
\begin{align}\label{eq:laplace-dirichlet-fibers}
    \begin{cases}
        \begin{aligned}
         -\Delta \phi_\ep&= 0 && \text{in $\Omega_\textup{R}$} \, , \\
        \phi_\ep                                &= 1 && \text{on $\Gamma_\textup{pyl}$}\, , \\
        \phi_\ep                                &= 0 && \text{on $\Gamma_\textup{eso}$}\, ,
        \end{aligned}
          \quad \textup{and} \quad
    \end{cases}
    \begin{cases} 
        \begin{aligned}
         - \Delta \phi_\fp&= 0 && \text{in $\Omega_\textup{R}$} \, , \\
        \phi_\fp                                &= 1 && \text{on $\Gamma_\textup{pyl}$}\ , \\
        \phi_\fp                                &= 0 && \text{on $\Gamma_\textup{fundus}$}\, .
        \end{aligned}
    \end{cases}
\end{align}
Here, $\phi_\ep$ and $\phi_\fp$ are scalar harmonic fields, whose gradients $\nabla \phi_\ep$ and $\nabla \phi_\fp$ define smooth vector fields aligned with the anatomical directions from esophagus to pylorus and fundus to pylorus, respectively. 
$\Gamma_\textup{pyl}$, $\Gamma_\textup{eso}$, and $\Gamma_\textup{fundus}$ are the Dirichlet boundaries reflecting the pyloric sphincter, the lower esophageal sphincter, and the fundus apex point, respectively (cf.,~\cref{fig:stomach_field_1} and~\cref{fig:stomach_field_2}). 
The fiber directions are computed as normalized gradients:
\begin{align}
\vec{f}_\textup{R}^\lf = \frac{\nabla \phi_\ep}{\norm{\nabla \phi_\ep}}\, \quad \text{and } \quad
\vec{f}_\textup{R}^{\cf} = \vec{n} \times \frac{ \nabla \phi_\fp}{\norm{\nabla \phi_\fp}} \,
,
\end{align}
where $\vec{f}_\textup{R}^\lf$ and $\vec{f}_\textup{R}^\cf$ are the longitudinal and circumferential fiber directions, respectively, and $\vec{n}$ corresponds to the outward normal direction. Note that, contrary to engineering intuition, the circumferential and longitudinal directions are not necessarily orthogonal. These directions are defined anatomically, based on experimental observations. To accurately represent them, a second coordinate system is introduced to capture the anatomical circumferential and longitudinal orientations.
The scalar fields and their associated Dirichlet boundary conditions used to construct the fiber directions are presented in~\cref{fig:stomach_field_1,fig:stomach_field_2}. The computed fiber directions are visualized in~\cref{fig:fiber_long,fig:fiber_cir}, whereas the outward normal vector is shown in~\cref{fig:fiber_perp}.

%%%%%%%%%%%%
\subsubsection{Intrinsic frequencies of ICCs and passive behavior of SMCs}
\label{sec:intrinsic-frequencies}
\iccs{} exhibit excitability as a fundamental electrophysiological property that governs their ability to generate and propagate electrical signals. This excitability is spatially heterogeneous across the gastric surface and is closely linked to each cell's intrinsic frequency. Thus, the heterogeneity in excitability gives rise to intrinsic frequency gradients that are essential for the coordinated initiation and propagation of slow waves. Through electrical coupling, regions with lower intrinsic frequency become entrained by the dominant pacemaker region, where excitability, and thus intrinsic frequency, is highest. 

Anatomically, the excitability distribution follows a distinct spatial pattern: \iccs{} located in the fundus (proximal stomach, above the pacemaker region) exhibit nearly no excitability and thus do not actively contribute to slow-wave initiation. The pacemaker region itself exhibits the highest intrinsic frequency and excitability. Moving away from the pacemaker, excitability and intrinsic frequency decrease gradually both along the longitudinal axis toward the pylorus and circumferentially from the lesser to the greater curvature.
In the present model, this spatial variation is captured by an excitability parameter $a^\ICC(\vec{X})$, which controls the local responsiveness of \iccs{}. We build on our previous work~\cite{brandstaeter2018a}, where a similar formulation was introduced for idealized geometries, and extend it here to a realistic stomach geometry. This excitability parameter depends on two scalar fields that define an intrinsic coordinate system aligned with physiologically meaningful axes of the stomach: $\xi_l(\vec{X}) \in [0,1]$, representing the longitudinal axis of the stomach, and $\xi_c(\vec{X}) \in [0,1]$, representing the circumferential coordinate from the greater to the lesser curvature. Both normalized control coordinates are constructed by solving independent Laplace-Dirichlet problems over the reference configuration $\Omega_\textup{R}$. 
The longitudinal harmonic field $\phi_\fp$ is defined in~\cref{eq:laplace-dirichlet-fibers}, while the circumferential field $\phi_\gl$ is obtained by solving
\begin{align}
\begin{cases}
    \begin{aligned}
     -\Delta \phi_\gl&= 0 && \text{in $\Omega_\textup{R}$}\, , \\
    \phi_\gl                                &= 1 && \text{on $\Gamma_\textup{lesser}$}\, , \\
    \phi_\gl                                &= 0 && \text{on $\Gamma_\textup{greater}$}\, ,
    \end{aligned}
\end{cases}
\end{align}
where $\Gamma_\textup{lesser}$ and $\Gamma_\textup{greater}$ are the Dirichlet boundaries describing the lesser and greater curvatures, respectively.

To support spatially varying excitability with physiologically meaningful regional distinctions, we define a threshold value $\phi^* \in (0,1)$, corresponding to an isocontour value of $\phi_\fp$. This value is automatically determined from anatomical features, such as points near the esophageal sphincter on the lesser curvature. Specifically, the geometric center and average radius of the esophageal sphincter outlet are computed. The algorithm then identifies the closest node along the lesser curvature whose radial distance from the center exceeds $1.3$ times the average outlet radius. The value of $\phi_\fp$ at this node is then assigned as the threshold $\phi^*$. Consequently, $\phi^*$ depends on the specific geometry under consideration; for the geometry in~\cref{sec:case2_simulated_gastric_peristalis}, this yields $\phi^* \approx 0.543$. Using this threshold, we define the normalized longitudinal coordinates $\xi_l$ in a piecewise manner as
\begin{align}
\label{eq:control_variable_long}
    \xi_l(\vec{X}) = \begin{cases}
    \begin{aligned}
      \frac{\phi_\fp-\phi^*}{1-\phi^*}\qquad& \text{if } \phi_\fp \geq \phi^*\, \quad  (\text{distal stomach})\, , \\
   \frac{\phi^*-\phi_\fp}{\phi^*}\qquad& \text{if } \phi_\fp < \phi^*\, \quad (\text{proximal stomach})\, ,
    \end{aligned}
\end{cases}
\end{align}
while the circumferential coordinate is defined as
\begin{align}
\label{eq:control_variable_cir}
    \xi_c(\vec{X}) =\phi_\gl\, .
\end{align}
\begin{figure}[tp]
    \centering
    \subfloat[]{%
        \includegraphics[trim={0cm 0cm 0cm 0cm},width=0.4\textwidth]{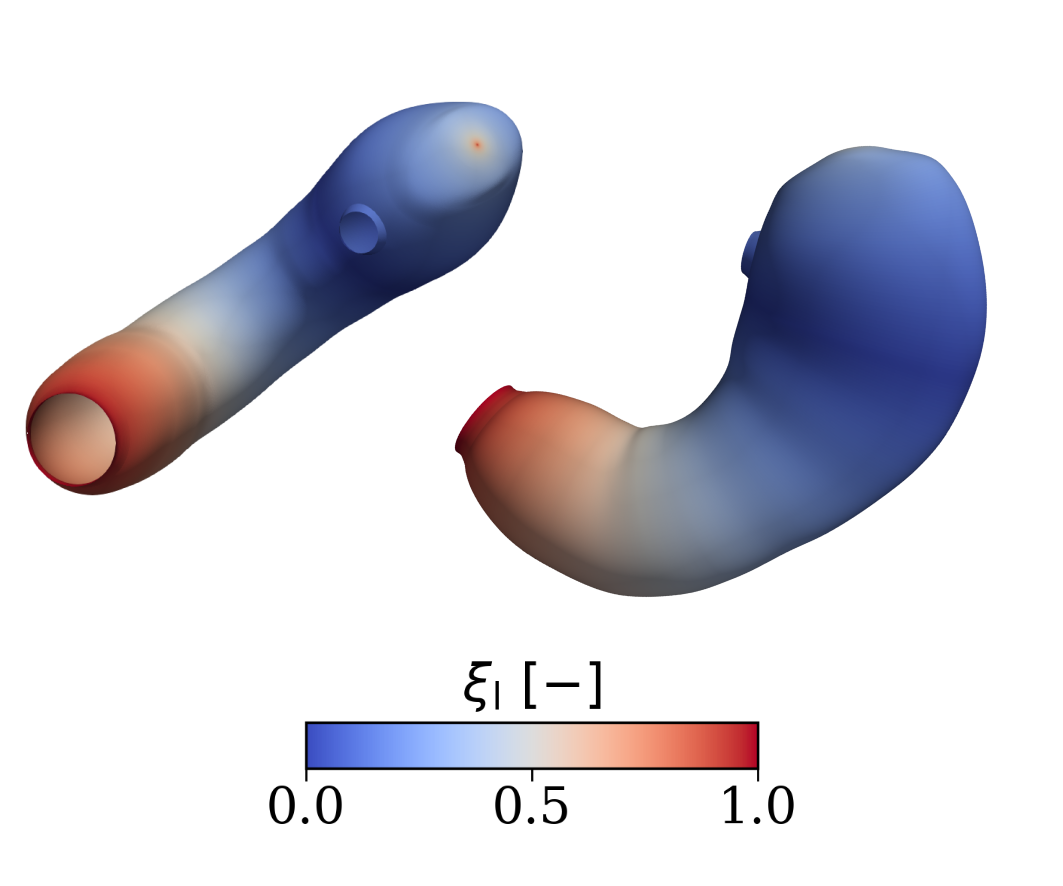}\label{fig:stomach_xi_l}}
        \hspace{2cm}
    \subfloat[]{%
        \includegraphics[trim={0cm 0cm 0cm 0cm},width=0.4\linewidth]{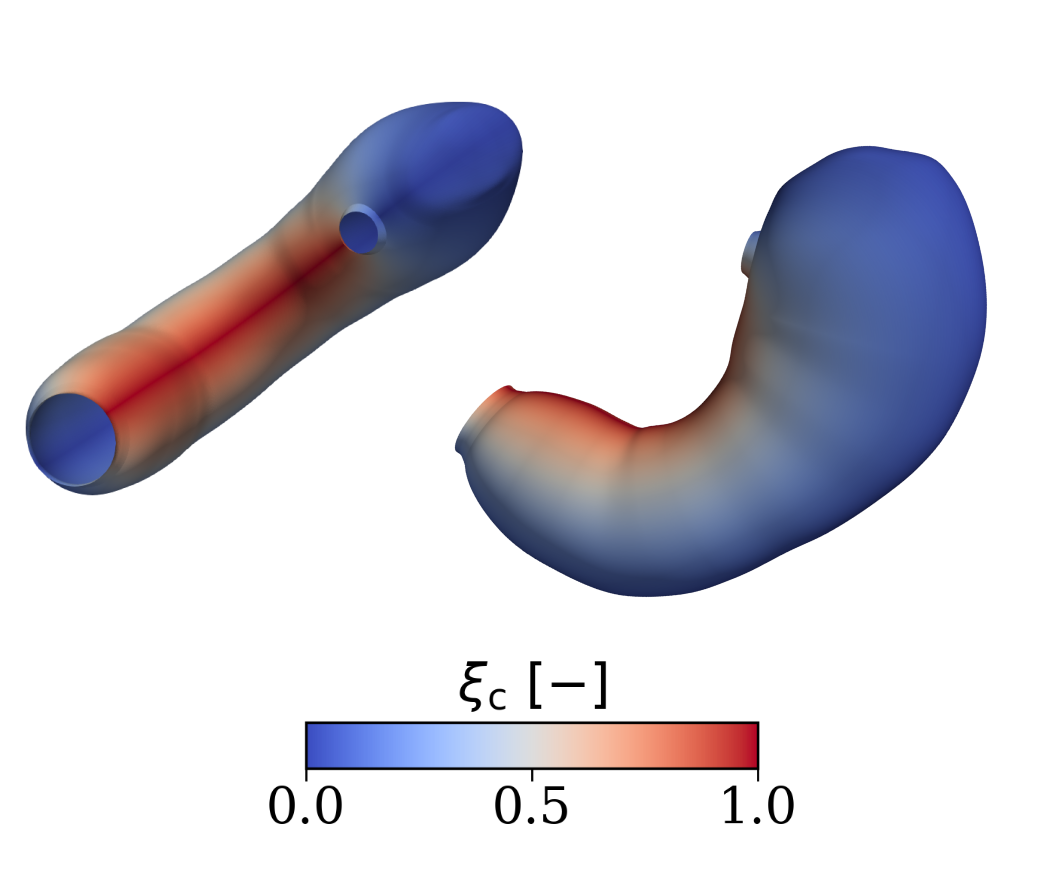}\label{fig:stomach_xi_c}}
    \caption{Control variables used for spatial parameter modulation visualized from two perspectives. 
    (a) The normalized longitudinal coordinate $\xi_l$, defined piecewise in~\cref{eq:control_variable_long}, distinguishes the proximal and distal regions of the stomach and allows longitudinal variation of model parameters.
    (b) The normalized circumferential coordinate $\xi_c$ follows the longitudinal field $\phi_\gl$ and allows circumferential parameter modulation.
    These coordinates provide a structured framework to spatially vary parameters such as excitability along anatomically meaningful axes.
    }\label{fig:control_variables}
\end{figure}
Although $\xi_l$ is piecewise-defined for the proximal and distal stomach, it is constructed to remain continuous across the transition at $\phi_\fp = \phi^*$. Both control variables $\xi_l$ and $\xi_c$ are visualized in~\cref{fig:control_variables} and serve to systematically modulate tissue properties in the model along the longitudinal and circumferential directions.
These coordinates allow continuous spatial modulation of parameters in both regions independently, while preserving anatomical and physiological relevance.
Accordingly, the excitability of \iccs{} is modeled as
\begin{align}
\label{eq:Exci-a}
    a^\ICC(\vec{X}) = a_\textup{min}^{\ICC,r}+(a_\textup{max}^\ICC - a_\textup{min}^{\ICC,r}) f_l^r(\xi_l)f_c^r(\xi_c)\, ,
\end{align}
where $r \in \{\textup{prox}, \textup{distal}\}$ indicates whether the current point $\vec{X}$ is in the proximal or distal region. The parameter $a_\textup{min}^{\ICC,r}$ denotes the minimal excitability in each region and $a_\textup{max}^\ICC$ is the global upper bound for excitability. The excitability parameter $a^\ICC$ enters the model equations via the inward current terms, where it modulates the ionic fluxes generated by \iccs{}, see~\cref{eq:InwardCurrent}.
The shaping functions $f_l(\xi_l)$ and $f_c(\xi_c)$ determine the longitudinal and circumferential variation of excitability in each region:
\begin{subequations}
\begin{align}
f_l(\xi_l) &= 1- \frac{c_l^r-1}{\exp{[-b_l^r]-1}} \, \Big(1-\exp{[b_l^r(\xi_l)^2]}\Big)\, ,\\
f_c(\xi_c) &= 1- \frac{c_c^r-1}{\exp{[-b_c^r]-1}} \, \Big(1-\exp{[b_c^r (\xi_c)^2]}\Big)\, ,
\end{align}
\label{eq:Exci_Gaussian}
\end{subequations}
where the parameters $b_l^r$, $b_c^r$, $c_l^r$, $c_c^r$ control the shapes and decay behavior in each direction (cf.~\cref{tab:excitability_parameters}).
The resulting excitability field is shown in~\cref{fig:a_i}. 
In contrast, \smcs{} do not actively trigger electrical signals. Therefore, the parameter for \smcs{}, $a^\SMC$, is set to zero ($a^\SMC(\vec{X}) = 0$), reflecting their passive behavior.

%%%%%%%%%%
\subsubsection{Regional contributions of ICCs and SMCs}\label{sec:regional-contributions-iccs-smcs}
The proximal region of the stomach is known to exhibit minimal electrical activity and does not generate peristaltic contractions~\cite{Grady-2021-GastricConductionReview}. 
To account for heterogeneous contributions of \iccs{} and \smcs{} in the normalized form of the monodomain formulation, we introduce dimensionless scaling functions $\chi^\ICC$ and $\chi^\SMC$, which represents the regional contributions of \icc{} and \smcs{}, respectively. 
For \iccs{}, which are excitable cells and the primary drivers of slow wave initiation and propagation, $\chi^\ICC$ defines the regional capacity to generate electrical signals. It varies smoothly across the domain to reflect known spatial activation patterns. To represent this, we use the same normalized control coordinates $\xi_l(\vec{X}) \in [0,1]$ and $\xi_c(\vec{X}) \in [0,1]$ (see~\cref{eq:control_variable_long,eq:control_variable_cir}). Consistent with the excitability formulation, $\chi$ is spatially modulated by the longitudinal and circumferential shaping functions $f_l(\xi_l)$ and $f_c(\xi_c)$ (see~\cref{eq:Exci_Gaussian}), allowing a physiologically meaningful heterogeneity between proximal and distal stomach regions. Specifically, it is defined as
\begin{align}
\label{eq:surface_to_volume_ratio}
    \chi^\ICC(\vec{X}) =  \begin{cases}
    \begin{aligned}
1.0
 \qquad& \text{if } \phi_\fp \geq \phi^*\, \quad (\text{distal stomach})\, ,\\
    f_l(\xi_l)f_c(\xi_l)
\qquad& \text{if } \phi_\fp < \phi^*\,  \quad (\text{proximal stomach})\,  .
    \end{aligned}
\end{cases}
\end{align}
The resulting heterogeneous distribution of $\chi^\ICC$ is illustrated in~\cref{fig:chi_i}. 

The same spatial weighting function is applied to the intracellular coupling term via the field $\chi^\textup{gap}(\vec{X})$, which represents the regional capacity of \iccs{} to transmit signals to \smcs{} through gap junctions. Specifically, we set $\chi^\textup{gap}(\vec{X}) = \chi^\ICC(\vec{X})$ to account for the reduced density of ICC–SMC gap junctions in the proximal region and the spatial specificity of their electrical coupling~\cite{Sanders-2016-RegulationSMCFunction,Grady-2021-GastricConductionReview}.
In contrast, for \smcs{} we assume a uniform distribution of $\chi^\SMC(\vec{X}) = 1$ throughout the domain, reflecting the assumption of spatially homogeneous \smc{} properties.
%%%%%%%%%%%%%%%%%%%%%%%%%%%%%%%%%%%%%%%%%%%%%%%%%%%%%%%%%%%%%%%%%%%%%%%%%%%%%%%%%%%%%
\subsubsection{Electrical diffusivity}
\label{sec:heterogenous_diffusivity}
The diffusion coefficients $\sigma^\ICC$ and $\sigma^\SMC$ represent the electrical coupling within \icc{} and \smc{}, respectively.
Experimental studies report spatial variation in gastric slow wave propagation velocity, with higher velocities observed near the greater curvature and lower velocities along the lesser curvature of the stomach~\cite{EgbujiGrady-2010-OriginPropagationSW,Hosseini2023}.
This observation aligns with anatomical considerations: peristaltic waves propagate along the centerline of the stomach, maintaining a trajectory perpendicular to it. 
To preserve this orientation, waves must traverse a longer path along the greater curvature than the lesser curvature, necessitating a higher conduction velocity at the greater curvature.
Previous studies have attributed the observed propagation patterns and associated conduction velocities primarily to intrinsic frequency gradients within gastric tissue. However, our findings suggest that this explanation is insufficient.
Instead, we demonstrate that tissue diffusivity $\sigma^\ICC$ plays a more significant role in determining the conduction velocity (cf.~\ref{ap:influence_of_diffusion_coefficient}).
To capture this effect, we introduce a heterogeneous distribution for $\sigma^\ICC$ and $\sigma^\SMC$, with increased diffusivity values near the greater curvature---accounting for the higher observed conduction velocities---and decreased diffusivity values near the lesser curvature, consistent with the slower velocities reported in that region.

To model this effect, we assume that the local conduction velocities correlates with the gradient magnitude of a harmonic field $\phi_\fp$, which encodes the longitudinal axis from the fundus to the pylorus. Specifically, the quantity $\norm{\nabla \phi_\fp}$ reflects the local rate of change along this axis. While this field does not mechanistically drive heterogeneity, it provides an anatomically meaningful surrogate to prescribe a spatially varying distribution for tissue diffusivity.
As a physiological reference, we prescribe a fixed diffusivity at the lesser curvature boundary $\Gamma_\textup{lesser}$:
\begin{align}
    \sigma^\ICC(\vec{X}) =\sigma_\textup{lesser}^i \, \quad \forall \,  \vec{X} \in \Gamma_\textup{lesser}\, 
\end{align}
and assume a linear relationship between the conduction velocity $v_c$ and the diffusion coefficient of \iccs{}, $\sigma^\ICC$,
\begin{align}
    v_c = m_\sigma \sigma^\ICC + c_\sigma \, , \qquad m_\sigma>0\, , \quad c_\sigma \in \mathbb{R}\, .
\end{align}
We postulate that $v_c$ is proportional to $\norm{\nabla \phi_\fp}$,
\begin{align}
    v_c(\vec{X}) = v_c (\vec{X}_\textup{lesser})\frac{\norm{\nabla \phi_\fp(\vec{X})}}{\norm{\nabla \phi_\fp(\vec{X}_\textup{lesser})}} \, ,
\end{align}
where $\vec{X}_\textup{lesser}$ denotes a point on the lesser curvature such that $\phi_\fp(\vec{X}_\textup{lesser})=\phi_\fp(\vec{X})$, i.e., both points lie on the same isocontour of the longitudinal field. The parameters $m_\sigma$ and $c_\sigma$ are calibrated by fitting conduction velocity and diffusion coefficients using a simplified one-dimensional simulation setup, while experimental conduction velocities are taken from literature. The calibrated parameters are summarized in~\cref{tab:baseline_parameters}.
Combining these relations and solving for $\sigma^\ICC(\vec{X})$ yields the closed‐form expression
\begin{align} 
    \sigma^\ICC(\vec{X}) =\frac{\norm{\nabla \phi_\fp(\vec{X}_\textup{lesser})}}{\norm{\nabla \phi_\fp(\vec{X})}}\left( \sigma^\ICC_\textup{lesser} + \frac{c_\sigma}{m_\sigma}\right)- \frac{c_\sigma}{m_\sigma}\, .
\end{align}
This formulation produces a spatially continuous and anatomically consistent distribution of $\sigma^\ICC$, capturing regional electrophysiological variability in gastric slow wave propagation (cf.~\cref{fig:sigma_i}). 

In alignment with~\cite{brandstaeter2018a}, we set the field of the \smc{} diffusion coefficient to $\sigma^\SMC(\vec{X}) = 0.1\sigma^\ICC(\vec{X})$.
This reflects the comparatively weaker electrical coupling among \smcs{} relative to \iccs{}.

%%%%%%%%%%%%%%%%%%%%%%%%%%%%%%%%%%%%%%%%%%%%%%%%%%%%%%%%%%%%%%%%%%%%%%%%%%%%
\subsection{Boundary and initial conditions}
\label{sec:boundary_conditions}
\begin{figure}[!t]
    \centering
        \includegraphics[width=0.8\linewidth]{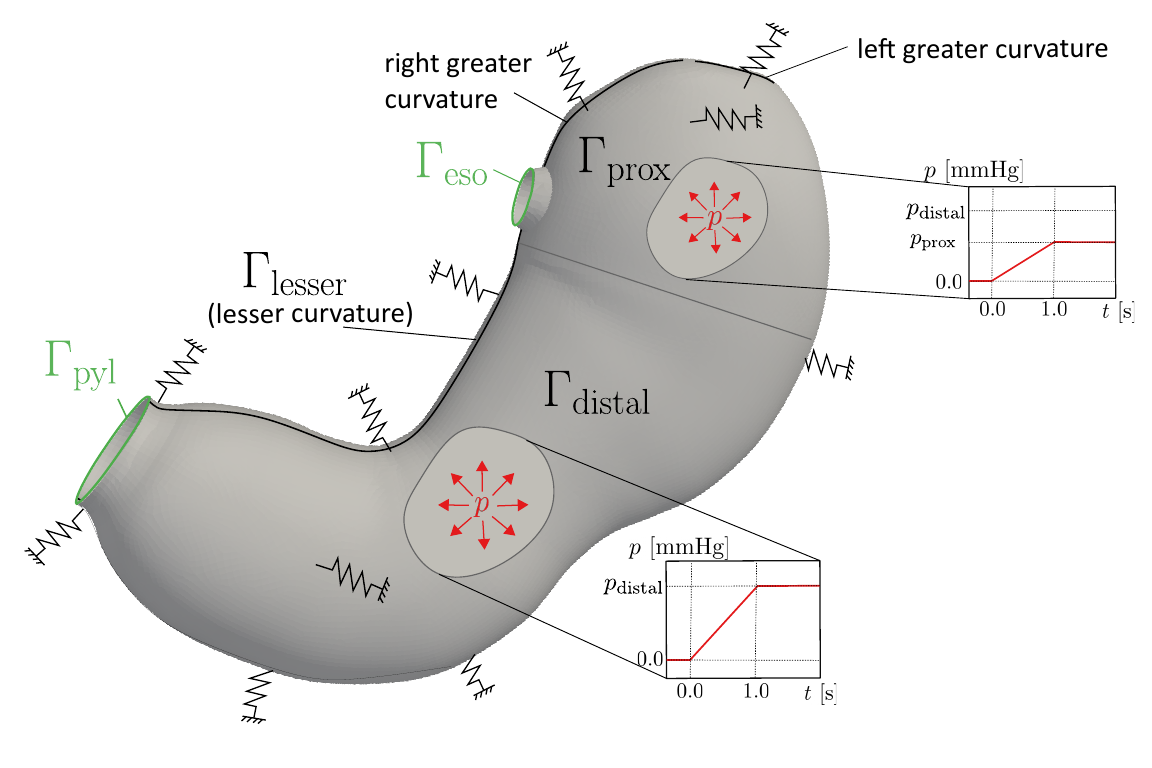}
        \caption{Boundary conditions applied to the stomach geometry. Spring elements act in the surface-normal direction to model surrounding tissue and organ support, with increased stiffness along the greater ($\Gamma_\textup{greater}$) and lesser curvature ($\Gamma_\textup{lesser}$). Note that $\Gamma_\textup{greater}$ includes both the left and the right greater curvature. Axial constraints are applied at the lower esophageal sphincter ($\Gamma_\textup{eso}$) and pyloric sphincter ($\Gamma_\textup{pyl}$). A spatially varying intraluminal pressure is applied across the surface to replicate physiological loading.}
    \label{fig:stomach_boundary_conditions}
\end{figure}
The definition of appropriate boundary conditions is critical in any simulation model, however, modeling the human stomach is particularly challenging due to its highly complex and dynamic mechanical environment.
The stomach is subject to multiple interacting constraints arising from its anatomical attachments, including the esophagus proximally, the duodenum distally, ligaments such as the omentum, and mechanical interactions with neighboring organs and tissues. Large deformations during gastric contractions, combined with interaction with internal fluids, complicate the accurate specification of boundary conditions. These factors introduce nonlinearities and spatially heterogeneous mechanical responses that must be carefully represented to ensure realistic simulations. An additional challenge arises from substantial inter-subject variability in gastric shape and orientation, which affects the nature of contact with surrounding tissues and organs. This variability results in patient-specific mechanical boundary conditions that are difficult to characterize experimentally but can be explored computationally.

To model these effects, we apply spring elements acting in the surface-normal direction across the stomach surface with spatially varying stiffness, see~\cref{fig:stomach_boundary_conditions}. Higher stiffness values are applied along the greater and lesser curvatures, as well as close to the pyloric sphincter, to reflect anatomical support to the omentum and the relatively stiff pylorus tissue. Transitions between stiffness regions are smoothed using sigmoid functions based on the harmonic field $\phi_\gl$ (see~\cref{sec:intrinsic-frequencies}) to avoid stress concentrations and ensure numerical stability.

At anatomical openings, namely the pyloric and lower esophageal sphincter, zero Dirichlet boundary conditions are imposed only in the longitudinal direction, restricting axial displacement while allowing radial and circumferential deformation. This configuration reflects the anatomical support provided by the adjoining duodenum and esophagus, which constrains motion along the longitudinal axis while permitting local wall deformation during peristalsis. In addition, spring-type boundary conditions are applied in the surface-normal direction to represent the elastic compliance of the pyloric sphincter region, preventing excessive displacement and ensuring controlled radial contraction. This combined approach reproduces the limited longitudinal motion observed in MRI studies~\cite{Hosseini2023} and preserves physiologically realistic deformation of the distal stomach.

To approximate the mechanical effect of digesta effects in the absence of fluid-structure interaction, we apply spatially varying pressure loads acting normal to the stomach wall. These loads emulate the physiological distension caused by chyme. Higher pressure values are applied in the distal stomach (\SI{25}{\mmHg}) to reflect its role in active mixing and digestion, while lower values are assigned in the proximal stomach (\SI{10}{\mmHg}) for its primary function in accommodation. These values lie within the physiological intragastric pressure range~\cite{Janssen-2011-IntragastricPressure,Wali-2013-PatientFullStomach} and, together with our prestressing procedure, ensure a realistic initial stress state of the gastric wall. This approach represents a first-step approximation of digesta effects, allowing us to focus on the electromechanical behavior. It still captures, to a certain degree, the essential mechanical influence of the chyme, thereby enabling stable and physiologically plausible electromechanical simulations.

For the electrophysiological problem, we assume electrical insulation and impose no-flux Neumann boundary conditions at the lower esophageal sphincter and pyloric sphincter
\begin{align}
    \nabla v^j \cdot \vec{n} = 0  \quad \text{on} \, \, \Gamma_\textup{eso} \cup \Gamma_\textup{pyl} \times [0, \infty) \quad \, \forall\, j \in \{\ICC, \ \SMC\},
\end{align}
where $\vec{n}$ denotes the outward unit normal. A complete list of the boundary and initial condition values is provided in~\cref{tab:boundary-conditions,tab:inital-conditions}.

%%%%%%%%%%%%%%%%%%%%%%%%%%%%%%%%%%%%%%%%%%%%%%%%%%%%%%%%%%%
\section{Numerical implementation}
\label{sec:numerical_implementation}

\subsection{General}
We implemented the coupled gastric motility model in~\cref{sec:electromechanical_modeling_of_gastric_peristalsis} into the open-source multiphysics simulation framework 4C~\cite{4C}. Both the scalar transport problem and the solid mechanics problem are addressed numerically via the finite element method. Linear triangular scalar transport elements and linear quadrilateral nonlinear finite shell elements, based on the $7$-parameter formulation, are used for the discretization. The implementation supports non-conforming meshes at the structure-scalar interface through a dual mortar method~\cite{Popp-2009-Mortar}, enforcing constraints via a Lagrange multiplier approach with node-wise penalty regularization. This approach ensures stable and accurate transfer of information between electrophysiology and mechanics meshes, while allowing independent mesh refinement. In particular, since the simulation cost is largely driven by the solid mechanical problem, a coarser mesh can be used for this discretizations, improving computational efficiency without compromising solution fidelity. For further details on the mortar methods, the reader is referred to~\cite{Popp-2009-Mortar,Popp-2010-DualMortar}.

\subsection{Convergence study}
\label{sec:convergence_study}
To verify the correct implementation and assess the numerical robustness of the proposed mathematical model, we conducted systematic convergence studies utilizing QUEENS~\cite{Biehler2025}, an open-source Python framework for solver-independent multi-query analyses of large-scale computational models. These studies aimed to determine optimal temporal and spatial discretization parameters for both the electrophysiological and mechanical subproblems, ensuring numerical accuracy and computational efficiency.

\subsubsection{Electrophysiology}
It is well established that the conduction velocity $v_c$ in reaction-diffusion systems is sensitive to the discretization parameters of the numerical scheme~\cite{brandstaeter2018a}, particularly the time step $dt$ and the mesh size $h$. We investigated the convergence behavior of $v_c$ in a $1$D gastric electrophysiology model. Mechanical coupling was excluded from this study to isolate electrophysiological properties. The computational domain spans the length of the stomach from the pacemaker region to the pylorus and is defined as a $1$D line with $x \in [\SI{0.0}{\mm}, \SI{250.0}{\mm}]$. No-flux boundary conditions are applied at both ends to reflect the electrically quiescent nature of the fundus and the pylorus. The intrinsic frequencies of \iccs{} are prescribed using a Gaussian-like function, as described in~\cref{sec:intrinsic-frequencies}, with parameters listed in~\cref{tab:excitability_parameters}.

To evaluate the influence of discretization on the simulated conduction velocity $v_c$ we varied the time step size $dt$ and the mesh size $h$ across the ranges \SI{31.25}{\ms} $\leq dt \leq $\SI{0.2}{\s} and \SI{15.625}{\micro\m} $\leq h \leq $\SI{1.0}{\mm}, respectively. Each parameter was halved iteratively, resulting in a total of $49$ simulations covering the parameter space. 
The conduction velocity $v_c$ was calculated based on the time $\Delta t$  required for the rising phase of the transmembrane potential $v^\ICC$ to propagate from $x_1 =$ \SI{120}{\mm} to $x_2 =$ \SI{130}{\mm}.

\begin{figure}[!t]
    \centering
    \subfloat[]{%
        \includegraphics[width=0.48\linewidth]{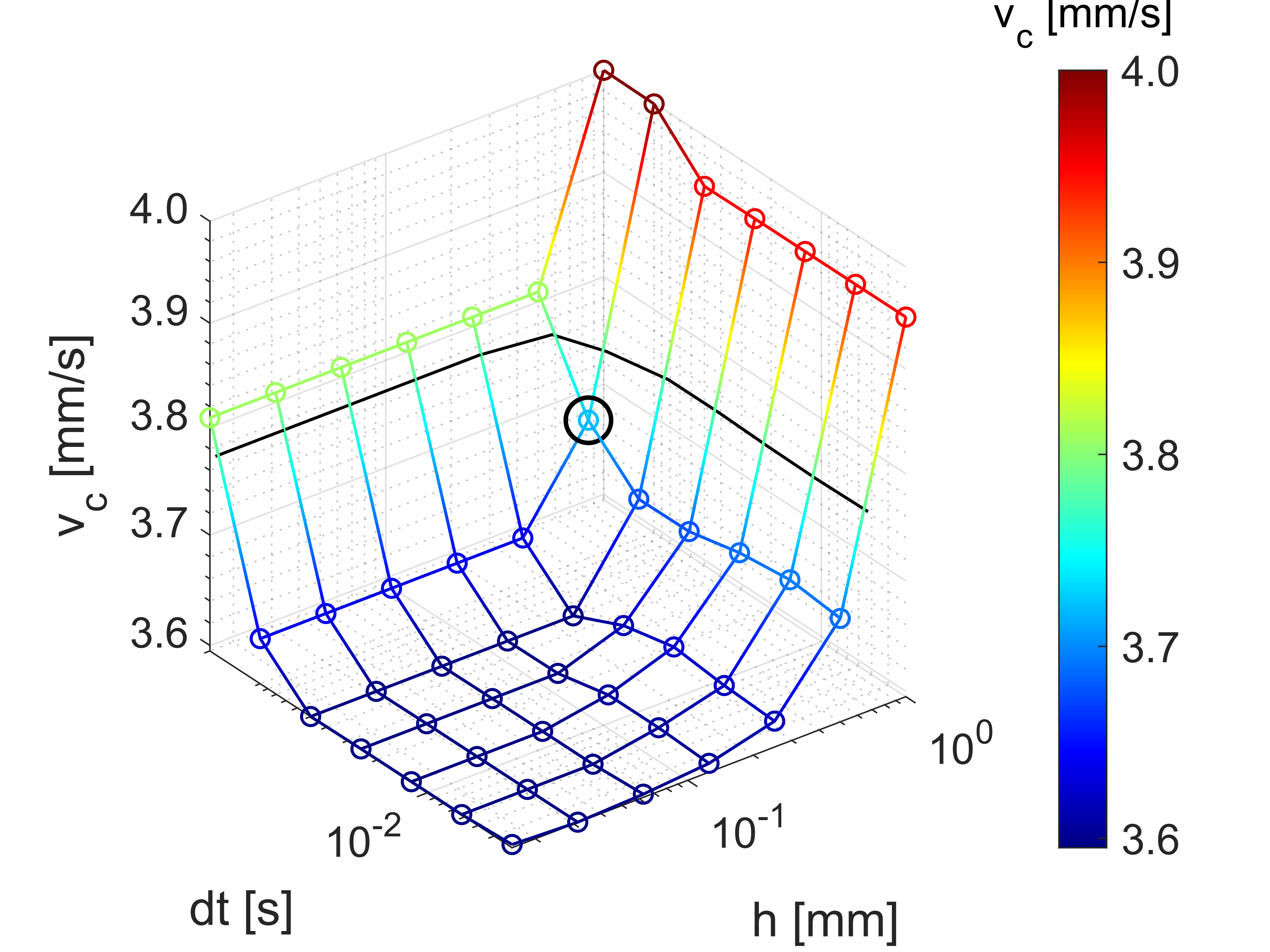}\label{fig:convergence_study_ep}}\quad
    \subfloat[]{%
        \includegraphics[width=0.48\linewidth]{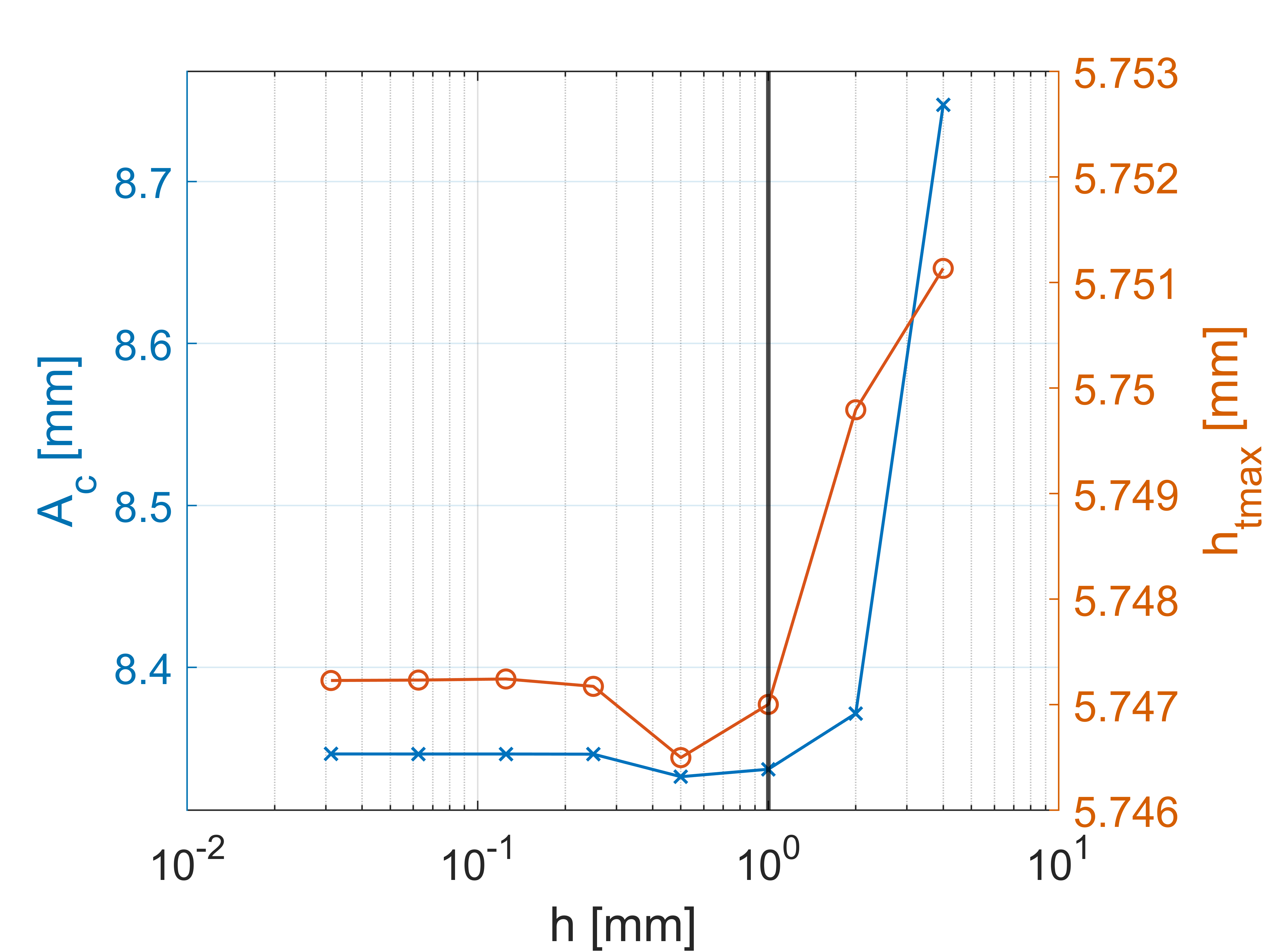}\label{fig:solid_conv_study_A_c}}
    \caption{Numerical convergence analysis of the electrophysiology and electromechanical models. (a) Convergence of conduction velocity $v_c$ in a $1$D line domain, evaluated for varying time step sizes $dt$ and mesh sizes $h$. The black contour indicates parameter combinations with less then $5\%$ relative error compared to a high-resolution reference solution ($dt$ = \SI{3.125}{\ms}, $h =$ \SI{15.625}{\micro\m}). The circle marks the discretizations chosen for all subsequent electrophysiology simulations. (b) Convergence of contraction amplitude $A_\textup{c}$ and maximum tissue thickness ${h_t}_{max}$ in a cylindrical wedge domain, evaluated for different solid mesh sizes $h$ using a static activation profile for electrophysiology. The black vertical line denotes the selected mesh size for all subsequent solid mechanics simulations.}  \label{fig:convergence_studies}
\end{figure}

To avoid boundary effects, all calculations were performed after \SI{450}{\s} of simulated time, ensuring a fully entrained slow wave pattern. The results, summarized in~\cref{fig:convergence_study_ep}, illustrate the convergence behavior of $v_c$ with decreasing time step and mesh size. The black contour line indicates a $5\%$ relative deviation from the reference value obtained with the finest spatio-temporal resolution in this analysis. Based on these results, we selected $dt=$ \SI{0.1}{\s} and $h=$ \SI{0.5}{\mm} to solve the electrophysiological problem in all subsequent examples. This choice provides a compromise between numerical accuracy and computational cost, maintaining relative errors below $5\%$. This analysis confirms that the selected time step and the mesh size yield conduction velocities within physiologically plausible ranges~\cite{EgbujiGrady-2010-OriginPropagationSW}.

\subsubsection{Coupled electromechanics}
To determine appropriate numerical parameters for the solid mechanics component of the coupled electromechanical model, we performed a second convergence analysis using a coupled gastric motility simulation. Building on the electrophysiological convergence study, we next assessed the mechanical response under coupled conditions. In contrast to the previous study, this analysis was carried out on a $2$D cylindrical wedge, as defined in~\cref{sec:FEMcomparisonSetup}, to capture physiologically relevant fiber orientations. The mechanical response was modeled using bilinear quadrilateral $7$-parameter shell elements. %, incorporating \ans{}, \eas{}, and \sdc{} enhancements to improve numerical performance. 
The constitutive framework follows the finite elasticity formulation presented in~\cref{sec:electromechanical_modeling_of_gastric_peristalsis}, incorporating  a constrained mixture model with active strain and prestress. All model parameters were adopted from~\cref{tab:realistic_stomach_parameters}, which summarizes the baseline values of the constrained mixture model. The parameters that were specifically modified here are listed in \cref{tab:parameters_cylinder_hcm}. In particular, identical material properties were applied in both the circumferential and longitudinal directions, using those defined for the longitudinal direction.

To isolate the numerical influence of the mechanical problem, the electrophysiological activation pattern was precomputed using the time step and spatial resolution identified in the first study ($dt=$ \SI{0.1}{\s} and $h=$ \SI{0.5}{\mm}). This precomputed activation was then coupled with the mechanical problem, in which only spatial resolution was varied over the range \SI{3.125}{\micro\m} $\leq h \leq $ \SI{4.0}{\mm}, while the mechanical time step remained consistent with that of the electrophysiology problem. 
We examined the influence of solid mesh size on the contraction amplitude $A_\textup{c}$ and the maximum stomach wall thickness $h_{t,\textup{max}}$, using the static deformation field. The contraction amplitude is defined as the range of normal displacement in the deformed configuration relative to the reference (resting) state:
\begin{align}
A_\textup{c} &= \max \left( u_n (\vec{x}) \right) - \min \left( u_n(\vec{x}) \right)\, , \quad \text{with } u_n (\vec{x})=\vec{u}(\vec{x}) \cdot \vec{n}(\vec{x})\, ,
\end{align}
where $ u_n(\vec{x})$ denotes the projection of the displacement vector $\vec{u}(\vec{x})$ onto the local surface outward normal $\vec{n}(\vec{x})$. Positive values of $u_n$ indicate inward motion (contraction), while negative values correspond to outward motion (distension).
Both quantities are evaluated in the central region of the cylindrical domain ($x \in [\SI{50.0}{\mm}, \SI{200.0}{\mm}]$) to minimize boundary effects. 
As shown in~\cref{fig:solid_conv_study_A_c}, the contraction amplitude converges to a stable value as the mesh is refined. Coarser meshes overestimate the magnitude of contraction, exhibiting noticeable deviations from the trend observed at finer resolutions. The results highlight the critical role of spatial discretization in accurately capturing the mechanical response and support the choice of a mesh size of \SI{1.0}{\milli\meter} for subsequent simulations.

%%%%%%%%%%%%%%%%%%%%%%%%%%%%%%%%%%%%%%%%%%%%%%%%%%%%%%%%%%%
\subsection{Comparison between different finite element types}
\label{sec:assesment-of-shell-elements}

\subsubsection{Membrane - solid - shell}
The choice of finite element formulation significantly impacts accuracy, computational efficiency, and ease of implementation. While $3$D solid elements capture detailed representations of mechanical behavior, they are prone to locking phenomena and pose challenges in meshing thin geometries. Ensuring a proper aspect ratio often requires a fine in-plane mesh, leading to increased degrees of freedom and higher computational costs. Moreover, solid meshes tend to become distorted under large deformations, potentially causing inaccurate or unstable simulations; in the worst case, this necessitates re-meshing to maintain solution fidelity~\cite{Liu2024}. To address these issues, $2$D manifold representations—membrane and shell elements—can be adopted. However, their comparative performance, particularly in gastrointestinal biomechanics, remains underexplored. 
This section evaluates three element formulations for gastric electromechanics: (i) a nonlinear finite membrane element model~\cite{Gruttmann-1992-membrane}, (ii) a $3$D solid eight-noded hexahedral (HEX$8$) elements using the F-Bar method to avoid volumetric locking in the case of near incompressibility, and (iii) the $7$-parameter shell element model introduced above. To mitigate locking pathologies (such as in-plane shear, transverse shear, membrane, 
Poisson thickness, and curvature thickness locking) associated with the finite shell element implementation, we employ both the \eas{}~\cite{AndelfingerRamm-1993-EAS,Büchter-1994-3DExtensionEAS,SimoRifai-1990-EAS} and the \ans{}~\cite{BatheDvorkin-1985-ANS} methods. The original $7$-parameter shell formulation can produce ill-conditioned stiffness matrices due to the thin-walled nature and large eigenvalue spectrum. To address this, we incorporate the scaled director conditioning approach by~\cite{Gee-2005-SDC}, a mechanically motivated preconditioner tailored for thin-walled structures discretized with continuum-based elements.

To illustrate performance differences between the above three structure models (membrane, $3$D solid, shell) we first present a numerical example that illustrates the distinct performance characteristics of each element type, focusing on their ability to reproduce contraction wave dynamics in a simplified gastric geometry.

\subsubsection{Setup}
\label{sec:FEMcomparisonSetup}
The computational domain is a cylindrical wedge with an opening angle of $0.563^\circ$. The cylinder's axis aligns with the $x$-axis. It measures \SI{250}{\mm} in length, has a radius of \SI{50.930}{\mm}, and a wall thickness of \SI{3.5}{\mm}. The computational meshes are generated using Coreform Cubit~\cite{coreform2023}.
 
A prior convergence study identified an ideal mesh size of \SI{0.5}{\mm} for the electrophysiology problem and \SI{1.0}{\mm} for the mechanical problem using $7$-parameter shell elements, ensuring converged results (see~\cref{sec:convergence_study}). In the present study, we adopt a uniform mesh size of \SI{0.5}{\mm} for both fields to maintain convergence and consistency between analyses. 
For the solid mechanics problem, three different element formulations are considered: $7$-parameter shell elements, membrane elements, and HEX$8$ solid elements. The in-plane mesh size is consistent across these formulations, while the solid element mesh is further discretized through the thickness with a sufficient number of elements to capture through-thickness variations. Specifically, the meshes comprise (i) \num{500} quadrilateral shell elements, (ii) \num{500} quadrilateral membrane elements, or (iii) \num{3000} HEX$8$ solid elements (\num{500} elements in the axial direction and six in the radial direction).
In each case, the use of a single element in the circumferential direction reflects the underlying rotational symmetry of the problem.
For the electrophysiology problem, a $2$D mesh of $500$ bilinear quadrilateral scalar transport elements is used consistently across all cases, assuming uniformity in the radial direction of the cylinder.

A unidirectional excitability distribution is applied (see~\cref{tab:excitability_parameters}), neglecting circumferential propagation to focus on the fully developed contraction wave state. No-flux boundary conditions are applied for the electrophysiology model at both ends of the cylinder. For the solid mechanics problem, zero Dirichlet boundary conditions are imposed axially at both ends, permitting only radial displacements. 

In contrast to the constrained mixture model in~\cref{sec:constrained_mixture}, this example uses a Neo-Hookean material with the following strain energy functions
\begin{align}
    \Psi_\textup{2D}= c_1 (\textup{tr}\rightCG_\textup{e}-3)\, ,\qquad
    \Psi_\textup{3D}= c_1 (\textup{tr}\rightCG_\textup{e}-3)+\frac{c_1}{c_2}(J_\textup{e}^{-2c_2}-1).
\end{align}
Here, $\Psi_\textup{2D}$ is used for simulations with incompressible membrane elements, while $\Psi_\textup{3D}$ corresponds to the coupled Neo-Hookean formulation (see~\cref{sec:passive-material}), and is applied to simulations involving $7$-parameter shell elements or solid HEX$8$ elements. Material constants are taken from~\cref{eq:strain_energy_coupled_neoHooke} and~\cref{tab:parameters_cylinder}.

The electrophysiology model is solved for \SI{496}{\s}, ensuring entrainment~(see~\cref{sec:case2_sw_entrainment_propagation}). To minimize transient effects from electrophysiology, we consider the activation at this final time point as a static input to the solid mechanics problem. The solid mechanics simulation is then performed for an additional \SI{1}{\s} using a time step size of \SI{0.02}{\s}. Prestress effects are not included here.

\subsubsection{Results}
\begin{figure}[!t]
    \centering
    \subfloat[]{%
\includegraphics[width=\linewidth]{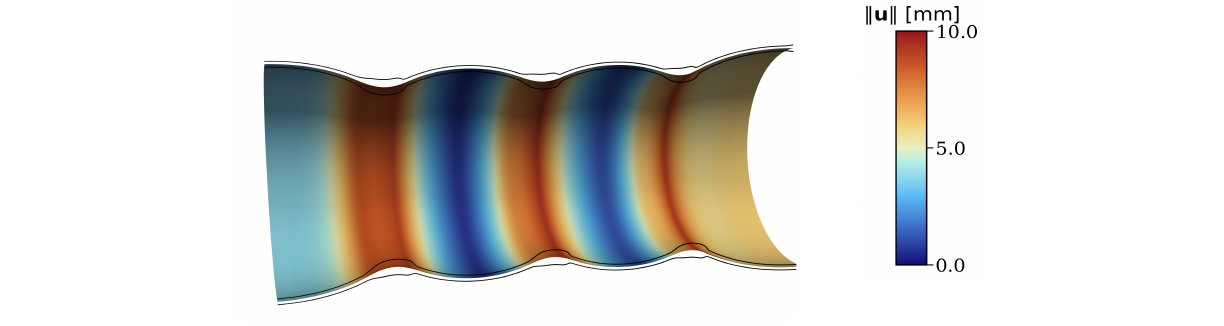}\label{fig:shell_rot_cyl}}\\
    \subfloat[radial displacements $d_r$ over $x$]{%
        \includegraphics[width=0.48\linewidth]{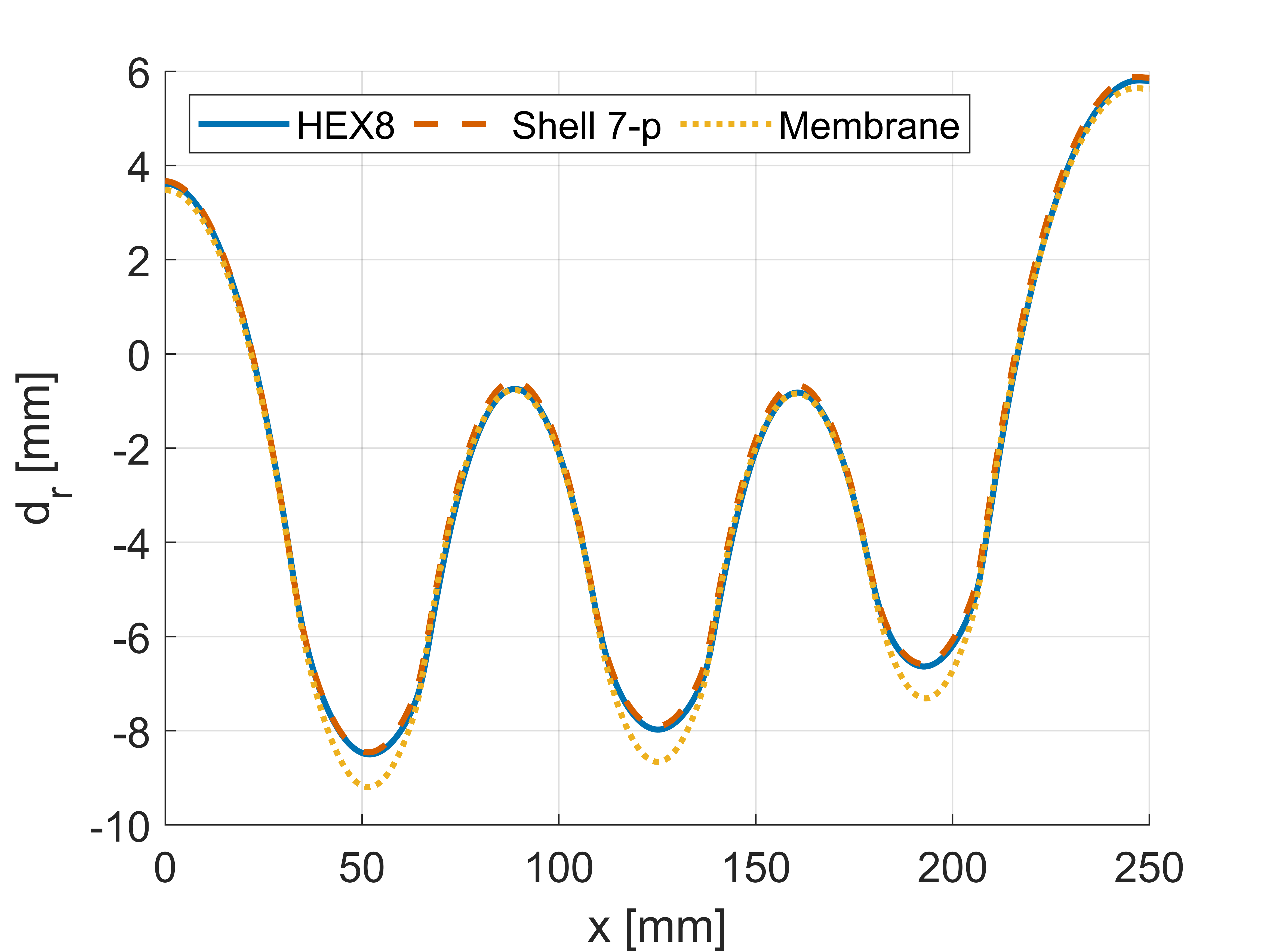}\label{fig:3d_mem_shell_displ}}\quad
    \subfloat[thickness $h_t$ over $x$]{%
        \includegraphics[width=0.48\linewidth]{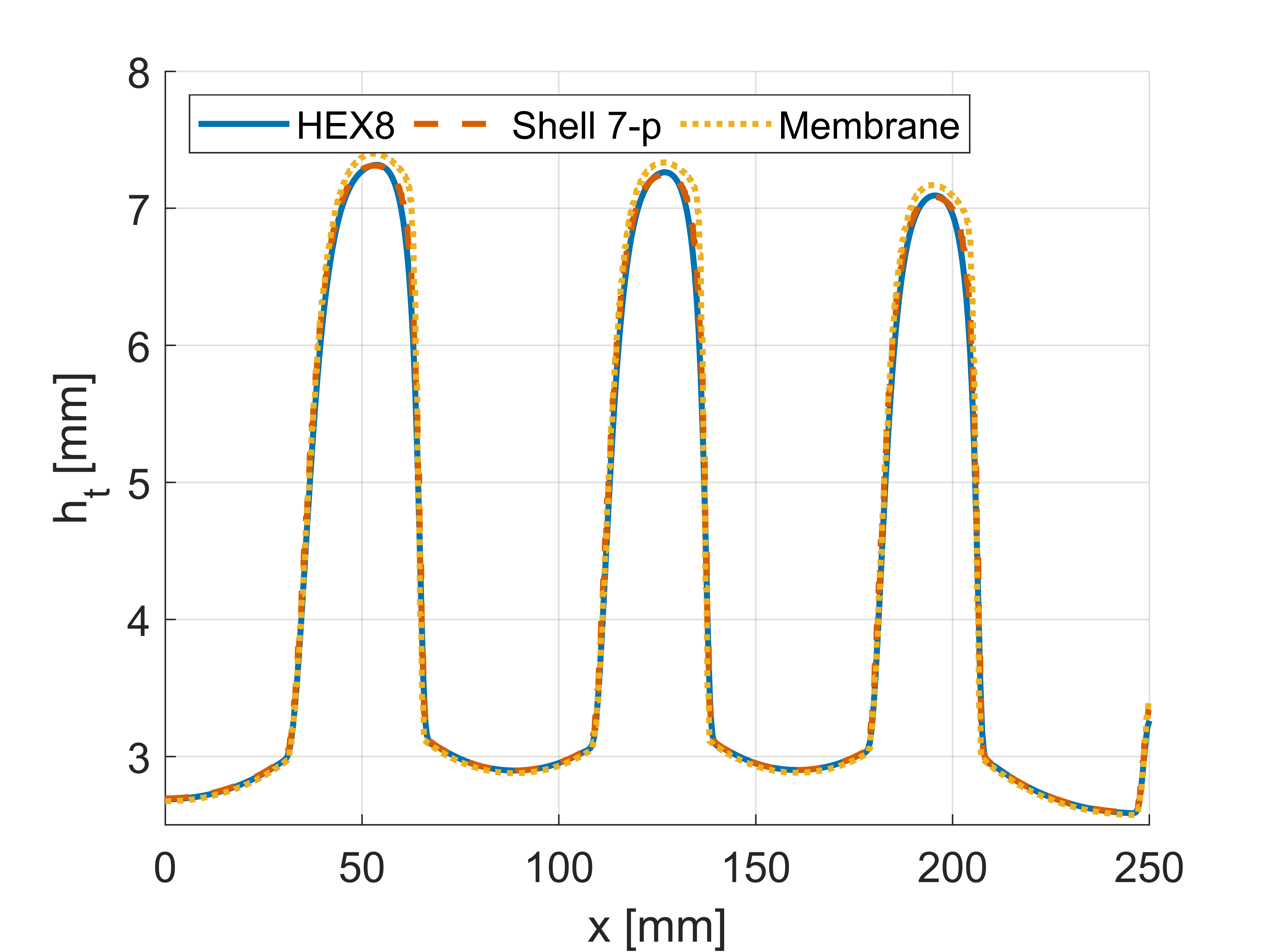}\label{fig:3d_mem_shell_thickness}}
    \caption{Comparison of finite element formulations for gastric electromechanics.
(a) Deformed configuration at $t=$ \SI{496}{\s} using $7$-parameter shell elements, visualized by displacement magnitude. The mesh is rotated about the $x$-axis for clarity. Two black lines indicate the reconstructed $3$D thickness using the shell director field.
(b-c) Axial profiles of radial displacement $d_r$ and wall thickness $h_t$ at the same time point for solid HEX$8$ F-Bar, $7$-parameter shell, and membrane elements. All element types reproduce the qualitative contraction state. Shell and solid formulations yield comparable results, while membrane elements, due to the lack of bending stiffness, exaggerate radial motion and thickness changes in the contraction zone.}  \label{fig:3d_vs_mem_vs_shell}
\end{figure}
\cref{fig:3d_mem_shell_displ} compares the radial displacements $d_r$ along the axial position $x$ for different element formulations. Displacements from the HEX$8$ model are extracted at the midsurface to enable comparison with $2$D formulations. 
The $7$-parameter shell and the HEX$8$ elements yield nearly identical displacement profiles, while membrane elements predict increased displacement magnitudes and exhibit notable thickness changes due to their inability to resist bending. These discrepancies stem from the kinematically enforced incompressibility in membrane formulations, which does not adequately capture out-of-plane bending modes.
The changes of the thickness further illustrate the limitations of membrane elements in reproducing complex $3$D deformation states (cf.~\cref{fig:3d_mem_shell_thickness}). In contrast, the shell elements capture bending effects, providing closer agreement with HEX$8$  elements, although slight differences persist due to the homogeneous thickness assumptions and locking treatment.  
\Cref{tab:elements_summary} summarizes both the accuracy and computational efficiency of the different element formulations. Maximum absolute errors of radial displacement and thickness relative to the HEX$8$ reference are reported, with spatial error distributions consistent with those shown in~\cref{ap:fig:absolute_errors_all}. Relative runtimes and the number of degrees of freedom (DOFs) are also listed. The shell formulation achieves comparable accuracy with roughly three times fewer DOFs and only $22$\% of the HEX$8$ runtime, representing an appealing compromise between accuracy and efficiency. We note that the HEX$8$ solution is employed as a high-fidelity numerical reference rather than an exact analytical solution. Our convergence study indicates that further mesh refinement produces only marginal changes in the results, confirming that the HEX$8$ mesh provides a sufficiently accurate reference for the comparisons presented here.
Solid elements nevertheless yield the most accurate kinematic representation, but their associated computational burden makes them impractical for large-scale or patient-specific simulations. Conversely, membrane elements offer substantial performance improvements but are susceptible to geometric instabilities such as buckling, particularly under higher contraction strengths or complex boundary conditions.
The shell formulation offers an advantageous compromise, combining geometric accuracy, bending and membrane capability, and numerical robustness with manageable computational cost. This balance makes the shell formulation particularly suitable for simulations of gastric motility, where thin-walled morphologies, nonlinear deformations, and large domains must be resolved efficiently. 

\begin{table}[htbp]
\centering
\caption{Comparison of different element formulations: degrees of freedom (DOFs), relative computational runtime, and maximum absolute errors of radial displacement $d_r$ and wall thickness $h_t$ along the longitudinal axis, relative to HEX$8$.}
\begin{tabular}{lcccc}
\toprule
Element formulation & Relative DOFs & Relative runtime & $\max \abs{d_r - d_r^\text{HEX8}}$  & $\max \abs{h_t - h_t^\text{HEX8}}$  \\
& [\%] & [\%] & [mm] & [mm] \\
\midrule
HEX$8$ (reference) & $100$ & $100$ & $0.0$ & $0.0$ \\
Membrane      & $14.2$ & $4$  & $0.713$ & $1.063$ \\
Shell $7$-p & $28.5$ & $22$ & $0.561$ & $0.657$ \\
\bottomrule
\end{tabular}
\label{tab:elements_summary}
\end{table}

%%%%%%%%%%%%%%%%%%%%%%%%%%%%%%%
\section{Examples}
\label{sec:examples}

\subsection{Influence of the smooth muscle contraction intensities}
\label{sec:influence_of_smc_contraction_intensities}
An open question in the physiology of gastric contraction waves is how the longitudinal and circumferential muscle layers coordinate to produce the large deformations observed in the human stomach during peristalsis.  
Competing hypotheses exist in the literature.
Some studies suggest that both muscle layers contract simultaneously~\cite{Lentle-2016-ContractileActivity} and with similar intensity~\cite{Sarna-1993-Contractions}, 
while others report regionally variable or reciprocal patterns,
in which the longitudinal muscle layer relaxes while the circumferential layer contracts~\cite{huizinga2009a}. Similar debates exist for other parts of the gastrointestinal tract~\cite{Lammers2002,Mittal2016}. 
Overall, muscle coordination mechanisms in the gastrointestinal system remain poorly understood.
For the first time, our model enables the testing of such hypotheses in the stomach through a coupled electromechanical computational framework.
 
\begin{figure}[!t]
    \centering
    \subfloat[]{%
        \includegraphics[width=0.48\linewidth]{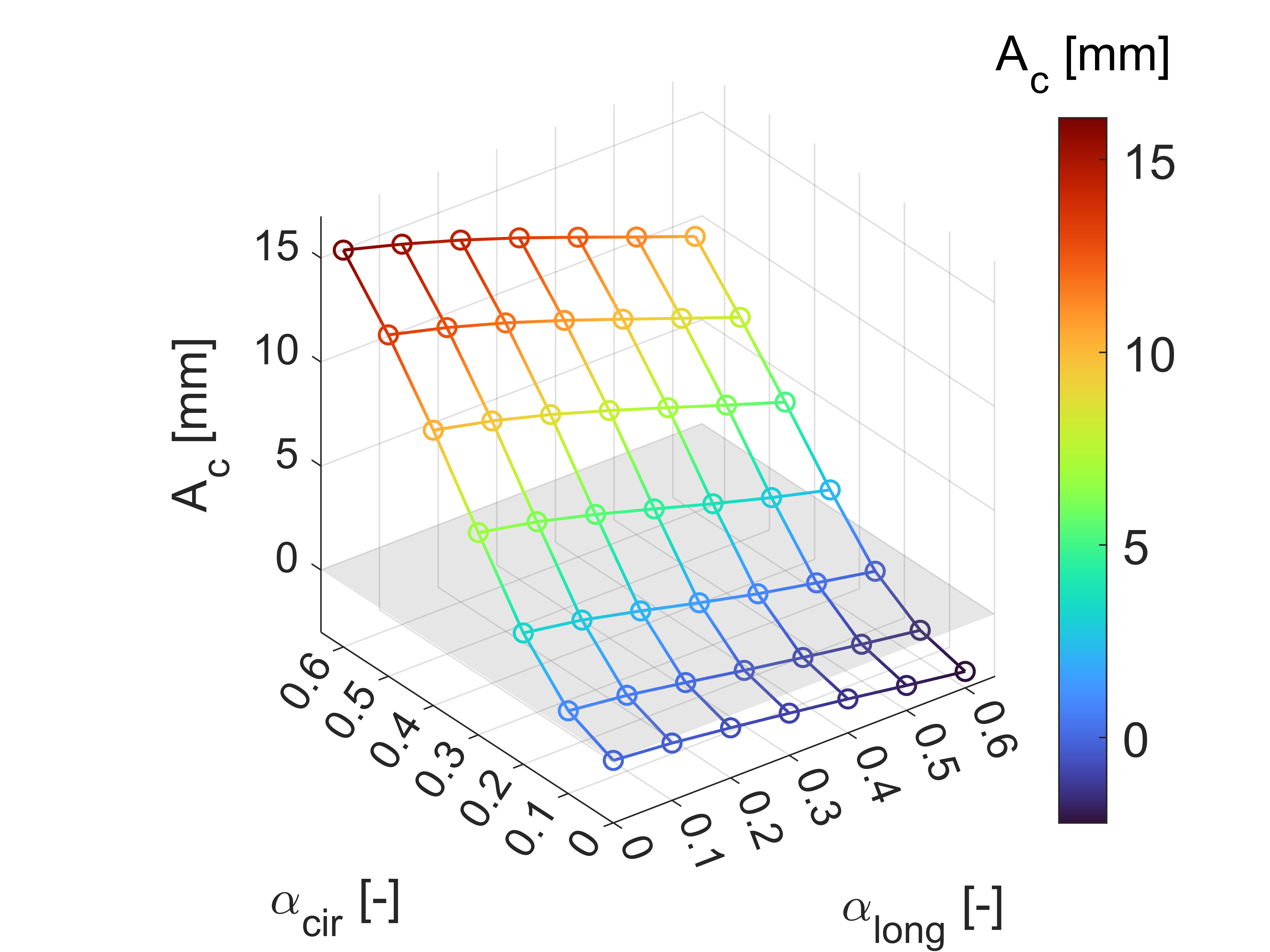}\label{fig:convergence_study_solid_contraction}}\quad
    \subfloat[]{%
        \includegraphics[width=0.48\linewidth]{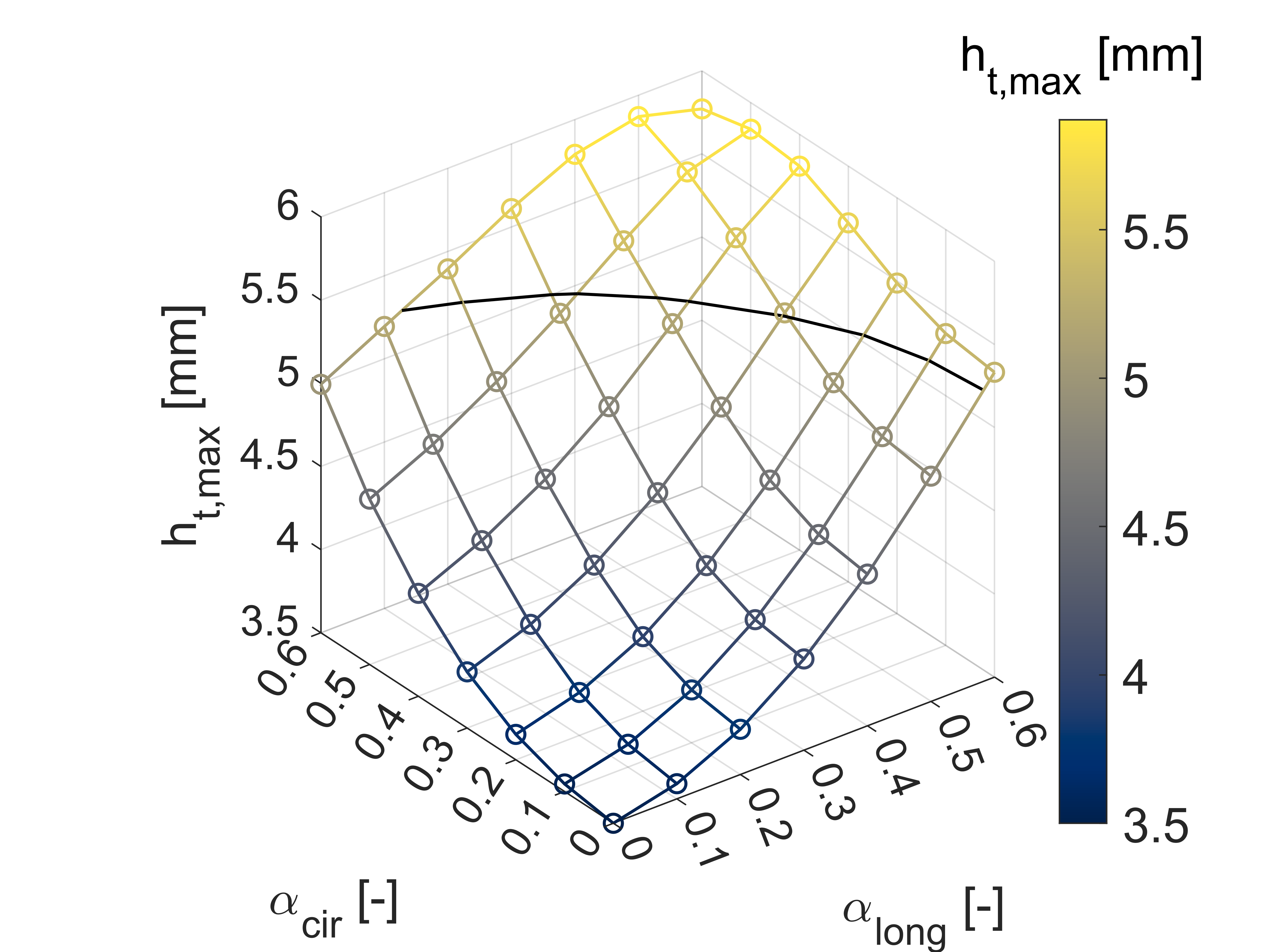}\label{fig:convergence_study_solid_thickness}}
    \caption{Radial contraction amplitude and tissue thickening under varying muscle activation intensities. 
    (a) Radial contraction amplitude $A_\textup{c}$ and (b) maximum tissue thickness $h_{t,\textup{max}}$  as functions of circumferential ($\alpha_\cf$) and longitudinal ($\alpha_\lf$) smooth muscle contraction intensities. In (b), the black contour line indicates the region where the tissue has increased by $50\%$ relative to its initial value.}  \label{fig:convergence_study_solid}
\end{figure}

To investigate the influence of active muscle contraction intensities in both fiber directions, we performed a parameter study in which the muscle fiber contraction intensities $\alpha_\lf$ and $\alpha_\cf$ were varied over the range $[0,0.6]$. The solid simulations utilized the numerical parameters identified from the previous study ($dt=$ \SI{0.1}{\s} and $h=$ \SI{1.0}{\mm}) and employed the same cylindrical wedge setup as before. For each pair of contraction intensities, we evaluated the resulting contraction amplitude $A_\textup{c}$ and the maximum tissue thickness $h_{t,\textup{max}}$ within the contraction zone. The results in~\cref{fig:convergence_study_solid} show that circumferential contraction has a dominant influence on radial deformation, whereas longitudinal contraction primarily contributes to wall thickening. For example, at $\alpha_\cf = 0.6$ and $\alpha_\lf=0.0$, the contraction amplitude reached $A_\textup{c}= $ \SI{16.0835}{\mm}. 

Furthermore, in several cases, $h_{t,\textup{max}}$ increased by more than $50\%$ relative to the initial wall thickness (black contour in~\cref{fig:convergence_study_solid_thickness}), indicating strong thickening under high longitudinal activation. 

The directional contribution of circumferential and longitudinal smooth muscle layers to overall gastric motility remains an open physiological question. Experimental quantification of these muscle fiber contractions is extremely limited, largely due to technical challenges in measuring active tension or wall thickening in vivo, especially in a complex anisotropic architecture of the gastric wall. To our knowledge, no conclusive experimental data are available quantifying the relative strength or intensity of longitudinal versus circumferential smooth muscle contractions in physiological conditions. In this context, in silico models such as the one presented here offer a powerful alternative to explore plausible physiological scenarios. Our simulation results demonstrate distinct mechanical roles for the circumferential and longitudinal muscle layers: circumferential contraction appears to be the primary driver of radial displacement (contraction amplitude), while longitudinal activation contributes more significantly to local wall thickening. 

Our results confirm that the balance of activation between the two fiber directions has meaningful implications for the mechanical efficiency and functional outcome of gastric motility.
Based on our findings, it is plausible that the circumferential fibers exhibit greater contraction intensities than the longitudinal ones to achieve effective luminal closure, especially in the context of slow wave-driven peristalsis. Accordingly, our choice of a slightly stronger circumferential contraction in subsequent simulations may reflect a mechanically favorable or even physiologically representative state.

%%%%%%%%%%%%%%%%%%%%%%%%%%%%%%%%%%%%%%%%%%%%%%%%%%%%%%%%%%%
\subsection{Electromechanical simulations of the stomach}
\label{sec:electromechanical_simulations_of_the_stomach}
In the following, we apply the proposed approach to a human stomach geometry to simulate coupled gastric electromechanics.

\subsubsection{Stomach setup}
\label{sec:case2_simulation_setup}
The geometry of our stomach is constructed using Coreform Cubit~\cite{coreform2023} by defining a central longitudinal path from the fundus apex to the pyloric sphincter, and define the surface geometry by ellipsoidal radii, which are shaped and scaled based on geometric comparisons of multiple \mri{} datasets of the human stomach. This approach allows us to capture the natural curvature and complex shape of the stomach, ensuring that the model reflects realistic anatomical features.
By combining these \mri-derived dimensions with an ellipsoidal cross-section, our model preserves key anatomical characteristics, such as the curvature of the stomach, while ensuring a smoothed surface and introducing simplifications around the openings to the duodenum and esophagus. This approach ensures anatomical fidelity in these regions without introducing numerical instabilities.

The mesh resolution was guided by the convergence analysis to ensure numerical accuracy.
The resulting finite element meshes comprises \num{210864} linear triangular scalar transport elements for the electrophysiology problem, and \num{17939} bilinear quadrilateral $7$-parameter shell elements for the solid problem. 
The boundary conditions applied follow those described in~\cref{sec:boundary_conditions}.
The fully distended stomach can accommodate approximately \SIrange[]{2}{4}{\litre}, with intragastric pressures reaching up to \SI{35}{\mmHg}. Under resting conditions, intragastric pressure typically remains below \SI{7}{\mmHg}~\cite{Janssen-2011-IntragastricPressure,Wali-2013-PatientFullStomach}. The stomach geometry used in our simulations provides an estimated volume of nearly \SI{2}{\litre}. Within this physiological range, we select pressure values of \SI{25}{\mmHg} in the distal stomach and \SI{10}{\mmHg} in the proximal stomach to represent a partially filled state.
Due to the lack of validated mixture-type material models for gastric tissue, simulation parameters are estimated to reflect physiological behavior~\cite{Liu2024} (see~\cref{tab:realistic_stomach_parameters}). The mass fractions are assumed based on analogies with cardiac~\cite{Gebauer-2023-CMMCardiac} and vascular tissue~\cite{Bellini-2014-MicrostructeArteialWall}. The muscle contraction intensities are set to $\alpha_\cf = 0.35$ and $\alpha_\lf = 0.1$, reflecting the relatively weaker contraction of the longitudinal muscle layer compared to the circumferential muscle layer (see~\cref{sec:influence_of_smc_contraction_intensities}). The stiffnesses of the spring-type boundary conditions are selected to approximate the mechanical support provided by the surrounding organs and tissues. 
The remaining parameters for the electrophysiology are listed in~\cref{tab:baseline_parameters}.

\subsubsection{Gastric slow wave entrainment and conduction velocity analysis}
\label{sec:case2_sw_entrainment_propagation}
\begin{figure}[!ht]
    \centering
        \includegraphics[trim={0cm 0cm 0cm 0cm},width=\linewidth]{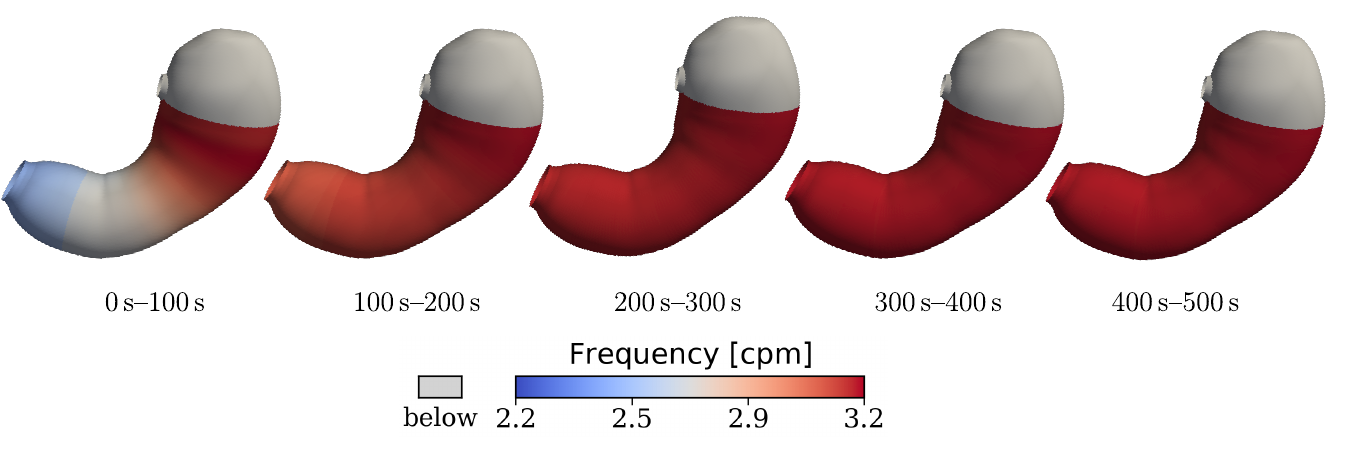}
         \caption{Spatiotemporal development of \icc{} frequencies from intrinsic to entrained state. The color bar applies to all panels, with grey indicating to areas without detectable activity (typically the proximal stomach). Initially (\SIrange[]{0}{100}{\s}, left),  frequencies decrease along the longitudinal direction and more gradually in the circumferential direction. By \SI{300}{\s}, frequencies stabilize across the domain and converge to that of the pacemaker region, indicating full entrainment.}
         \label{fig:frequencies_stomach}    
\end{figure}
\begin{figure}[!ht]
    \centering
     \includegraphics[trim={0cm 0 0cm 0},width=\linewidth]{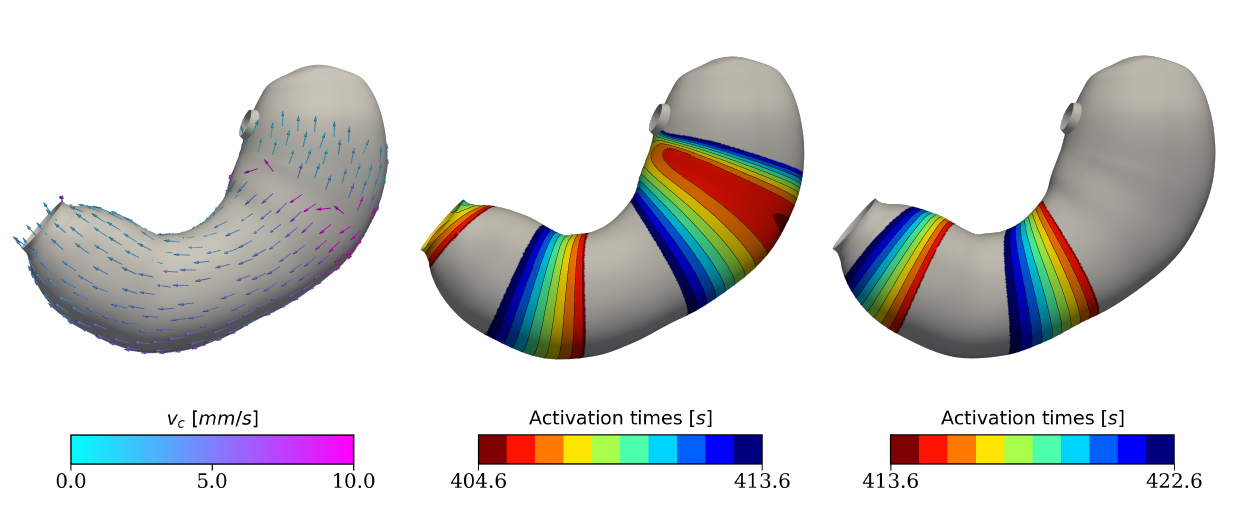}
     \caption{Computed conduction velocity $v_c$ distribution and propagation of entrained \icc{} slow waves. Left: $v_c$ field obtained post-simulation using a triangulation-based estimation method. Slow waves originate at the pacemaker region and propagate predominantly longitudinally, while proximal waves dissipate rapidly. Spatial heterogeneity in the diffusion coefficient induces a conduction velocity gradient, producing a curved propagation path approximately orthogonal to the longitudinal axis. Middle, Right: Isochrone maps of simulated activation times (intervals: \SI{1}{\s}). The middle panel covers $t =$ \SI{404.6}{\s} to $t =$ \SI{413.6}{\s}, and the right panel $t =$ \SI{413.6}{\s} to  $t =$ \SI{422.6}{\s}, illustrating the spatiotemporal wave progression.}%
    \label{fig:stomach-cv}
\end{figure}
To assess whether the proposed model reproduces key features of gastric electrophysiology, we study the propagation of slow waves over a \SI{500}{\s} time interval.  
The primary objectives are to evaluate whether the system exhibits frequency entrainment and to characterize the conduction velocity and propagation pattern of the resulting slow wave fronts.
We compute the local oscillation frequencies from the transmembrane potential $v^\ICC$ using a peak-detection algorithm applied over five successive \SI{100}{\s} time windows. Only peaks with amplitudes above a threshold of $0.7$ are considered. Frequency estimation is performed for each node of the computational domain using the open-source Python library SciPy~\cite{scipy}.

\Cref{fig:frequencies_stomach} shows the spatial distribution of these frequencies for each time window. During the initial \SIrange[]{0}{100}{\s} window (left), the frequency field exhibits a pronounced gradient along the longitudinal direction and a smaller decrease along the circumferential direction. This spatial heterogeneity directly reflects the underlying excitability field $a^\ICC$ (see~\cref{sec:intrinsic-frequencies}), which represents regional differences in ICC intrinsic frequencies. At the beginning of the simulation, local activation patterns reflect these imposed excitability gradients, producing spatially distinct oscillation patterns. Diffusive coupling then promotes synchronization, gradually reducing the heterogeneity. By \SI{300}{\s}, the electrical oscillations of the \iccs{} synchronize to a nearly constant frequency, matching the highest intrinsic frequency of the network, approximately \SI{3.2}{\cpm}. This behavior indicates robust entrainment across the tissue.

To further quantify the slow wave propagation, we computed the conduction velocity vector $\vec{v}_c$ of a slow wave front in the entrained system using a triangulation-based method~\cite{Cantwell-2015-Triangulation}, as detailed in~\ref{ap:estimation_of_conduction_velocity}. The resulting vector fields, shown in \cref{fig:stomach-cv}, confirm the aborally directed propagation originating from the pacemaker region. The pattern is anisotropic, with wave fronts spreading more rapidly circumferentially than longitudinally. Estimated velocity magnitude values range from \SI[per-mode = symbol]{4}{\mm\per\s} to \SI[per-mode = symbol]{6}{\mm\per\s} at the greater curvature and from \SI[per-mode = symbol]{2}{\mm\per\s} to \SI[per-mode = symbol]{3}{\mm\per\s} at the lesser curvature. In the pacemaker region, higher values of up to \SI[per-mode = symbol]{10}{\mm\per\s} are observed. These results confirm that spatially varying diffusion coefficients yield physiologically consistent conduction velocity gradients, in agreement with experimental observations~\cite{Grady-2021-GastricConductionReview}.

To complement these velocity analyses, we generated isochrone maps that represent the activation times of the slow wave across the stomach surface, 
allowing the interpretation of activation patterns directly from simulation data, as comparable to techniques used in experimental gastric mapping~\cite{Grady-2021-GastricConductionReview}. The isochrone plots (see~\cref{fig:stomach-cv}) reveal multiple simultaneous slow waves, typically two to three waves propagating concurrently, reflecting the entrained, periodic activity of the gastric tissue. The spatial arrangement of the isochrones illustrates smooth and continuous physiological wavefront progression consistent with the anisotropic velocity fields, demonstrating orderly aboral propagation along the longitudinal axis toward the pylorus. The spacing between isochrones bands varies regionally, with wider spacing near the greater curvature corresponding to faster conduction, and tighter spacing near the lesser curvature indicating slower propagation. These features align well with experimental gastric electrophysiology data, further supporting the physiological relevance of our heterogeneous diffusion model.

Importantly, the generated slow wave follows the longitudinal axis through the entire distal stomach and reaches the pylorus. This behavior reflects physiological propagation patterns and marks a key improvement over previous computational models, which often fail to sustain longitudinal wave propagation across the full distal region.

\subsubsection{Prestress}
\label{sec:case2_prestress}
We now analyze the prestressing procedure, with a focus on determining the prestretch tensor of the ground matrix. To this end, the pressure and prestretch values of the mixture of \smcs{} and collagen fibers are linearly ramped up over $10$ time steps, followed by the iterative prestress algorithm to determine the ground matrix prestretch tensor (see~\cref{sec:prestress}). The algorithm terminated once the maximum Euclidean norm of the nodal displacements falls below $10^{-3}$. This tolerance value, identified by~\cite{Gebauer-2023-CMMCardiac}, offers a balance between accuracy and computational efficiency. 

\begin{figure}[!h]
    \centering
\includegraphics[width=\linewidth]{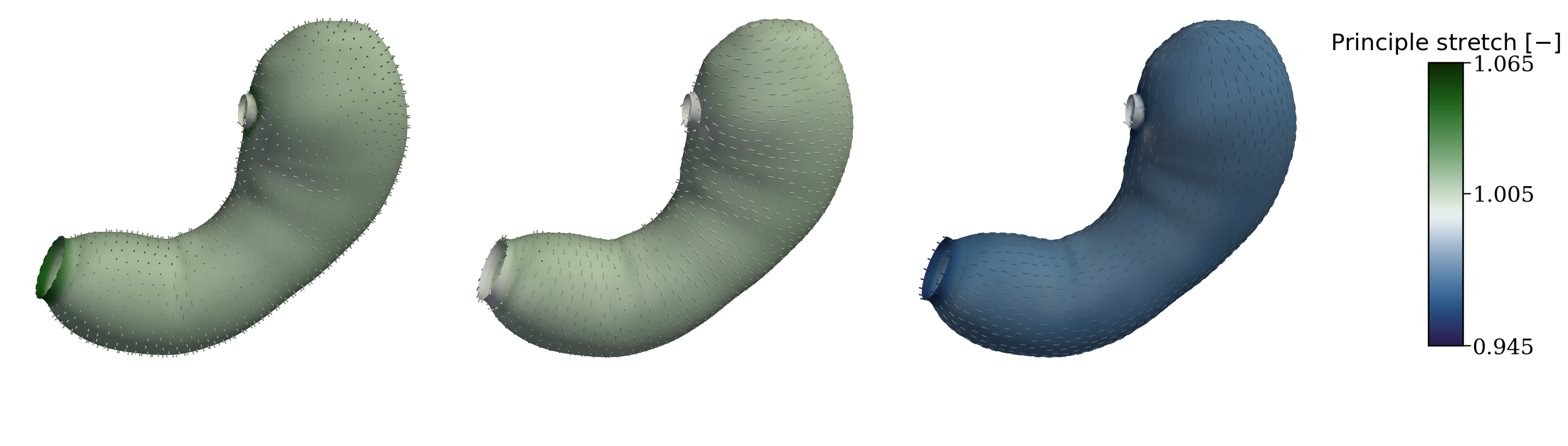} 
         \caption{Visualization of the ground matrix prestretch along principal directions. Left: First principal stretch (largest expansion). Middle: Second principal stretch (intermediate expansion/compression). Right: Third principal stretch (largest compression).}
    \label{fig:prestress_elastin}
\end{figure}

\Cref{fig:prestress_elastin} shows the spatial distribution of principal stretches in the ground matrix after convergence. The orientations of the principal stretches align with the embedded \smc{} and collagen fiber families, which share the ground matrix. The largest expansion (first principal stretch) occurs near the esophageal and duodenal ends, where Dirichlet boundary conditions are imposed. The ground matrix experiences expansion both in the wall-normal direction and within the fiber plane. Expansion is more dominant in the thickness direction, especially in areas of low curvature, and shifts toward circumferential directions in regions of higher curvature to accommodate luminal pressure and mechanical constraints from adjacent tissues.
Spring boundary conditions impose spatially varying external forces, resulting in compressive ground matrix response in the longitudinal direction, especially near the pyloric and duodenal attachments. These regions are influenced by displacement constraints and anchoring forces. The resulting ground matrix prestretch field reflects the mechanical balance between tissue microstructure, boundary conditions, and external loading.

\subsubsection{Coupled gastric motility}
\label{sec:case2_simulated_gastric_peristalis}
\begin{figure}[tb!]
    \centering
    \includegraphics[trim={0 0 0 0cm},width=\linewidth]{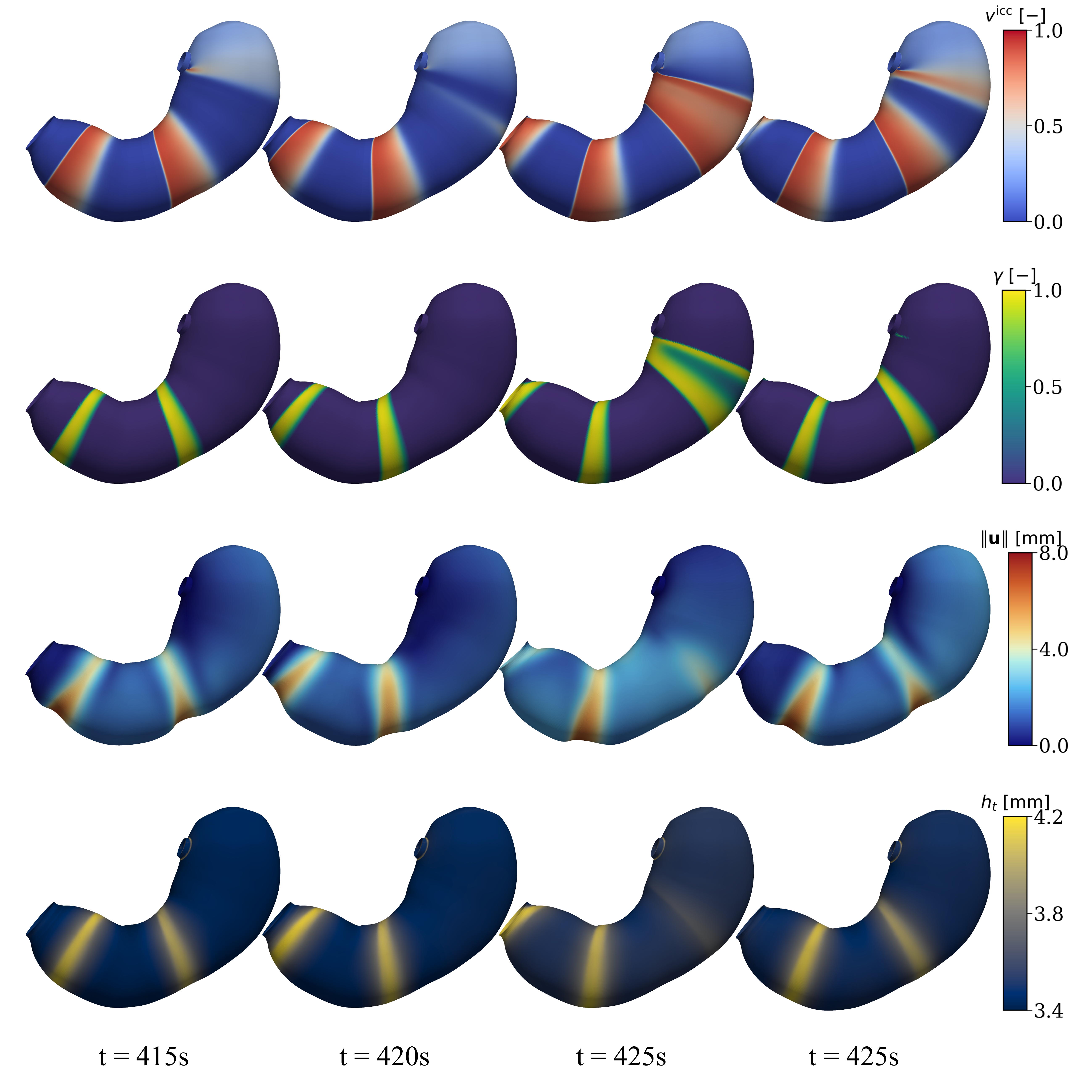}
    \caption{Temporal evolution of key characteristic parameters of the coupled electromechanical stomach model. The first row shows the normalized \icc{} transmembrane potential $v^\ICC$, illustrating the spatiotemporal activation patterns of the slow waves. The second row depicts the activation variable $\gamma$, representing smooth muscle contraction state. The third row presents the displacement magnitude $\norm{\mat{u}}$ visualizing the resulting deformation of the stomach wall. The fourth row displays the local stomach wall thickness $h_t$, indicating tissue thickening during contraction. Color bars correspond to each variable in the respective rows.}  \label{fig:stomach_electromechanics}
\end{figure}
We now investigate the coupled electromechanical simulation of gastric motility. The first row of~\cref{fig:stomach_electromechanics} shows the spatial distribution of the normalized transmembrane potential of \icc{} $v^\ICC$ at representative time points following complete entrainment of the network. At $t=$ \SI{415}{\s}, the pacemaker region is in its refractory phase, just before activation. Shortly thereafter, a new slow wave emerges at the pacemaker and propagates both longitudinally and circumferentially through the stomach wall. The wave fronts advances more rapidly in the circumferential direction, meeting at the lesser curvature and forming a closed activation ring. The waves directed toward the fundus dissipate within the proximal region, whereas the aborally traveling waves propagate toward the pylorus and attenuate near the sphincter. 
The resulting slow wave propagation pattern is periodic and stable, with two to three waves present simultaneously and an average inter-wave spacing of approximately \SI{65}{\mm}, consistent with experimental observations~\cite{Grady-2021-GastricConductionReview}. 
This spatial pattern reflects the successful reproduction of the native slow wave behavior in the distal stomach.

Coupling this electrophysiology model to the prestressed solid mechanical model yields spatial distributions of the activation parameter $\gamma$ and displacement magnitude $\norm{\vec{u}}$, shown in the second and third rows of~\cref{fig:stomach_electromechanics}, respectively. The mechanical responses are tightly synchronized with the propagating slow waves, showing regionally confined contractions that follow the electrical signals of \iccs{}.
To reproduce physiological observations, we prescribe stronger active contraction in the circumferential than in the longitudinal fiber direction ($\alpha_\cf=0.35$ and $\alpha_\lf=0.1$). This anisotropic activation leads to realistic, large peristaltic contractions, particularly in the distal stomach, where contraction amplitudes reach up to \SI{8}{\mm}, in agreement with recent in vivo measurements~\cite{Hosseini2023,Wang2024}.

The fourth row of~\cref{fig:stomach_electromechanics} shows temporal variations in gastric wall thickness $h_t$. Local increases in $h_t$ coincide with propagating contraction fronts, arising from the incompressibility of soft tissue and the active strain kinematics. The resulting deformation pattern is consistent with reported estimates for gastric wall deformation, although robust in vivo measurements remain limited. The overall mechanical behavior confirms the model’s ability to capture physiological gastric motility patterns on a realistic stomach geometry.
The coupled simulation is run on our in-house cluster ($16$ nodes with $2\times$ Intel Xeon Gold $6230$ Cascade Lake CPUs, $52$ cores, $376$ GB RAM), using $38$ cores on a single node. A representative fully coupled, entrained, and prestressed electromechanical simulation of \SI{50}{\second} real time ($2500$ time steps) required approximately \SI{2.89e+05}{\second} of wall-clock time ($\approx$ \SI{80}{\hour}). Profiling data indicate that the solid mechanics solver accounts for about $76\%$ of the total runtime, while the electrophysiology solver contributes roughly $24\%$, and input/output and setup operations are negligible ($<1\%$), see~\cref{tab:runtime_profile}. This computational distribution highlights the dominant cost of the nonlinear mechanics problem and supports the adopted strategy of using a finer electrophysiology mesh coupled to a coarser mechanical discretization through the dual-mortar interface, enabling accurate yet computationally efficient electromechanical coupling.

\begin{table}[htbp]
\centering
\caption{Profiling summary of a representative fully coupled, entrained, and prestressed electromechanical simulation of \SI{50}{\second} real time (2500 time steps). 
The total wall-clock runtime is approximately \SI{2.89e+05}{\second} ($\approx$ \SI{80}{\hour}). 
Relative cost fractions illustrate the computational load distribution across solvers and operations.}
\begin{tabular}{lcc}
\toprule
Component & Wall-clock time [s] & Relative cost [\%] \\
\midrule
Solid mechanics solver & \SI{2.187e5}{} & $75.7$ \\
Electrophysiology solver & \SI{7.001e4}{} & $24.2$ \\
Input/Output and setup operations & \SI{1e5}{} & $<1$ \\
\midrule
\textbf{Total} & \SI{2.889e5}{} & $100$ \\
\bottomrule
\end{tabular}
\label{tab:runtime_profile}
\end{table}

%%%%%%%%%%%%%%%%%%%%%
\section{Conclusions}
\label{sec:conclusions}
Compared to our previous electromechanical model~\cite{brandstaeter2018a}, the work represents a substantial extension from a conceptual prototype to a comprehensive, coupled whole-organ simulation framework for gastric electromechanics. The earlier formulation was restricted to simplified geometries, homogeneous material properties, and small, unphysiological contraction amplitudes, arising from the limitations of membrane elements. In contrast, the current framework integrates anatomically realistic geometry, advanced constitutive modeling, and spatially heterogeneous physiological parameterization to capture coordinated gastric motility with high fidelity.
This study presents a robust and extensible computational framework for simulating coupled gastric electromechanics, capable of capturing the large, anisotropic deformations appearing during peristaltic activity on anatomically realistic geometries. A key contribution is the integration of a nonlinear, rotation-free $7$-parameter shell formulation within the open-source multiphysics code $4$C. 
Unlike membrane elements, which lack bending stiffness and may suffer from numerical instabilities during large deformations or in regions of high curvature as common in gastric peristalsis, the proposed shell formulation ensures numerical stability while maintaining significantly lower computational cost than full $3$D solid formulations. It thus offers a good balance between numerical cost, stability, and geometric fidelity for organ-scale applications.

Building on our previous electromechanical model~\cite{brandstaeter2018a}, we extended the framework to include an anisotropic constrained mixture material model, prestress initialization, and non-uniform boundary conditions. These enhancements allow us to simulate physiologically realistic slow wave propagation and motility patterns, as validated by comparison to experimental data on contraction amplitudes~\cite{Schulze-2006-ImagingStomach,Hosseini2023}, conduction velocities~\cite{Wang2024}, and regional frequencies~\cite{Grady-2021-GastricConductionReview}. 
Specifically, the simulated contraction amplitudes (\SIrange[]{5}{8}{\mm}), conduction velocities (\SIrange[per-mode = symbol]{2}{6}{\mm\per\s}), and slow-wave frequencies ($\approx$ \SI{3.2}{\cpm}) fall within reported physiological ranges, supporting the biological plausibility of the model. Isochronal activation maps reproduce key experimental features, including circumferential spread near the pacemaker and longitudinal progression toward the pylorus~\cite{du2013c,OGrady-2012-AbnormalInitiationSW}, demonstrating physiologically consistent slow-wave dynamics. Although gastric motility disorders are common, no direct quantitative clinical assessment of gastric wall mechanics or electrical coordination exists, as tools such as scintigraphy and electrogastrography provide only indirect measures. Physics-based computational models thus offer a means to bridge this diagnostic gap by providing mechanistic insight and quantitative predictions otherwise inaccessible.

Extensive convergence analyses demonstrate numerical robustness, and application to realistic stomach geometries demonstrates the framework's suitability for whole-organ studies.

A key novelty lies in the spatially heterogeneous parameterization of the model properties, such as the excitability of \iccs{}, the diffusion coefficient, or the fiber directions. These were assigned using smooth harmonic fields obtained from solving Laplace-Dirichlet problems aligned with anatomical axes, enabling the construction of continuous and anatomically meaningful parameter distributions. While effective, the specific parameter values remain empirically defined. This underscores the potential for future integration of data-driven methods, as pioneered in cardiac modeling~\cite{Barone-2020-EstimationCardiacConductivities}, to improve personalization and physiological fidelity.

A central insight from our parametric study is that computational modeling can elucidate physiological mechanisms that remain experimentally inaccessible. Specifically, the relative roles of circumferential and longitudinal muscle layers in gastric wall deformation are poorly characterized in vivo. Our in silico results suggest that a stronger circumferential contraction may be mechanically advantageous for effective contraction, supporting the potential physiological relevance of such configurations. This highlights the broader potential of computational models not only for prediction, but also for hypothesis generation in the absence of direct experimental data. In silico models provide a unique opportunity to investigate gastric motility processes that are difficult to assess through in vitro or in vivo approaches~\cite{Liu2024}. There is substantial inter-individual variability in stomach anatomy, including differences in shape, orientation, and spatial relation to surrounding organs and tissues. These factors result in varying mechanical boundary conditions, which may correlate with distinct contraction patterns. Such variability is difficult to capture experimentally, but can be systematically explored through computational modeling to better understand its influence on gastric biomechanics. These models can also aid in evaluating the influence of gastric geometry and contractile patterns on digestion, as well as the mechanical interplay between muscle layers and their contribution to wall deformations.

While this work represents a major step forward in gastric modeling, several limitations remain. The electrophysiology model is currently phenomenological and one-way coupled, excluding mechanosensitive feedback mechanisms known to be relevant for \icc{} and \smc{} function~\cite{Kraichely-2007-MechanosensitiveIonChannels,Huizinga-2018-StretchCurrents}. Further, fluid–structure interactions are not explicitly modeled. A spatially uniform intragastric pressure is used to approximate the mechanical load on the gastric wall, providing a first-order approximation of wall mechanics but neglecting local fluid effects, shear stresses, and feedback between fluid motion and tissue deformation. This simplification may limit the ability to capture fully physiologically accurate patterns of intraluminal flow and wall deformation, representing an important avenue for future research~\cite{Brandstaeter2019,Liu2024}. In addition, although our framework, combining the $7$-parameter shell formulation with the constrained mixture model, efficiently captures large deformations and tissue mechanics, it represents the gastric wall in a homogenized manner. Intramural heterogeneity arising from the multilayered structure and its fiber dispersion %, and local myenteric plexus effects 
is not explicitly resolved, which limits the reproduction of intramural mechanics. Digesta transport is not explicitly modeled, and functional roles of anatomical regions such as the fundus, the pyloric sphincter, and the lower esophageal sphincter are not modeled in detail, despite their mechanically significance.

Future extensions could focus on incorporating stretch-activated currents~\cite{huizinga2009a,Joshi2021}, stress-assisted diffusion~\cite{Cherubini2017}, fractional diffusivity~\cite{Cusimano2019}, multiscale biophysical ion models~\cite{Mah2020,Athavale-2024-RatGastricSlowWave}, fluid~\cite{Palmada2023} and fluid–solid interaction~\cite{Fuchs2021}. Also contact mechanics could be investigated. Recent studies on the intestine~\cite{Djoumessi2025} indicate that contact interactions could be important for accurately capturing gastric motility. Parameter refinement will also benefit from integration with imaging and experimental datasets through uncertainty quantification and global sensitivity analyses~\cite{Brandstaeter2021a,Wirthl2023}.

In summary, this work provides a general, numerically robust, and physiologically informed modeling framework for gastric electromechanics on a realistic human geometry. Beyond methodological innovation, the framework provides a step toward connecting computational modeling with clinical applications by offering a quantitative tool to analyze normal and pathological gastric function. The modular design of the numerical framework allows for straightforward incorporation of more detailed electrophysiological cell models, alternative constitutive formulations, or more advanced boundary conditions, offering a flexible platform for future investigations. Furthermore, such simulations can support clinical interpretation of motility imaging, assist in virtual testing of pacing therapies, and provide mechanistic insight into disorders such as gastroparesis or functional dyspepsia. Integration with clinical imaging and electrogastrography could enable patient-specific model calibration and non-invasive assessment of gastric function, paving the way toward translational and personalized modeling of gastrointestinal motility. By addressing key limitations of prior approaches, it lays the foundation for future in silico studies of the human stomach and its disorders.

\section*{CRediT authorship contribution statement}
\textbf{Maire S. Henke:} Methodology, Software, Validation, Investigation, Formal analysis, Writing – Original Draft, Writing – Review \& Editing, Visualization. \textbf{Sebastian Brandstaeter:} Conceptualization, Methodology, Software, Formal analysis, Validation, Writing – Original Draft, Writing – Review \& Editing, Visualization, Supervision. \textbf{Sebastian L. Fuchs:} Methodology, Software, Writing – Review \& Editing. \textbf{Roland C. Aydin:} Conceptualization, Funding acquisition.
\textbf{Alessio Gizzi:} Conceptualization, Methodology, Writing – Original Draft, Writing – Review \& Editing, Supervision, Funding acquisition. \textbf{Christian J. Cyron:} Conceptualization, Methodology, Writing – Original Draft, Writing – Review \& Editing, Supervision, Funding acquisition.

\section*{Acknowledgements}
The authors thank Renate Sachse for valuable discussions on shells and Matthias Mayr for fruitful discussions on the Laplace–Dirichlet problems. They also thank Amadeus Gebauer for the insightful discussions on mesh conversions, constrained mixture theory and prestress.  Furthermore, the authors thank Fabiola Rienäcker for valuable discussions on medical image reconstruction.
The authors gratefully acknowledge the computing resources provided by the Data Science \& Computing Lab at the University of the Bundeswehr Munich.
This work was funded by the Deutsche Forschungsgemeinschaft (DFG, German Research Foundation) – 350481011, 469698389.
AG acknowledges the ERC Consolidator Grant support from the European Union’s Horizon Europe research and innovation programme under grant agreement No. 101170592 — MiGEM.

\section*{Data availability statement}
All the building blocks of the electromechanical finite element model developed and applied in this study are openly available in 4C Multiphysics at \url{https://github.com/4C-multiphysics/4C} \cite{4C}. Additional software and data that support the findings of this study are available from the corresponding authors upon reasonable request.

\section*{Declaration of competing interest}
The authors declare that they have no known competing financial interests or personal relationships that could have appeared to influence the work reported in this paper.

\numberwithin{figure}{section}
\renewcommand{\thefigure}{\thesection.\arabic{figure}}
\numberwithin{table}{section}
\renewcommand{\thetable}{\thesection.\arabic{table}}
\appendix
\section{Strain energy functions}
\label{ap:strain_energy_functions}
The specific form of the strain energy function $W^i$ for each constituent $i$ follows standard hyperelastic formulations and is briefly summarized below.

\subsection{Circumferential and longitudinal collagen and smooth muscle fibers}
Collagen and \smcs{} fiber families ($i \in \{\cf, \lf\})$ are modeled as quasi-one-dimensional structures that primarily resist stretch along their respective fiber directions. The strain energy function per unit mass for the $i$-th fiber family is given by the exponential form~\cite{HolzapfelGasserOgden-2012-ConstArterial}
\begin{align}
    W^{i} = \frac{k_1^i}{2k_2^i} \{ \exp{[k_2^i({I}_{4\textup{e}}^i-1)^2]}-1\}  \, ,
\end{align}
where $k_1^i$ is a stiffness-like and $k_2^i$ a non-dimensional material parameter.
The anisotropic fourth pseudo-invariant ${I}_{4\textup{e}}^i$ is the square of the elastic stretch $\lambda_\textup{e}^i$ of the fibers. If the fibers are aligned in direction $\vec{f}_\textup{R}^i$ in the reference configuration and in direction $\vec{f}^i$ in the deformed current configuration, the elastic part of the fiber deformation gradient is
\begin{align}
    \defgrad_\textup{e}^i = \lambda_\textup{e}^i\, \vec{f}^i \otimes \vec{f}_\textup{R}^i + \frac{1}{\sqrt{\lambda_\textup{e}^i}} \left( \mat{I} - \vec{f}^i \otimes \vec{f}_\textup{R}^i \right)\,, \qquad i \in \left\{ \cf,\, \lf \right\}.
\end{align}

\subsection{Ground matrix}
The remaining structural constituent, the ground matrix  ($i=\gm)$, primarily composed of elastin, is modeled as a three-dimensional decoupled isotropic Neo-Hookean material with an isochoric-deviatoric split~\cite{Holzapfel-2001-NonlinearSolidMechanics}.
The corresponding isotropic passive strain energy function for the ground matrix reads
\begin{align}
\label{eq:strain_energy_coupled_neoHooke}
    W^\gm=c_1 (\overline{{I}}^\gm_{1\textup{e}}-3) + \frac{c_1}{c_2}((J_\textup{e}^\gm)^{-2c_2}-1) 
    \,,\quad \textrm{with} \quad 
    c_1 =\frac{\mu^\gm}{2}, \quad c_2=\frac{\nu^\gm}{1-2\nu^\gm} \, ,
    %\frac{\kappa^\gm}{2} (J_e^\gm-1)^,
\end{align}
where material constants are function of the shear modulus $\mu^\gm$ and the Poisson's ratio $\nu^\gm$, while the first modified invariant of the elastic right Cauchy-Green tensor $\rightCG_\textup{e}^\gm= {\defgrad_\textup{e}^\gm}^T\defgrad_\textup{e}^\gm$ is given by $\overline{{I}}^\gm_{1\textup{e}}=(J_\textup{e}^\gm)^{-\nicefrac{2}{3}}\textup{tr}(\rightCG_\textup{e}^\gm)$. The isochoric part of the elastic deformation gradient of the ground matrix is defined as
$
\overline{\defgrad}_\textup{e}^\gm = (J_\textup{e}^\gm)^{-\nicefrac{1}{3}} \defgrad_\textup{e}^\gm \, ,
$
with a penalty formulation used to enforce near incompressibility $J_\textup{e}^\gm = \det(\defgrad_\textup{e}^\gm) = 1$.

\section{Electrophysiological cell model}
\label{ap:electrophysiology_cell_model}
Each cell type employs a modified version of the phenomenological two-variable Mitchell-Schaeffer model~\cite{MitchellSchaeffer-2003-TwoCurrentModel,Djabella-2008-CellModel,brandstaeter2018a} to characterize ionic exchanges via two state variables: the normalized transmembrane voltage $v^j \in [0,1]$ and the recovery (gating) variable $h^j \in [0,1]$ for cell type $j \in \{\ICC\, ,\ \SMC\}$.
Considering the monodomain formulation in \cref{eq:RDsys}, the time evolution of the state variables is governed by the following system of ordinary differential equations 
\begin{align}
\label{eq:ElectrophysiologyODE}
    \left\{ \begin{aligned} 
        I_\textup{ion}^j(v^j, h^j, \vec{X})&=I_\textup{in}^j(v^j, h^j, \vec{X})+I_\textup{out}^j(v^j, h^j),\\
        \dfrac{ h^j}{t}&=\frac{h_\infty^j-h^j}{\tau^j(v^j)}\, , 
    \end{aligned} \right.
\end{align}
where $I_\textup{in}^j$ and $I_\textup{out}^j$ represent the inward and outward ionic currents, respectively. Note that in the special case of an isolated cell not connected to neighboring cells by diffusion of electrical potential or currents between \smcs{} and \iccs{}, the ionic current $I_\textup{ion}^j$ equals the rate of change of the transmembrane potential, that is $I_\textup{ion}^j = dv^j / dt$.
The voltage-dependent time constant $\tau^j$ and the steady-state value $h_\infty^j$ are defined by
\begin{align}\tau^j(v^j)&=\frac{\tau_\textup{open}^j\tau_\textup{close}^j}{\tau_\textup{open}^j+h_\infty^j(v^j)(\tau_\textup{close}^j-\tau_\textup{open}^j)}\, ,\\
    h_\infty^j(v^j)&=1-H(v^j-v_\textup{gate}^j)\, ,
\end{align}
where $\tau_\textup{open}^j$ and $\tau_\textup{close}^j$ are time constants related to subcellular ionic dynamics, and 
\begin{align}
    H(v^j-v_\textup{gate}^j)  =
   \begin{cases}
        0, &  v^j\leq v_\textup{gate}^j\\
        1, &  v^j>v_\textup{gate}^j \, 
    \end{cases}
\end{align}
is a Heaviside function that determines the sign of the steady-state gating variable based on a voltage threshold value $v_\textup{gate}^j$.

To mitigate numerical instabilities, the Heaviside function can be approximated by a smooth hyperbolic tangent function
\begin{align}
    1-H(v^j-v_\textup{gate}^j) \approx  \frac{1}{2}\Big(1-\textup{tanh}\Big(\frac{v^j-v_\textup{gate}^j}{\eta_\textup{gate}}\Big)\Big) \, ,
\end{align}
where $\eta_\textup{gate}^j$ serves as a smoothing parameter. As in $\eta_\textup{gate}^j \rightarrow 0$, the hyperbolic tangent function converges to the Heaviside function.

The inward and outward currents are expressed using phenomenological cubic functions~\cite{brandstaeter2018a}
\begin{subequations}
\begin{align}
    \label{eq:InwardCurrent}
    I_\textup{in}^j(v^j, h^j,\vec{X})&=\frac{h^j}{\tau_\textup{in}^v}(v^j+a^j)(v^j+a^j-\lambda^j)(1-v^j)\, ,\\
    I_\textup{out}^j(v^j, h^j)&=-\frac{v^j}{\tau_\textup{out}^j} \, ,
\end{align} 
\end{subequations}
where $a^\ICC(\vec{X})$ is a spatially dependent model parameter that defines the intrinsic frequency of \iccs{}. In contrast, \smcs{} do not actively trigger electrical signals, thus the parameter is set to zero, denoted as $a^\SMC=0$. The parameter $\lambda^j$ further represents a secondary excitability parameter of \iccs{} and \smcs{}.

\section{Estimation of slow wave conduction velocities}
\label{ap:estimation_of_conduction_velocity}
\begin{figure}[!t]
    \centering
     \subfloat[]{%
        \includegraphics[trim={0 0 0 0},width=0.3\textwidth]{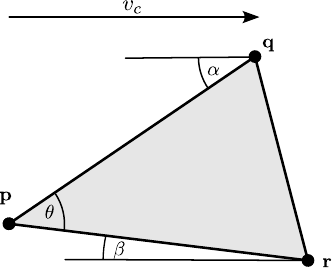} \label{fig:triangulation_magnitude}}
        \qquad
    \subfloat[]{%
        \includegraphics[trim={0 0 0 0},width=0.4\textwidth]{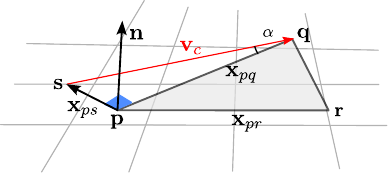}\label{fig:triangulation_direction}}\\
         \caption{Diagram illustrating (a) the calculation of conduction velocity and (b) the determination of the velocity direction vector, ensuring that the velocity vector remains coplanar with the triangle vertices $\vec{p}$, $\vec{q}$, and $\vec{r}$.}
         \label{fig:triangulation}
\end{figure}
Using elementary trigonometric relations, the activation times of three points can be used to estimate the average conduction velocity and direction, assuming the wavefront is locally planar~\cite{Cantwell-2015-Triangulation}. This method is illustrated in~\cref{fig:triangulation}. For a \SI{30}{\s} window, we extract activation times at all surface nodes of the discretized geometry, defined as the time of the transmembrane potential peak in \iccs{}. These times are identified directly from the simulation results using the open-source software SciPy~\cite{scipy}.

Consider a triangular element with vertices $p$, $q$, and $r$. The notation is as follows: the vector from point $p$ to $q$ is denoted as $\vec{x}_{pq}$, with magnitude $\norm{\vec{x}_{pq}}$, representing its length, and $t_{pq}$ denotes the activation time difference between $p$ and $q$. The interior angle $\theta$ between the sides of the triangle is given by the cosine rule
\begin{align}
    \theta=\arccos{\Bigg(\frac{\norm{\vec{x}_{pq}}^2+\norm{\vec{x}_{pr}}^2-\norm{\vec{x}_{qr}}^2}{2\norm{\vec{x}_{pq}}\norm{x_{pr}}}\Bigg)}\, .
\end{align}
The local conduction velocity $v_c$ and wavefront incidence angles can then be related by
\begin{align}
    \cos\beta = \cos(\theta-\alpha), \quad v_c = \frac{\abs{\alpha}\cos\alpha}{t_a},  \quad v_c = \frac{\abs{\beta}\cos\beta}{t_b}\, ,
\end{align}
where the angle $\alpha$ is obtained from 
\begin{align}
    \tan\alpha=\frac{\nicefrac{t_{pr}\norm{\vec{x}_{pq}}}{t_{pq}\norm{\vec{x}_{pr}}}-\cos\theta}{\sin\theta}\, .
\end{align}
Solving for $\alpha$ gives the direction of activation, and the velocity $v_c$ can subsequently be determined. 
To extent the method to surfaces embedded in the three-dimensional space, we compute the velocity direction vector $\vec{v}_c$ constrained to the plane of the triangle. The local surface normal is determined as
\begin{align}
    \vec{n}= \frac{\vec{x}_{pr}\times \vec{x}_{pq}}{\norm{\vec{x}_{pr}\times \vec{x}_{pq}}} \, .
\end{align}
As shown in~\cref{fig:triangulation_direction}, the extrusion from point $\vec{p}$ perpendicular to $\vec{x}_{pq}$ intersects the velocity direction at point $\vec{s}$, defined by
\begin{align}
    \vec{x}_{ps}= (\vec{n} \times \vec{x}_{pq})\tan\alpha\, .
\end{align}
The local velocity vector is then
\begin{align}
     \vec{v}_c= \vec{x}_{pq}-\vec{x}_{ps} \, ,
\end{align}
while the conduction velocity value is $v_c= \norm{\vec{v}_c}$.

\section{Spatial distribution of absolute errors for different element formulations}
\label{ap:absolute_errors}

To complement the maximum error values reported in \cref{tab:elements_summary}, we present the spatial distribution of absolute errors along the longitudinal axis. The absolute error is computed as
\begin{align}
    \Delta d_r(x) &= \abs{d_r(x) - d_r^\text{HEX8}(x)}, \\
    \Delta h_t(x) &= \abs{h_t(x) - h_t^\text{HEX8}(x)},
\end{align}
where $d_r^\text{HEX8}$ and $h_t^\text{HEX8}$ are the reference displacement and thickness profiles obtained from the HEX$8$ solid formulation. 
\begin{figure}[!htbp]
    \centering
    \subfloat[Radial displacement absolute error $\Delta d_r(x)$]{%
        \includegraphics[width=0.48\linewidth]{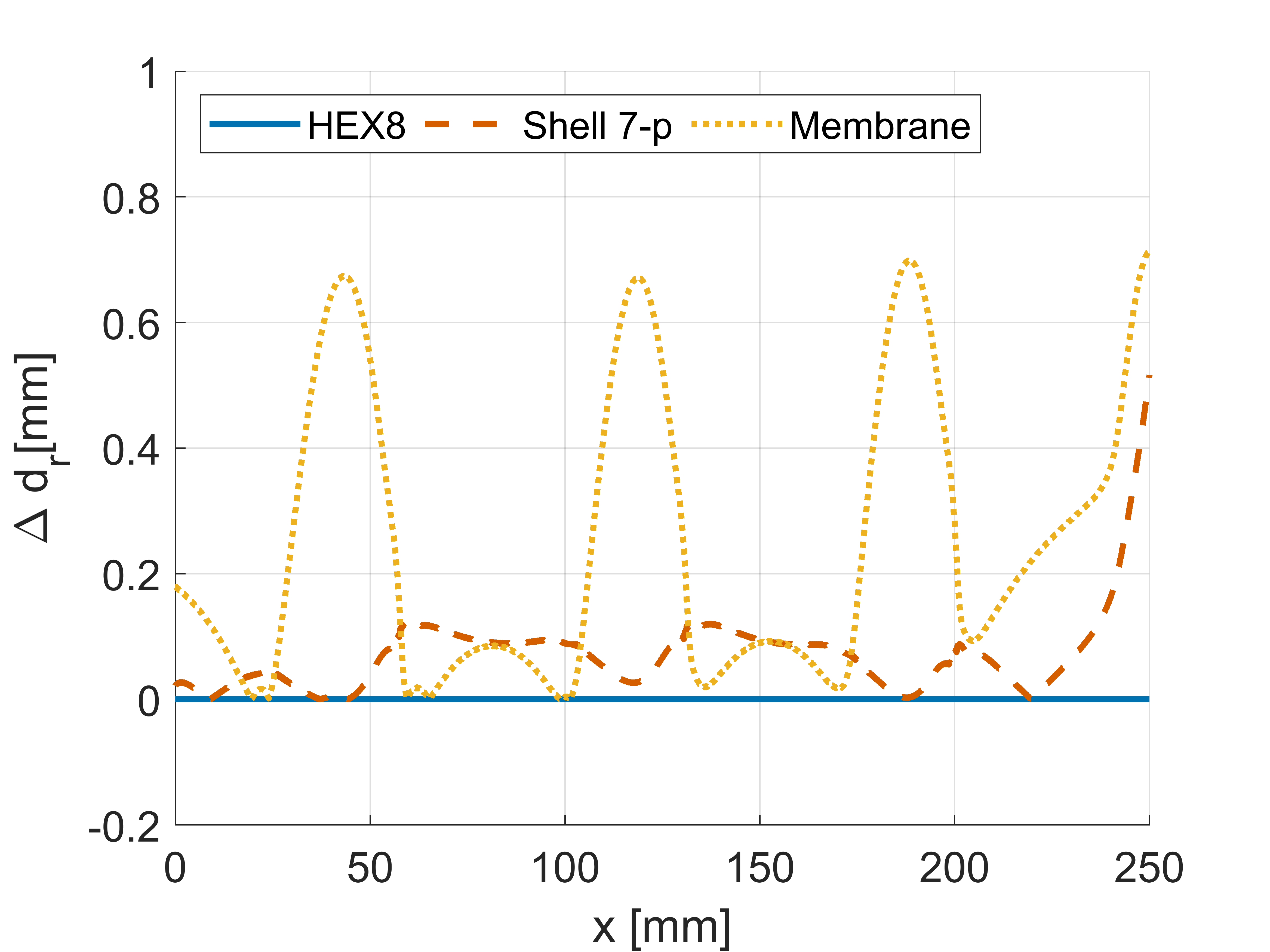}\label{ap:fig:disp_error}}\quad
    \subfloat[Thickness absolute error $\Delta h_t(x)$]{%
        \includegraphics[width=0.48\linewidth]{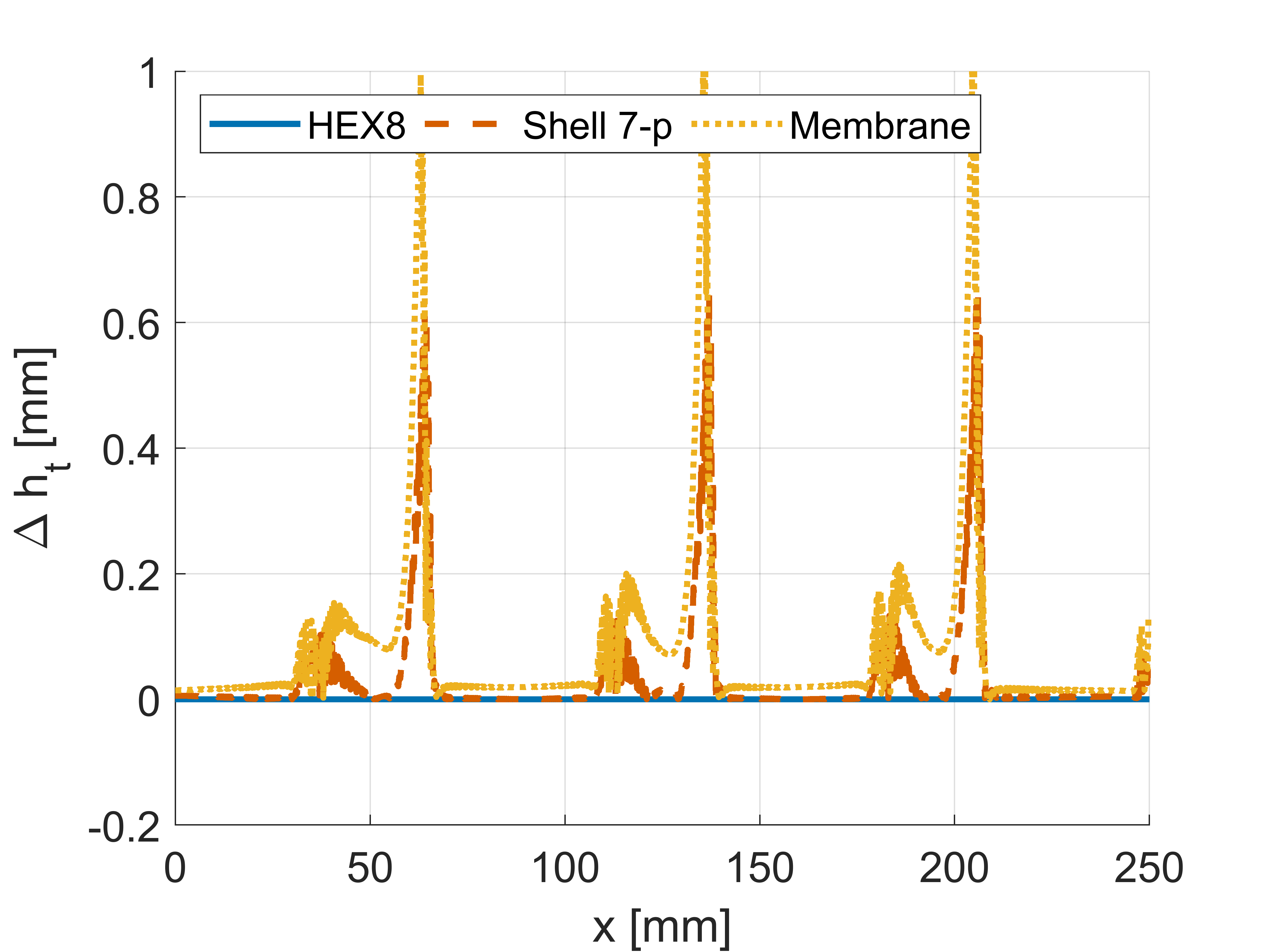}\label{ap:fig:thickness_error}}
    \caption{Spatial distribution of absolute errors along the longitudinal axis for membrane and $7$-parameter shell elements relative to the HEX$8$ reference solution. Membrane elements show larger errors in regions of high curvature and bending, while shell elements closely follow the reference solution with minor deviations.}
    \label{ap:fig:absolute_errors_all}
\end{figure}

The shell formulation closely matches the HEX$8$ reference, with small deviations mostly occurring near boundaries where locking treatments affect the solution. Membrane elements show significantly larger errors, particularly in regions where out-of-plane bending dominates, consistent with the trends observed in \cref{fig:3d_mem_shell_displ,fig:3d_mem_shell_thickness}.

\section{Influence of the diffusion coefficient on the propagation pattern}
\label{ap:influence_of_diffusion_coefficient} 
\begin{figure}[h]
    \centering
        \includegraphics[trim={0 0 0cm 0},width=0.4\textwidth]{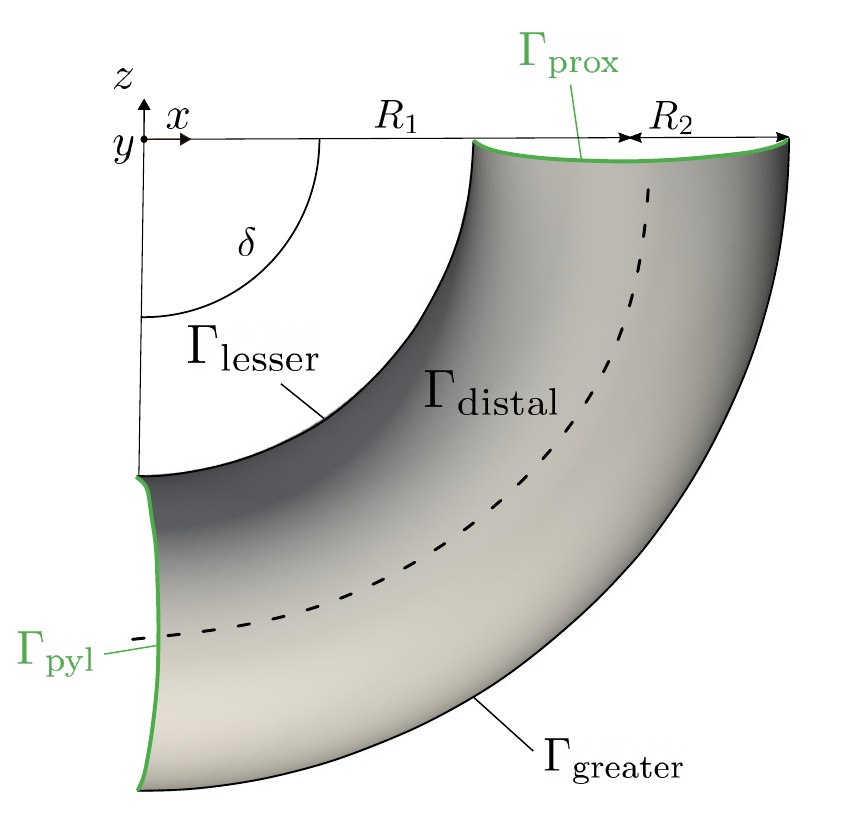}
   \caption{Sketch of the computational domain of the torus with opening angle $\delta=90^\circ$, major radius $R_1=$ \SI{159.155}{\mm}, minor radius $R_2=$ \SI{50.93}{\mm}, and wall thickness of $H_t=$ \SI{3.5}{\mm}. No-flux Neumann boundary conditions are applied at the pyloric sphincter $\Gamma_\textup{pyl}$ and pacemaker line $\Gamma_\textup{prox}$.}
\label{fig:torus_sketch}
\end{figure}
We model the functional distal stomach, excluding the electrically quiescent fundus region, which does not exhibit typical peristaltic contractions~\cite{Angeli-2014-ImprovedGutFunction,Athavale-2024-RatGastricSlowWave}, using an idealized quarter-torus geometry as shown in~\cref{fig:torus_sketch}. The torus features two openings representing the pyloric sphincter ($\Gamma_\textup{pyl}$) and the connection to the proximal stomach ($\Gamma_\textup{prox}$). The geometry, generated using Coreform Cubit~\cite{coreform2023}, is dimensioned to mimic anatomical proportions of the human stomach. 
We choose the mesh sizes based on the convergence study shown in \cref{sec:convergence_study}. 
Consequently, the mesh consists of \num{373006} linear triangular scalar transport elements for the electrophysiology problem.
\begin{figure}[!h]
    \centering
    \subfloat[]{\includegraphics[trim={0cm 0 0cm 0},width=0.45\textwidth]{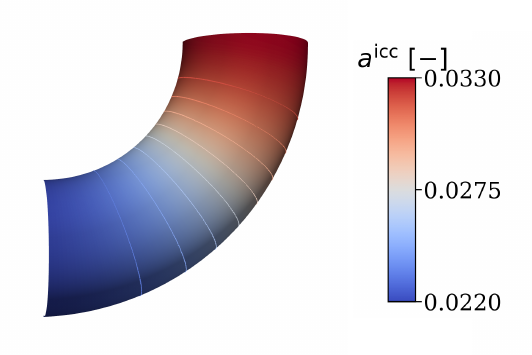}\label{fig:torus_exci}}
     \subfloat[]{\includegraphics[trim={0cm 0 0cm 0},width=0.45\textwidth]{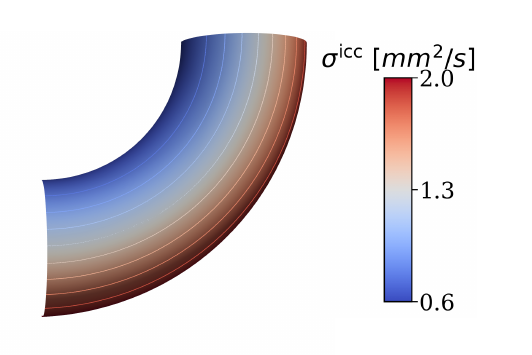}\label{fig:torus_sigma_i}}
     \caption{Contour plots of spatial distributions of (a) the excitability parameter $a^\ICC$, and (b) the \icc{} diffusion coefficient $\sigma^\ICC$.}%
     \label{fig:torus_spatial_plots}
\end{figure}
\begin{figure}[!h]
    \centering
    \includegraphics[trim={0 0 0 0},width=0.9\textwidth]{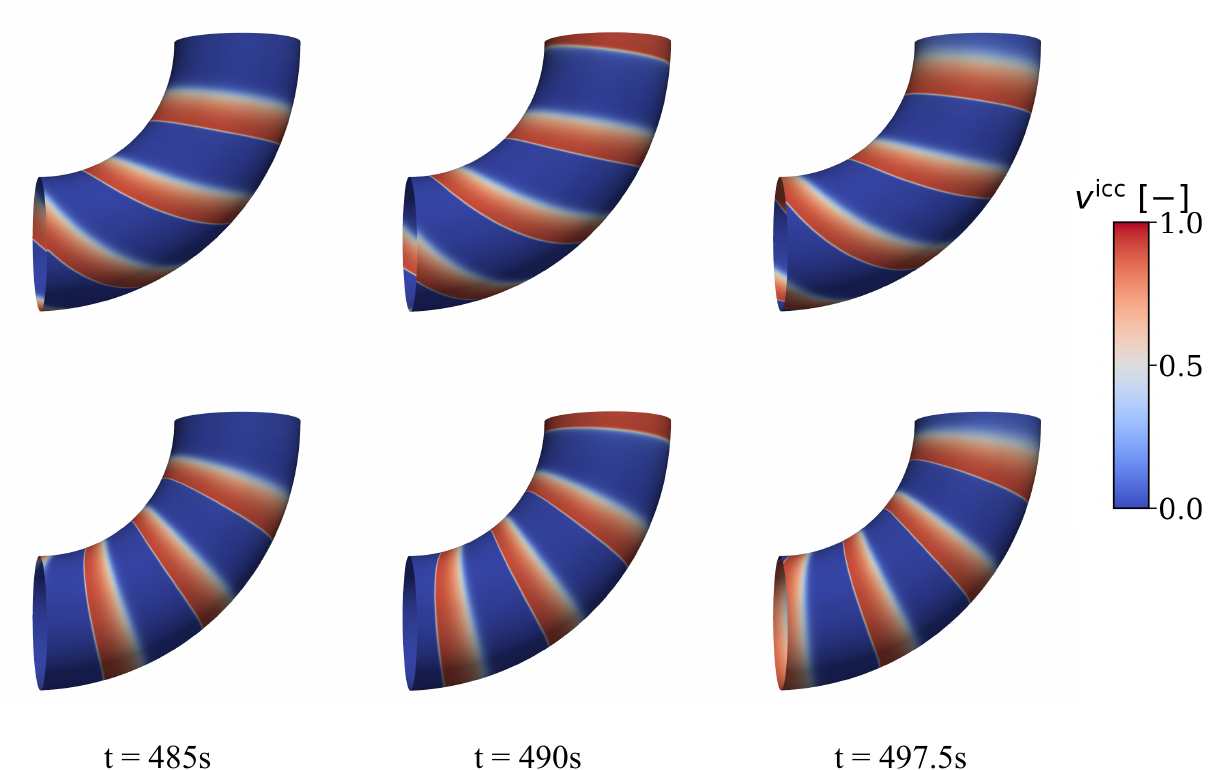}
    \caption{Collection of spatial plots illustrating the influence of different spatial distributions of the diffusion coefficient of \iccs{} and \smcs{} $\sigma^j$ on the propagation of normalized transmembrane potential $v^\ICC$. The first row shows the case of a homogeneous distribution, the second row the case of a spatially varying distribution. The colorbar applies to all subplots.}  \label{fig:influence_of_diffusion_coefficient}
\end{figure}
\begin{figure}[!h]
    \centering    \includegraphics[width=0.8\textwidth]{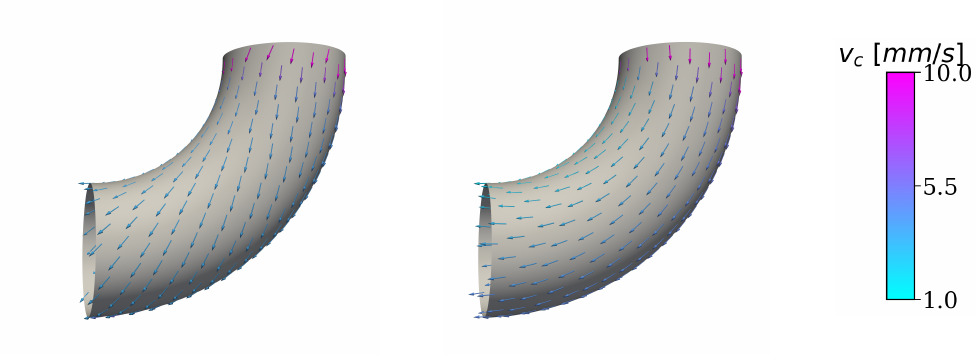}
     \caption{Estimated conduction velocity $v_c$ and propagation patterns of entrained \icc{} slow waves obtained with the triangulation method. Left: Case with spatially homogeneous diffusion coefficients. Right: Case with spatially  heterogeneous diffusion coefficients. In both simulations, excitability is unidirectional. The heterogeneous case produces a spatial gradient in $v_c$, enabling wave propagation approximately orthogonal to the longitudinal axis.}%
    \label{fig:torus_cv}
\end{figure}
We compare electrophysiology simulations using two distributions of the \icc{} diffusion coefficient $\sigma^\ICC$: homogeneous and heterogeneous (as described in~\cref{sec:heterogenous_diffusivity} and depicted in~\cref{fig:torus_sigma_i}). In both cases, a unidirectional excitability parameter distribution $a^\ICC$ (see~\cref{tab:excitability_parameters} and~\cref{fig:torus_exci}) is used to isolate the effects of the diffusion coefficients. Used parameters are listed in~\cref{tab:baseline_parameters}.
\Cref{fig:influence_of_diffusion_coefficient} shows the entrained \icc{} transmembrane potential $v^\ICC$ during propagation at selected time points.
In the homogeneous case, slow waves do not emerge simultaneously at the pacemaker line despite uniform excitability, instead depolarization occurs first at the pacemaker region.
Ring-shaped waves then travel aborally, but the conduction velocity remains nearly constant over the whole domain.

When $\sigma^\ICC$ is heterogeneously distributed, the propagation follows the expected pattern: waves originate at the pacemaker and spread more rapidly in the circumferential direction. Initially, the excitability distribution guides propagation, but the curved geometry of the tissue increasingly shapes the wavefront. The slow wave fronts align approximately perpendicular to the longitudinal axis of the torus due to the higher conduction velocity at the greater curvature. 

\Cref{fig:torus_cv} depicts the resulting differences. Both setups initiate waves at the pacemaker, forming ring-shaped depolarization pattern. However, only the heterogeneous $\sigma^\ICC$ distribution produces a conduction velocity gradient sufficient to align wave fronts with geometry curvature. 
This demonstrates that modulating intrinsic frequencies through the excitability parameter alone is insufficient, and spatially varying diffusion coefficients are required to reproduce physiologically realistic slow wave propagation in curved geometries.

\section{Model parameters}
\begin{table}[H]
\centering
\caption{Parameters defining the spatial distribution of the excitability $a^\ICC$ according to~\cref{eq:Exci-a}. The functions $f_\eta$ for $\eta \in\{l,c\}$ are given by~\cref{eq:Exci_Gaussian}.
}
\label{tab:excitability_parameters}
\begin{tabular}{cccc}
\hline
\multicolumn{1}{c}{\textbf{Parameter}} &
\multicolumn{1}{c}{$\xi_l$} &
\multicolumn{2}{c}{$\xi_c$} \\
\cmidrule{3-4}
 & & 
\multicolumn{1}{c}{\textbf{Unidirectional}} &
\multicolumn{1}{c}{\textbf{Anisotropic}}\\
\hline
$b_\eta$ & \SI{3.1622}{} &\SI{3.1622}{}
   &\SI{0.1}{} \\
$c_\eta$ & \SI{0.0000}{} &\SI{1.0000}{}
   &\SI{0.9}{} \\
\hline
\end{tabular}
\end{table}%
%%%%%%%%%%%%%%
\begin{table}[H]
\centering
\caption{Baseline electromechanical model parameters used in all simulations unless stated otherwise.}
\label{tab:realistic_stomach_parameters}
\begin{tabular}{llll}
\hline
\multicolumn{1}{l}{\textbf{Name}} &
\multicolumn{1}{l}{\textbf{Parameter}} &
\multicolumn{1}{l}{\textbf{Value}} &
\multicolumn{1}{l}{\textbf{Unit}} \\
\hline
\multicolumn{1}{l}{\textit{\underline{Geometry:}}}\\
Initial thickness of tissue & $H_t$ &
  \SI{3.50000}{} &
  \SI{}{\mm} \\
\multicolumn{1}{l}{\textit{\underline{Material parameters:}}} \\
Collagen and \smcs{}: \\
\quad Mass fractions &$\xi^\cf= \xi^\lf$ &
  \SI{3.50000e-1}{} &
  -- \\
\quad Fiber prestretch &$\lambda_\textup{e}^\cf= \lambda_\textup{e}^\lf$ &
  \SI{1.10000}{} &
  -- \\  
\quad Fung exponential parameters:  & $k_1^\cf$ &
  \SI{3.12000e-1}{} &
\SI[per-mode = symbol]{}{\J\per \kg} \\
& $k_2^\cf$ &
  \SI{1.60780e1}{} &
  -- \\
& $k_1^\lf$ &
  \SI{8.46700}{} &
  \SI[per-mode = symbol]{}{\J\per \kg} \\
& $k_2^\lf$ &
  \SI{3.14300}{} &
  -- \\
Ground matrix: \\
\quad Mass fraction &$\xi^\gm$ &
  \SI{3.00000e-1}{} &
  -- \\
\quad  Shear modulus & $\mu^\gm$ &
  \SI{7.00000e1}{} &
  \SI[per-mode = symbol]{}{\J\per \kg} \\
 \quad  Poisson's ratio & $\nu^\gm$ & 
  \SI{0.499}{} &
  \SI{}{} \\
Longitudinal intensity of active contraction& $\alpha_\lf$ &
  $\{$\SI{0.0}{}-\SI{0.6}{}$\}$ &
  -- \\
Circumferential intensity of active contraction& $\alpha_\cf$ &
  $\{$\SI{0.0}{}-\SI{0.6}{}$\}$ &
  -- \\
\hline
\end{tabular}
\end{table}
%%%%%%%%%%%%%%%%%
\begin{table}[!h]
\centering
\caption{Modified parameter values used in the convergence study. All other parameters are identical to those in~\cref{tab:realistic_stomach_parameters}}
\label{tab:parameters_cylinder_hcm}
\begin{tabular}{ccccc}
\hline
$\chi^\ICC=\chi^\SMC$ [--]  & $\sigma^\ICC$  [\si{\mm^2\per\s}]& $\sigma^\SMC$  [\si{\mm^2\per\s}]& $\alpha_\lf=\alpha_\cf$ [--] & $p$ [\si{\mmHg}] \\
\hline
 \SI{1.0}{} & \SI{1.2}{} & \SI{0.12}{} & \SI{0.5}{} & 25.0 \\
\hline
\end{tabular}
\end{table}
%%%%%%%%%%%%%%
\begin{table}[H]
\centering
\caption{Model parameters used for the comparison of finite element formulations.}
\label{tab:parameters_cylinder}
\begin{tabular}{ccccccc}
\hline
 $\chi^\ICC=\chi^\SMC$ [--] & $\sigma^\ICC$ [\si{\mm^2\per\s}] & $\sigma^\SMC$ [\si{\mm^2\per\s}] & $\alpha_\lf=\alpha_\cf$ [--] & $p$ [\si{\N\per\mm^2}]& $\mu$  [\si{\N\per\mm^2}]& $\nu$ [--] \\
\hline
\SI{1.0}{} & \SI{1.2}{} & \SI{0.12}{} & \SI{0.4}{} & \SI{3.0}{} & \SI{60.0}{} & \SI{0.499}{} \\
\hline
\end{tabular}
\end{table}
%%%%%%%%%%%%%%%%%%%%
\begin{table}[H]
\centering
\caption{Initial conditions values used for the electromechanics simulations.
}
\label{tab:inital-conditions}
\begin{tabular}{llll}
\hline
\multicolumn{1}{l}{\textbf{Name}} &
\multicolumn{1}{l}{\textbf{Parameter}} &
\multicolumn{1}{l}{\textbf{Value}} &
\multicolumn{1}{l}{\textbf{Unit}} \\
\hline 
 Transmembrane voltage of \icc{} and \smcs{} & $v^j(t=0)$ &
 \SI{3.41438e-2}{}&
  -- \\
 Recovery variable of \icc{} and \smcs{} & $h^j(t=0)$ &
 \SI{6.78747e-1}{}&
  -- \\
  \hline
\end{tabular}
\end{table}
%%%%%%%%%%%%%%%%
\begin{table}[H]
\centering
\caption{Boundary conditions used for the electromechanics simulations of the realistic stomach. The values correspond to the description in~\cref{sec:boundary_conditions}.
}
\label{tab:boundary-conditions}
\begin{tabular}{llll}
\hline
\multicolumn{1}{l}{\textbf{Name}} &
\multicolumn{1}{l}{\textbf{Parameter}} &
\multicolumn{1}{l}{\textbf{Value}} &
\multicolumn{1}{l}{\textbf{Unit}} \\
\hline 
Intraluminar pressure values:&  &
    &
    \\
\quad
Distal stomach & $p_\textup{distal}$ &
  \SI{25.0000}{} &
  \SI{}{\mmHg} \\
\quad
Proximal stomach & $p_\textup{prox}$ &
  \SI{10.0000}{} &
  \SI{}{\mmHg} \\
Spring stiffness values:& &
   &
     \\
\quad Pyloric sphincter & $k_\textup{pyl}$ &
  \SI{3.00000}{}&
  \SI[per-mode = symbol]{}{\kPa\per\mm}  \\
\quad Greater curvature & $k_\textup{greater}$ &
  \SI{2.00000e-3}{} &
  \SI[per-mode = symbol]{}{\kPa \per \mm}  \\
\quad Lesser curvature  & $k_\textup{lesser}$ &
  \SI{2.00000e-3}{} &
  \SI[per-mode = symbol]{}{\kPa\per\mm}  \\  
\quad Remaining stomach surface & $k$ &
  \SI{2.00000e-5}{} &
  \SI[per-mode = symbol]{}{\kPa\per\mm}  \\
  \hline
\end{tabular}
\end{table}
%%%%%%
\begin{table}[H]
\centering
\caption{Simulation parameter used to study gastric electromechanics.}
\label{tab:baseline_parameters}
\begin{tabular}{llll}
\hline
\multicolumn{1}{l}{\textbf{Name}} &
\multicolumn{1}{l}{\textbf{Parameter}} &
\multicolumn{1}{l}{\textbf{Value}} &
\multicolumn{1}{l}{\textbf{Unit}} \\
\hline
\multicolumn{1}{l}{\textit{\underline{ICC:}}}\\
Maximum primary excitability  & $a^\ICC_\textup{max}$ &
  \SI{3.31600e-2}{} &
  -- \\
Minimum primary excitability (distal) & $a^{\ICC,\textup{distal}}_\textup{min}$ &
  \SI{2.23075e-2}{} &
  -- \\  
Minimum primary excitability (proximal) & $a^{\ICC,\textup{prox}}_\textup{min}$ &
  \SI{0.00000}{} &
  -- \\    
Secondary excitability parameter & $\lambda^\ICC$ &
  \SI{1.25000e-2}{} &
  -- \\
Time constant of inward current& $\tau^\ICC_\textup{in}$ &
  \SI{2.29274e-2}{} &
  \SI{}{\s} \\
Time constant of outward current & $\tau^\ICC_\textup{out}$ & \SI{4.70719e-1}{} &
  \SI{}{\s} \\
Time constant of gate opening & $\tau^\ICC_\textup{open}$ &
  \SI{9.23200}{} &
  \SI{}{\s} \\
Time constant of gate closing & $\tau^\ICC_\textup{close}$ &
  \SI{4.77082}{} &
  \SI{}{\s} \\
Gating voltage & $v^\ICC_\textup{gate}$ &
  \SI{1.03825e-1}{}  &
  -- \\
Numerical smoothing parameter & $\eta^\ICC_\textup{gate}$ &
  \SI{4.50362e-2}{} &
  -- \\ 
  Minimum diffusion coefficient of ICC& $\sigma_\textup{lesser}^\ICC$ &
  \SI{6.00000e-1}{} &
  \SI[per-mode = symbol]{}{\mm^2\per \s} \\  
Slope in velocity–diffusion relation & $m_\sigma$ & \SI{1.62000}{}  & \SI[per-mode = symbol]{}{\mm\per \s} \\
Intercept in velocity–diffusion relation & $c_\sigma$ & \SI{2.15500}{}  & \SI[per-mode = symbol]{}{\mm^2\per \s} \\
\multicolumn{1}{l}{\textit{\underline{SMC:}}} \\
Excitability parameter & $a^\SMC$ &
  \SI{0.00000}{} &
  -- \\
Secondary excitability parameter & $\lambda^\SMC$ &
  \SI{1.25000e-2}{} &
  -- \\
Diffusion coefficient of \smc{}& $\sigma^\SMC$ &
  \SI{0.1}{}$\sigma^\ICC$ &
  \SI[per-mode = symbol]{}{\mm^2\per \s} \\  
Time constant of inward current & $\tau^\SMC_\textup{in}$ &
  \SI{1.14637e-1}{} &
  \SI{}{\s} \\
Time constant of outward current & $\tau^\SMC_\textup{out}$ &
  \SI{4.70719e-1}{} &
  \SI{}{\s} \\
Time constant of gate opening & $\tau^\SMC_\textup{open}$ &
  \SI{9.23200}{} &
  \SI{}{\s} \\
Time constant of gate closing & $\tau^\SMC_\textup{close}$ &
  \SI{4.77082}{} &
  \SI{}{\s} \\
Gating voltage & $v^\SMC_\textup{gate}$ &
  \SI{1.03825e-1}{} &
  -- \\
 Numerical smoothing parameter & $\eta^\SMC_\textup{gate}$ &
  \SI{4.50362e-2}{} &
  -- \\ 
\multicolumn{1}{l}{\textit{\underline{Tissue electrophysiology:}}} \\
Resistance of gap junctions & $D_{\text{gap}}$ &
  \SI{5.00000e-1}{} &
  \SI{}{1\per\s} \\
Circumferential bound of~\cref{eq:Exci_Gaussian} (proximal) & $c_\lf^\textup{prox}$ &
  \SI{9.00000e-1}{} &
  -- \\
Circumferential spread of~\cref{eq:Exci_Gaussian} (proximal) & $b_\cf^\textup{prox}$ &
  \SI{1.00000e1}{} &
  -- \\
Longitudinal bound of~\cref{eq:Exci_Gaussian} (proximal) & $c_\lf^\textup{prox}$ &
  \SI{0.00000}{} &
  -- \\
Longitudinal spread of~\cref{eq:Exci_Gaussian} (proximal)  & $b_\lf^\textup{prox}$ &
\SI{3.16220e2}{} &
-- \\
\multicolumn{1}{l}{\textit{\underline{Electromechanics:}}} \\
Calcium dynamics &$\beta_1$ &
  \SI{1.00000e1}{} &
  -- \\
Opening dynamics of VDCC & $\beta_2$ &
 \SI{1.00000e1}{}&
  -- \\
Threshold activation parameter of VDCC & $v^\textup{thr}$ &
 \SI{2.50000e-1}{}&
  -- \\
\hline
\end{tabular}
\end{table}
%%%%%%%%%%%%%%%%%%%

\clearpage
\bibliographystyle{elsarticle-num} 
\bibliography{literature}

\end{document}